\documentclass[ ]{elsarticle}
\usepackage[utf8]{inputenc}
\usepackage{amsmath}
\usepackage{amsfonts}
\usepackage{amssymb}
\usepackage{indentfirst}
\usepackage{graphicx}
\usepackage[outdir=figures/]{epstopdf}

\graphicspath{{figures/}}
\usepackage{multirow}
\usepackage{subcaption}

\usepackage{cleveref}
\title{ A dynamically load-balanced parallel $ p $-adaptive implicit  high-order flux reconstruction method for under-resolved turbulence simulation }
\author[1]{Lai Wang}
\author[2]{Matthias K. Gobbert}
\author[1]{Meilin Yu}
\address[1]{Department of Mechanical Engineering, University of Maryland, Baltimore County}
\address[2]{Department of Mathematics and Statistics, University of Maryland, Baltimore County }

\begin{document}
\maketitle
\section*{Abstract}
We present a dynamically load-balanced parallel  $ p $-adaptive implicit high-order flux reconstruction method for under-resolved turbulence simulation. The high-order explicit first stage, singly diagonal implicit Runge--Kutta (ESDIRK) method is employed to circumvent the restriction on the time step size. The pseudo transient continuation is coupled with the matrix-free restarted generalized minimal residual (GMRES) method to solve the nonlinear equations at each stage, except the first one,  of ESDIRK. We use the spectral decay smoothness indicator as the refinement/coarsening indicator for $ p $-adaptation. A dynamic load balancing technique is developed with the aid of the open-source library ParMETIS. The trivial cost, compared to implicit time stepping, of mesh repartitioning and data redistribution enables us to conduct $ p $-adaptation and load  balancing every time step. An isentropic vortex propagation case is employed to study the impact of element weights used in  mesh repartitioning on parallel efficiency.  We apply the
$ p$-adaptive solver for implicit large eddy simulation (ILES) of the
transitional flows over a cylinder when Reynolds number (Re) is $ 3900 $ and 
the SD7003 wing when Re is $ 60000 $. Numerical experiments demonstrate that a significant reduction in the run time (up to 70\%) and total number of solution points (up to 76\%) can be achieved with $ p $-adaptation. 

\section*{Keywords}
Dynamic load balancing; dynamic $ p $-adaptation; implicit large eddy simulation; implicit high-order flux reconstruction; matrix-free GMRES.
\section{Introduction}
Recent decades have witnessed tremendous developments in high-order computational fluid dynamics (CFD) methods, such as discontinuous Galerkin methods (DG)~\cite{CockburnShu89,cockburn1990runge,bassi1997high,cockburn2001runge, bassi2005discontinuous,NDG08,gassner2013accuracy,uranga2011implicit}, spectral difference methods (SD)~\cite{liu2006spectral,zhou2010implicit,castonguay2010simulation,Yu2011}, classic compact finite difference methods~\cite{galbraith2008implicit,garmann2013comparative}, finite difference summation by parts (SBP) operators~\cite{boom2013time,carpenter2014entropy,svard2014review}, and flux reconstruction/correction procedure via reconstruction methods (FR/CPR)~\cite{huynh2007flux,huynh2009reconstruction,wang2009unifying,vincent2011new, romero2016simplified, wang2017compact,witherden2014pyfr}. Low-dissipation and low-dispersion properties of high-order methods have made them attractive for implicit large eddy simulation (ILES) of turbulent flows. 
It has been reported in~\cite{fernandez2017subgrid} that DG-based ILES can outperform subgrid-model-based LES for transitional flows and wall bounded flows.  The dissipation of high-order methods on low-frequency large-scale flow features is trivial, and it is only significant on high wavenumbers/frequencies. Therefore, the truncation error of high-order methods is considered as an implicit subgrid model for turbulence simulation.
We note that high-order methods are prone to suffer from instabilities due to aliasing errors in under-resolved turbulence simulation, especially when the spatial polynomial degree exceeds two. It has been shown that with proper de-aliasing techniques, ILES using high-order methods has promising capabilities in under-resolved turbulence simulation~\cite{gassner2013accuracy}. Four popular types of stabilization approaches for under-resolved turbulence simulation can be found in the literature, including (a) the split form~\cite{gassner2013skew,gassner2016split}, (b) over integration~\cite{kirby2003aliasing,mengaldo2015dealiasing}, (c) artificial viscosity via spectral vanishing viscosity~\cite{tadmor1989convergence,karamanos2000spectral,pasquetti2006spectral}, and (d) polynomial filtering~\cite{gottlieb2001spectral,hesthaven2008filtering,fischer2001filter}. 
In this study, we employ nodal polynomial filtering proposed  by Fisher and Mullen~\cite{fischer2001filter} when de-aliasing is needed.

A uniformly high-order spatial discretization in the entire flow field for under-resolved turbulence simulation of certain problems, such as wall bounded turbulent flows at high Reynolds numbers, is very expensive. As a matter of fact, the high-order spatial discretization is only needed in the near wall region, such as the turbulent boundary layer, and the wake region where vortex shedding dominates. Collocation schemes favor a straightforward implementation of $ p $-adaption,  which has the potential to significantly decrease computational cost.  In the literature, three major groups of adaptation methodologies can be found. The first one is the feature-based adaptation~\cite{gassner2015space,tugnoli2017locally,naddei2018comparison}, the second one is the truncation-error-based or discretization-error-based adaptation~\cite{hartmann2002adaptive,gao2011residual,kompenhans2016comparisons}, and the last one is the output-based or adjoint-based adaptation~\cite{venditti2003anisotropic,wang2009adjoint,fidkowski2011review,fidkowski2011output}.
Feature-based adaptation methods 
are usually \textit{ad hoc} and heavily rely on empirical parameters; however, their ease of implementation and reasonable robustness make them a good choice for adaptation. Truncation-error based approaches usually use the correction from either an additional coarser mesh or a lower-order discretization to estimate the local discretization error, which can serve as the adaptation indicator. A comparison of several   feature-based and discretization-error-based adaptation indicators is conducted by Naddei et al.~\cite{naddei2018comparison}. 
Adjoint-based adaptation methods are popular for engineering purposes since engineers are more interested in output functionals, such as, lift and drag. Their superiority over the former two approaches has been demonstrated for steady problems. However, 
the computational cost of adjoint-based adaptation, especially for unsteady turbulence simulation, can be large.  Recently, Bassi et al.~\cite{bassi2019entropy} employed an efficient entropy-adjoint-based~\cite{fidkowski2010entropy} $ p $-adaptive DG solver to conduct scale-resolving turbulence simulation.  In this study, we will evaluate the performance of  a feature-based adaptation method~\cite{naddei2018comparison}, which employs the spectral decay smoothness indicator~\cite{persson2006sub} as the refinement/coarsening indicator when it is applied to under-resolved turbulence simulation.  

Explicit high-order Runge-Kutta (RK) methods~\cite{cockburn1989tvb, gottlieb2001strong} have been widely applied to unsteady flow simulation. 
$ p $-adaptation will naturally lead to $ p $-enrichment in near wall regions where the elements are usually clustered, thus worsening the Courant–Friedrichs–Lewy (CFL) condition when explicit methods are employed. Implicit time integrators can essentially circumvent the CFL restriction that explicit methods have. Diagonally implicit RK methods~\cite{kennedy2016diagonally}  and backward differentiation formula (BDF) methods are among the most popular implicit time integration methods. Recently, linearly implicit Rosenbrock methods have become popular for under-resolved turbulence simulation~\cite{bassi2015linearly,wang2019comparative}. Matrix-based implicit methods are notorious for the large memory consumption.  Therefore, the matrix-free implementation~\cite{ wang2019comparative, franciolini2017efficiency} is usually employed to   reduce memory usage~\cite{knoll2004jacobian} for massive turbulence simulation.
Though Rosenbrock methods can be potentially more efficient than ESDIRK methods~\cite{wang2019comparative}, we employ ESDIRK methods in this study since they are more robust than Rosenbrock methods in the context of matrix-free implementation with an element-Jacobi preconditioner.

Given that the time step size of an implicit time integrator can be relatively large, dynamic adaptivity is desired to track the rapid change of turbulence features. Consequently, the work loads on all processes in parallel simulation will be imbalanced once $ p $-adaptation takes place. The difference of numbers of degrees of freedom on different processes can be over 500\% for a simple isentropic vortex propagation problem~\cite{wang2019jacobian}. Hence, a dynamic load balancing technique is of crucial importance for the parallel efficiency of $ p $-adaptive methods. Existing publications regarding $ p $-adaptive high-order methods for turbulence simulation use the mean flow field to conduct the adaptation without dynamic adaptivity~\cite{tugnoli2017locally,bassi2019entropy}. 
We utilize the open-source library ParMETIS~\cite{karypis2011metis} to achieve dynamic load balancing for $ p $-adaptation. Technical details on  using ParMETIS for dynamic adaptivity are presented in Section~\ref{adaptation_hpc} for interested readers. In our numerical experiments, we find that the cost of mesh repartitioning and  data redistribution is trivial compared to that of implicit time stepping, which enables the $ p $-adaptation to be conducted every time step to significantly decrease the run time.

\textit{Contributions}. We develop a dynamically load balanced $ p $-adaptation technique for parallel implicit high-order flux reconstruction solution of unsteady Navier--Stokes equations. The implementation regarding the dynamic load balancing technique is presented in detail. We discuss the impact of weight calculation for each element on the parallel efficiency of $ p $-adaptation when implicit time integrators are used. The $ p $-adaptive solver is applied to  under-resolved turbulence simulation of the transitional flow over an infinite cylinder at $ \text{Re} = 3900 $ and  the transitional flow over the SD7003 wing at $ \text{Re}=60000 $. Compared to the $ p $-uniform spatial discretization, the $ p $-adaptive FR method can save up to $ 70\% $ computational cost   when $ p\le3 $ in our experiments and provide favorable numerical predictions. We expect more savings when $ p $ is higher as demonstrated in a simple experiment of the isentropic vortex propagation problem. 

\textit{Organization}. The remainder of the paper is organized as follows. Section~\ref{backgroud} provides the mathematical background of the governing equations, the spatial discretization,  and the time integration. Section~\ref{adaptation_hpc} introduces the $ p $-adaptation algorithm with the spectral decay smoothness indicator, and then explains the parallel mesh partitioning technique with ParMETIS. A simple example is employed to demonstrate the impact of weight calculation of each element on the parallel efficiency. Applications of the $ p $-adaptive solver to under-resolved turbulence simulation are presented in Section~\ref{numerical_results}. In Section~\ref{CFW}, we draw conclusions from the current work.
\section{Background} \label{backgroud}

\subsection{Governing equations} \label{governing_eqs}
Using Einstein summation convention, the compressible  Navier--Stokes equations can be written as 

\begin{equation}\label{mass}
\frac{\partial \rho}{\partial t} + \frac{\partial( \rho u_j)}{\partial x_j} = 0,
\end{equation}
\begin{equation}\label{momentum}
\frac{\partial (\rho u_i)}{\partial t} + \frac{\partial (\rho u_j u_i+\delta_{ji}p)}{\partial x_j} = \frac{\partial \tau_{ji}}{\partial x_j},
\end{equation}
\begin{equation}\label{energy}
\frac{\partial(\rho E) }{\partial t}+\frac{\partial(\rho  u_j H)}{\partial x_j} = \frac{\partial(u_i\tau_{ij}-K_j)}{\partial x_j},
\end{equation}
where $i= 1 ,\dots ,n_d$, and  $ n_d $ is the dimension number. Herein, $ \rho $ is the fluid density, $ u_i $ is the velocity component, $ p $ is the pressure, $  E=\frac{p/\rho}{\gamma-1}+\frac{1}{2} u_k u_k $ is the specific total energy, $ H=E+\frac{p}{\rho} $ is the specific total enthalpy, $ \tau_{ij} $ is the viscous stress and $ K_j $ is the heat flux. Note that only in this section, $ p $ refers to pressure; in other sections, $ p $ stands for polynomial degree. $\gamma$ is the specific heat ratio defined as $ \gamma = C_{p}/C_{v} $, where $ C_{p} $ and $ C_{v} $ are  specific heat capacity at constant pressure and volume, respectively. In this study, $\gamma$ is set as 1.4. The ideal gas law $ p=\rho RT $ holds, where $ R $ is the ideal gas constant and $ T $ is the temperature. The viscous stress tensor and heat flux vector are given by
\begin{equation}
\tau_{ij} = 2 \mu \left\{S_{ij}-\frac{1}{3}\frac{\partial u_k}{\partial x_k}\delta_{ij}\right\},
\end{equation}
\begin{equation}
K_j = -\frac{\mu C_p}{\text{Pr}}\frac{\partial T}{\partial x_j},
\end{equation}
where $ \mu  $ is the fluid dynamic viscosity, Pr is the molecular Prandtl number, and the strain-rate tensor $ S_{ij} $ is defined as 

\begin{equation}
S_{ij} = \frac{1}{2}\left(\frac{\partial u_i}{\partial x_j}+\frac{\partial u_j}{\partial x_i}\right).
\end{equation}	
In this study, $ \mu $ and Pr are treated as constants.

\subsection{The FR/CPR method}\label{flux_reconstruction}
For completeness, a brief review of the FR/CPR method~\cite{wang2009unifying} is presented in this section.
A symbolic form of the compressible Navier--Stokes equations~\eqref{mass}, \eqref{momentum}, \eqref{energy} is written as 
\begin{equation} \label{ns_equation_symbolic}
\frac{\partial \boldsymbol{q}}{\partial t} + \nabla \cdot \boldsymbol{f} = 0,
\end{equation}
which is defined in domain $ \Omega $. $ \Omega $ is partitioned into $ N $ non-overlapping and conforming elements $ \Omega_e $, where $ e=0,1,\ldots,N-1  $. 
After multiplying each side by the test function $ \boldsymbol{\vartheta} $ and integrate over $ \Omega_e$, one obtains
\begin{equation} \label{eqn:NS_weighted}
\int_{\Omega_e}\frac{\partial  \boldsymbol{q}_{e}}{\partial t}\,\boldsymbol{\vartheta }\,dV + \int_{\Omega_e}\boldsymbol{\vartheta}\,\nabla \cdot \boldsymbol{f}_e\,dV =0.
\end{equation}
On applying the integration by parts and divergence theorem, Eq.~\eqref{eqn:NS_weighted} reads

\begin{equation} \label{eqn:IBP_1}
\int_{\Omega_e}\frac{\partial \boldsymbol{q}_{e}}{\partial  t}\,\boldsymbol{\vartheta }\,dV + \int_{\partial \Omega_e} \boldsymbol{\vartheta}\,\boldsymbol{f}_e\cdot \boldsymbol{n}\, dS -\int_{\Omega_e}\boldsymbol{f}_e\cdot\nabla  \boldsymbol{\vartheta }\,dV=0,
\end{equation}
where $ \boldsymbol{n} $ is the outward-facing unit normal vector of the faces of the element $ \Omega_e $. In the discrete form, we assume $ \boldsymbol{q}_{e}^h $ is the approximate solution in element $ \Omega_e $. The solution and the test function belong to the polynomial space of degree $ k $, i.e., $  \boldsymbol{q}_{e}^h \in p^k $ and $  \boldsymbol{\vartheta}^h \in p^k $. To ensure conservation, $ \boldsymbol{f}_e\cdot \boldsymbol{n} $ in Eq.~\eqref{eqn:IBP_1} is replaced with $ \boldsymbol{f}^{com}_{\boldsymbol{n}} $, the common flux in the normal direction of the element surfaces. Eq.~\eqref{eqn:IBP_1} then reads 
\begin{equation}\label{numerical_ibp}
\int_{\Omega_e} \frac{\partial \boldsymbol{q}_{ e}^h}{\partial t}\,\boldsymbol{\vartheta }^h\,dV + \int_{\partial \Omega_e} \boldsymbol{\vartheta}^h\,\boldsymbol{f}^{com}_{\boldsymbol{n}}\, dS -\int_{\Omega_k}\boldsymbol{f}_e^h\cdot\nabla  \boldsymbol{\vartheta }^h\,dV=0.
\end{equation}
After applying integration by parts and divergence theorem again to the last term of Eq.~\eqref{numerical_ibp}, one obtains
 
\begin{equation}\label{penalty}
\int_{\Omega_e} \frac{\partial \boldsymbol{q}_{ e}^h}{\partial t}\,\boldsymbol{\vartheta}^h\,dV + \int_{\Omega_e}\boldsymbol{\vartheta}^h\,\nabla \cdot \boldsymbol{f}_e^h\,dV+ \int_{\partial \Omega_e} \boldsymbol{\vartheta}^h\,[ \boldsymbol{f} ]\, dS=0,
\end{equation}
where $ [\boldsymbol{f}]=\boldsymbol{f}^{com}_{\boldsymbol{n}} - \boldsymbol{f}^{loc}_{\boldsymbol{n}} $ with $ \boldsymbol{f}^{loc}_{\boldsymbol{n}} = \boldsymbol{f}_e^h\cdot \boldsymbol{n}$.
In the FR/CPR method, the correction field $ \boldsymbol{\delta}_e \in p^k $ is defined as~\cite{wang2009unifying}
\begin{equation}
\int_{\partial \Omega_e} \boldsymbol{\vartheta }^h\,[\boldsymbol{f} ]\, dS = \int_{\Omega_e} \boldsymbol{\vartheta }^h \, \boldsymbol{\delta}_e\,dV.
\end{equation}
Therefore, Eq.~\eqref{penalty} can be expressed as
\begin{equation}
\int_{\Omega_e}\left( \frac{\partial \boldsymbol{q}_{ e}^h}{\partial t} + \nabla\cdot\boldsymbol{f}_e^h+\boldsymbol{\delta}_e\right)\, \boldsymbol{\vartheta}^h\, dV =0.
\end{equation}
The differential form can then be employed as
\begin{equation} \label{eqn:FR_diff}
 \frac{\partial \boldsymbol{q}_{ e}^h}{\partial t} + \mathbb{P}\left(\nabla\cdot\boldsymbol{f}_e^h\right)+\boldsymbol{\delta}_e =0.
\end{equation}
Herein,  $ \mathbb{P} \left(\nabla\cdot\boldsymbol{f}_e^h\right)$ is the projection of the flux divergence $ \left(\nabla\cdot\boldsymbol{f}_e^h\right)$, which may not be a polynomial, onto an appropriate polynomial space. We note that Eq.~\eqref{eqn:FR_diff} can be directly derived from the differential form; their equivalence has been established in~\cite{yu2013connection}.
Specifically, for quadrilateral and hexahedral elements, the correction field can be obtained by means of the tensor product of the one dimensional correction polynomials; for triangular and tetrahedral elements, the readers are referred to~\cite{wang2009unifying,williams2013energy}. Only hexahedral elements are considered in this study. 

The Roe approximate Riemann solver~\cite{roe1981approximate} is used to calculate the common inviscid fluxes at the cell interfaces in their normal directions as
\begin{equation}\label{invis_com}
\boldsymbol{f}_{\boldsymbol{n},inv}^{com} = \frac{\boldsymbol{f}_{\boldsymbol{n},inv}^{+}+\boldsymbol{f}_{\boldsymbol{n},inv}^{-}}{2}- \boldsymbol{R}|\boldsymbol{\varLambda}|\boldsymbol{R}^{-1}\frac{\boldsymbol{q}^{+} -\boldsymbol{q}^{-} }{2},
\end{equation}
where superscripts `$-$' and `$+$' denote the left of right side of the current interface, the subscript $ \boldsymbol{n} $ is the unit normal direction from left to right, $\boldsymbol{\varLambda}$ is a diagonal matrix consisting of the eigenvalues of the preconditioned Jacobian $\partial \boldsymbol{f}_{\boldsymbol{n}} /  \partial \boldsymbol{q}$, and $\boldsymbol{R}$ consists of the corresponding right eigenvectors evaluated with the averaged values. 
The common viscous fluxes at the cell interfaces are $ \boldsymbol{f}_{n,vis}^{com} = \boldsymbol{f}_{vis}(\boldsymbol{q}^{+},\nabla \boldsymbol{q}^{+} ,\boldsymbol{q}^{-},\nabla \boldsymbol{q}^{-} ) $. Here, we need to define the common solution $ \boldsymbol{q}^{com}  $ and common gradient $ \nabla \boldsymbol{q}^{com} $ at the cell interface. 
On simply taking average of the primitive variables, we get 
\begin{equation}\label{com_q}
\boldsymbol{q}^{com} = \frac{\boldsymbol{q}^{+}+\boldsymbol{q}^{-}}{2}.
\end{equation}
The common gradient is computed as 
\begin{equation}\label{com_grad_q}
\nabla \boldsymbol{q}^{com} = \frac{\nabla\boldsymbol{q}^{+}+\boldsymbol{r}^{+}+\nabla\boldsymbol{q}^{-}+\boldsymbol{r}^{-}}{2},
\end{equation}
where $ \boldsymbol{r}^{+} $ and $ \boldsymbol{r}^{-} $ are the corrections to the gradients on the interface. The second
approach of Bassi and Rebay (BR2)~\cite{bassi2005discontinuous} is used to calculate the corrections. 

\subsection{ESDIRK methods with pseudo transient continuation}\label{ESDIRK_methods}

The ESDIRK methods for the 3D compressible Navier-Stokes equation~\eqref{ns_equation_symbolic} can be written as 
\begin{equation}\label{ESDIRK_ns}
\begin{cases}
\boldsymbol{q}^{n+1} = \boldsymbol{q}^{n}+\Delta t\,
\sum_{i=1}^{s}b_i\boldsymbol{R}(\boldsymbol{q}^i),\\
\boldsymbol{q}^i = \boldsymbol{q}^n, i=1,\\
\boldsymbol{q}^i = \Delta t\, \omega\, \boldsymbol{R}(\boldsymbol{q}^i)+\boldsymbol{q}^n+\Delta t\, \sum_{j=1}^{i-1}a_{ij}\boldsymbol{R}(\boldsymbol{q}^j),i=2,\dots,s,
\end{cases}	
\end{equation}
where $ i$ is the stage number, $ s $ is number of total stages, $ n $ denotes the physical time step, and $ \boldsymbol{R} = -\nabla\cdot\boldsymbol{f} $. The second-order, three-stage ESDIRK2~\cite{kennedy2016diagonally}, and fourth-order, six-stage ESDIRK4~\cite{bijl2002implicit} methods are employed in this paper. A comparative study of different implicit time integration methods can be found in~\cite{wang2019comparative}. In every stage except the first one, a nonlinear system is to be solved, which can be expressed as 	
\begin{equation}\label{ESDIRK_conservative}
\boldsymbol{F}(\boldsymbol{q}^{i}) = \left(-\frac{1}{\omega\,\Delta t}\boldsymbol{q}^i +\boldsymbol{R}(\boldsymbol{q}^i)\right)+\frac{1}{\omega\,\Delta t}\left(\boldsymbol{q}^n+\Delta t\sum_{j=1}^{i-1}a_{ij}\boldsymbol{R}(\boldsymbol{q}^j)\right),\,i=2,\dots,s.
\end{equation}

The pseudo transient continuation for the $ i $-th stage reads

\begin{equation}\label{pseudo_transient}
\frac{\boldsymbol{q}^{m+1,i}-\boldsymbol{q}^{m,i}}{\Delta \tau}=\boldsymbol{F}(\boldsymbol{q}^{m+1,i}),
\end{equation}
where $ m $ is the iteration step. 
Eq.~\eqref{pseudo_transient} can be linearized as 
\begin{equation}\label{linearized_pseudo_transient_final}
\left(\frac{1}{\omega \,\Delta t}+\frac{1}{\Delta \tau}-\frac{\partial \boldsymbol{R}}{\partial\boldsymbol{q} }\right)^m\Delta \boldsymbol{q}^{m,i} = \boldsymbol{F}(\boldsymbol{q}^{m,i} ),
\end{equation}
where $ \Delta \boldsymbol{q}^{m,i}  = \boldsymbol{q}^{m+1,i} -\boldsymbol{q}^{m,i}  $.
We employ the successive evolution relaxation (SER) algorithm~\cite{mulder1985experiments} to update the pseudo time step size as  
\begin{equation}\label{ser}
\Delta \tau ^{0} = \Delta \tau_{init},\ \text{and}\
\Delta \tau^{m+1} = \min\left(\Delta \tau^{m}\frac{||\boldsymbol{F}||_{L_2}^{m-1}}{||\boldsymbol{F}||_{L_2}^{m}}, \Delta \tau_{max} \right).
\end{equation}
In all the numerical experiments conducted in this study, we set the convergence tolerance of relative residual of the pseudo transient continuation as $ tol_{rel}^{pseudo} = 10^{-4} $.

We use the restarted GMRES framework in the  portable, extensible toolkit for
scientific computation (PETSc) package~\cite{petsc-user-ref} with user-defined functions to do the matrix-vector product and preconditioning. In Krylov subspace methods, the Jacobian matrix only appears in the matrix-vector product. A finite difference approximation of  the matrix-vector product reads
\begin{equation}
\left(\frac{\partial \boldsymbol{R}}{\partial\boldsymbol{q}}\right)\boldsymbol{X} = \frac{\boldsymbol{R}(\boldsymbol{q}+\varepsilon \boldsymbol{X})-\boldsymbol{R}(\boldsymbol{q})}{\varepsilon}+O(\varepsilon),
\end{equation}
 where $ \varepsilon = 10^{-6} $ in this study. The element-Jacobi preconditioner, i.e., the inverse of the diagonal blocks of $ \left(\frac{1}{\omega \Delta t}+\frac{1}{\Delta \tau_{max}}-\frac{\partial \boldsymbol{R}}{\partial\boldsymbol{q} }\right) $, is used for left preconditioning in this study. The preconditioner is only evaluated once at the starting stage of each physical time stepping.

The pseudo transient continuation is an inexact Newton's method. Therefore, we assign a relatively large tolerance to the GMRES solver, i.e., $ tol_{rel}^{gmres} = 10^{-1}$, to save computational cost~\cite{wang2019comparative}. However, for stiff problems, $ tol_{rel}^{gmres} = 10^{-1}$ may lead to divergence of the pseudo transient continuation~\cite{wang2019implicit}. In this case, we will decrease $ tol_{rel}^{gmres} $ to  $ 10^{-2} $. If not specifically mentioned, the restart number is 60 for all numerical experiments.
We note that the performance of the element-Jacobi preconditioner will quickly deteriorate as $ \Delta \tau $ increases to large values $ \Delta \tau \gg \Delta t $. 
Therefore, in the pseudo transient continuation, we do not increase the pseudo time step $ \Delta \tau $ to large values to ensure that the  relative tolerance of GMRES $ tol_{rel}^{gmres} $ can always be met within 100 iterations. Otherwise, we will decrease the current pseudo time step size by half and redo the current pseudo-time iteration. In this study, we set $ \Delta \tau_{max} $ as $ \Delta \tau_{max} = O(\Delta t) $ and $ \Delta \tau_{min} $ is usually one magnitude smaller than $ \Delta t $. The authors have developed a $ p $-multigrid solver for coarsely-resolving simulation of  low-Mach-number turbulent flows in~\cite{wang2019implicitpmultigrid}. Applying the $ p $-multigrid solver as a preconditioner for Newton-Krylov method will be our future work. 

\section{Dynamically load-balanced $ p $-adaptation for high performance computing}\label{adaptation_hpc}
\subsection{$ p $-adaptation using spectral decay smoothness indicator}\label{adaptation}
The spectral decay smoothness indicator has been successful used to detect trouble cells for shock-capturing~\cite{persson2006sub}. It is defined as 
\begin{equation}
\eta_k = \frac{\left\|s_p-s_{p-1}\right\|_{L^2}}{\left\|s_p\right\|_{L^2}}
\end{equation}
for one element. $ \left\|\cdot\right\|_{L^2} $ is defined as 
\begin{equation}
 \left\|\cdot\right\|_{L^2} = \frac{\sum_{\xi = 1}^{p+1}\sum_{\eta= 1}^{p+1}\sum_{\zeta = 1}^{p+1}\left[(\cdot)^2\omega_\xi \omega_\eta \omega_\zeta |\boldsymbol{J}|_{\xi,\eta,\zeta} \right]}{\sum_{\xi = 1}^{p+1}\sum_{\eta= 1}^{p+1}\sum_{\zeta = 1}^{p+1}\left[\omega_\xi \omega_\eta \omega_\zeta |\boldsymbol{J}|_{\xi,\eta,\zeta} \right]}
\end{equation}
for a hexahedral element, where $ \boldsymbol{J} $ is the Jacobian matrix of the coordinate transformation from a physical element to the standard element, $ |\boldsymbol{J}| $ is the determinant of $ \boldsymbol{J} $, and $ \omega_{\xi/\eta/\zeta} $ are the quadrature weights in the $ \xi/\eta/\zeta $ directions, respectively. $ s_{p-1} $ is obtained by projecting the solution from the degree $ p $ polynomial space to the degree $ p-1 $ space. 
With the spectral decay smoothness indicator,
the adaptation procedure can be achieved as follows:
\begin{itemize}
	\item calculate the smoothness indicator of every element;
	\item adjust the polynomial degree of every element according to the adaptation criteria;
	\item limit the difference of polynomial degrees at non-conforming interfaces to one;
	\item project or prolong the solutions when the polynomial is decreased or increased, respectively.
\end{itemize}
The adaptation criteria we employ in the present study are organized as follows:
\begin{itemize}
	\item increasing the polynomial degree by one when $ \eta_k>\nu_{max}\eta_{k,max} $;
	\item decreasing the polynomial degree by one when $ \eta_k<\nu_{min}\eta_{k,max} $.
\end{itemize}
Herein, $ \nu_{max} $ and $ \nu_{min} $ are problem dependent.  The polynomial degree $ p $ of an element in the flow field is  $ p\in[p_{min},p_{max}] $, where $ p_{min} $ and $ p_{max} $ are the minimum and maximum polynomial degree, respectively. If not specifically mentioned, we choose momentum in the $ x $ direction, i.e., $ \rho u $, as the variable for smoothness indicator calculation; $\nu_{max} = 0.1  $ and $ \nu_{min} = 0.001 $. In this study, all adaptive solvers will have $ p_{min} = 1 $. When the description adaptive $ p^k $ FR or $ p^k $ FR with $ p $-adaptation is used, we are referring to the adaptive FR method with  $ p_{min} = 1 $ and $ p_{max} = k $. Some preliminary results using the current $ p $-adaptation method to solve 2D unsteady Navier--Stokes equations have been presented in~\cite{wang2019jacobian}. Since no dynamic load balancing was employed there, the differences of the  numbers of degrees of freedom on different processors in parallel simulation can be over 500\% for a simple isentropic vortex propagation problem. In this study, we propose to develop a dynamic load balancing strategy for parallel simulation with $ p $-adaptation.

\subsection{Implementation of parallel mesh partitioning}
\begin{figure}		
	\centering

	\includegraphics[width=\textwidth]{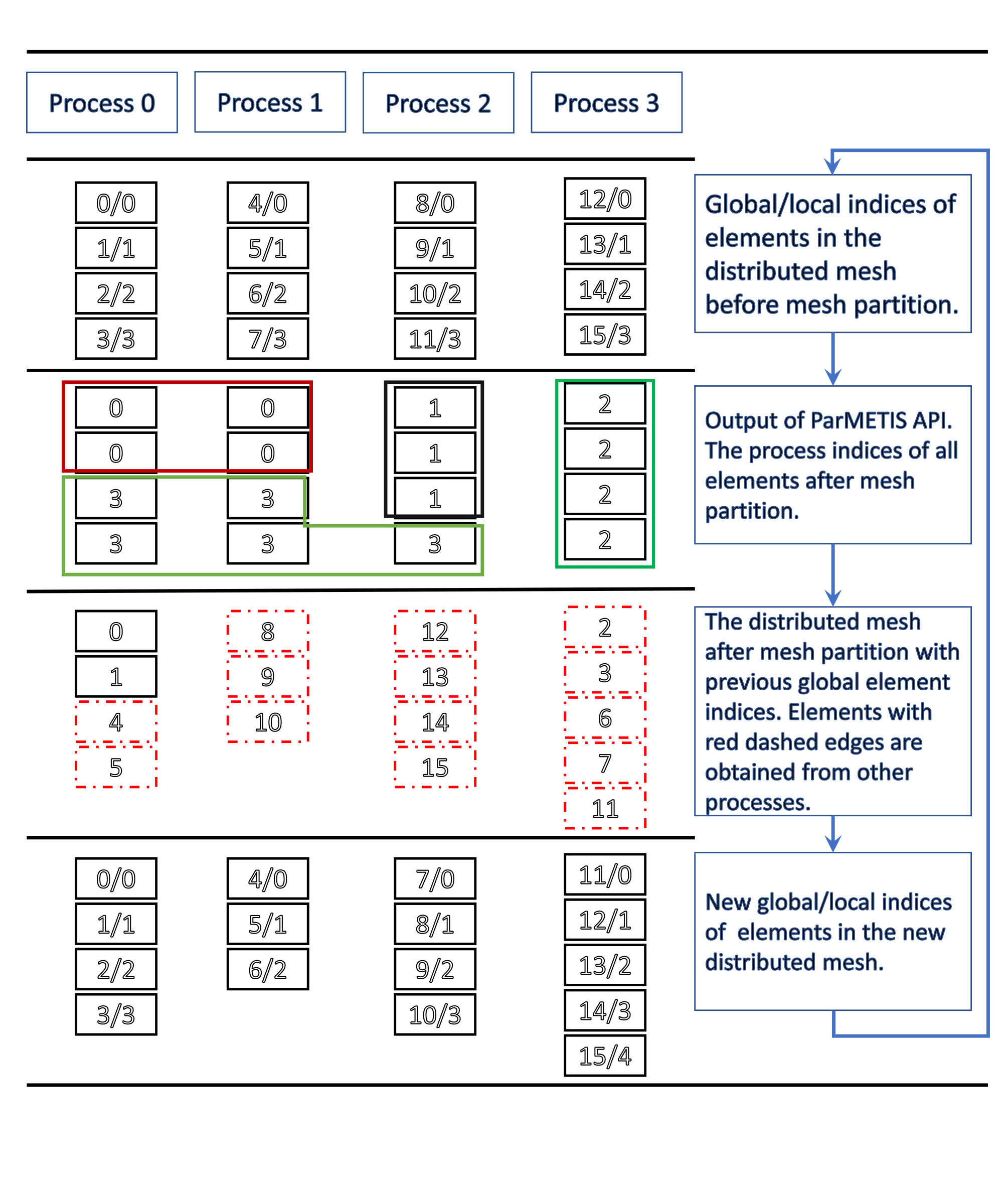}

	\caption{Illustration of the parallel mesh partitioning using the ParMETIS API.}	
	\label{parmetis_illustration}	
\end{figure}

To achieve dynamic load balancing, \verb|ParMetis_V3_AdaptiveRepart()| 
in the open source library ParMETIS~\cite{karypis2011metis} is employed for efficient parallel mesh partitioning.
This application programming interface (API) is particularly developed to repartition locally adapted mesh in parallel computing.  It allows one to use $ n_{proc} $ processes to partition the mesh into $ n_{part} $ parts. In this work, $ n_{proc} = n_{part} $ is used to assure the load of repartitioning is balanced among all processes. A distributed mesh is required as one of the inputs of \verb|ParMetis_V3_AdaptiveRepart()|. We employ \verb|METIS_PartMeshDual()| in serial  METIS~\cite{karypis2011metis} to partition the mesh to obtain the initial distributed mesh and no weights are assigned to any elements. 
On using \verb|ParMetis_V3_AdaptiveRepart()|, each element of the unstructured mesh is regarded as a vertex in the graph.
An illustration of the parallel mesh partitioning is presented in Figure~\ref{parmetis_illustration} to explain the technical details.
Following  C++ convention, all indices start from 0. There are $ n_{proc} $ processes and each process possesses one mesh partition. Assume the $ i $-th process has $ n_{i}^e $ elements. The global index of the $ j $-th local element in the $ i $-th process must be $ \text{Index}(j,i)= \sum_{m=0}^{i-1}n_m^e+j $ to ensure that the distributed mesh is a legal input of \verb|ParMetis_V3_AdaptiveRepart()|.
One output of \verb|ParMetis_V3_AdaptiveRepart()| is an array of size $ n_{i}^e $ which stores the process indices of the local elements after parallel mesh partitioning. As shown in the second row in Figure~\ref{parmetis_illustration}, before the parallel mesh partitioning, each process has four elements. Process 0 has Elements 0 to 3, Process 1 has Elements 4 to 7, etc. After parallel mesh parititioning, the index of  the process that an element belongs to is stored locally.  Due to change of element weights resulting from $ p $-adaptation, elements could appear to be `randomly' distributed to all processes. In other words, some processes will possess a part of the elements that they have before partitioning and some will obtain all elements from other processes. As shown in the fourth row in Figure~\ref{parmetis_illustration}, Process 0 has four elements and two of them,   Elements 4 and 5, are obtained from Process 1. Process 3 needs to fetch Elements 2, 3 from Process 0, Elements 6, 7 from Process 1, and Element 11 from Process 2. To make the new distributed mesh as a legal input for \verb|ParMetis_V3_AdaptiveRepart()| in the next parallel mesh partitioning, one needs to reorganize the global element indices as illustrated in the last row of Figure~\ref{parmetis_illustration}. Corresponding CFD data should also be reorganized following the mapping between the old and new global element indices. 

For data redistribution, we use a collect-and-distribute strategy.
We utilize \verb|MPI_Allgather()|   to gather all the conservative variables on all the processes and each process will fetch the  corresponding working variables from the collected data pool. The aforementioned randomness of  redistributing elements to all processes leads to the fact that when it comes to data redistribution, the elements on many of the processes could be totally different from those before the parallel mesh partitioning. This implies that the cost of a process-to-process communication strategy to exchange CFD data could possibly be close to that of the collect-and-distribute stategy. Besides, the coding complexity of the process-to-process strategy is overwhelming.  In our numerical experiments, we have found that the total amount of run time needed for mesh partitioning and data redistribution is trivial when compared to that of implicit time-stepping. We would like to clarify that in the spatial and temporal solvers, communication among all processes is done in a process-to-process manner to maximize efficiency.

An important input of  \verb|ParMetis_V3_AdaptiveRepart()| is the weight of each element in the distributed mesh.
For the FR discretization, the number of solution points $ n_{sp} $ within a element is $ (p+1)^3 $ for a hexahedral element. And there are five equations at each solution point to be solved in three dimensional problems. Hence, the number of degrees of freedom in one element is $ n_{dof} = 5n_{sp} $. For a hexahedral element, all operations of the FR/CPR methods are indeed conducted dimension by dimension. Therefore, we roughly estimate the computational complexity of one-time residual evaluation as $ O\left(n_{dof} \cdot (p+1)\right) $. When implicit time integrators are employed, the cost of one-time residual evaluation is trivial compared to three major parts, (a)~evaluating the element Jacobian matrix,(b)~calculating the element-Jacobi preconditioner and (c)~solving the nonlinear/linear equations using Newton/Krylov methods.
When evaluating the element Jacobian matrix, the finite difference approach is used. For each element, there will be $ n_{dof} $ times the residual evaluation and the computational complexity is $ O(n_{dof}^2 \cdot (p+1)) $. We use lower-upper (LU) decomposition to invert the element Jacobian matrix to obtain the element-Jacobi preconditioner.  The computational complexity of is
$ O(n_{dof}^3) $.
In the matrix-free implementation of the GMRES solver, the approximation of  matrix-vector product  and preconditioning will contribute to the computational cost dominantly. For the matrix-vector product  approximation, the computational cost  will be that of one-time residual evaluation. The left preconditioning is used in our approach and the complexity  is $O(n_{dof}^2)$.  Overall, the complexity of the Newton-Krylov solver is $ G(O(n_{dof}^2)+  O\left(n_{dof} \cdot (p+1)\right) $, where $ G $ is the total number of  GMRES iterations in the pseudo transient continuation. $ G $ is highly problem dependent. Thus, three candidates to calculate the weight of each element, namely, $ \omega_e=n_{dof}/5=n_{sp} $, $ \omega_e=n_{sp}^2 $, and $ \omega_e=n_{sp}^3 $, can be investigated.

Though we can sketch the computational complexity within each element to pursue an optimal candidate for weight calculation, the parallel performance is only directly related to the output distributed mesh of ParMETIS. 
In our numerical experiments, we observe that when the disparity of element weights is excessively large, e.g., $ w_e =n_{sp}^3 $, $ p_{min} =2 $, and $ p_{max}=5 $, the output mesh will lead to degraded parallel efficiency and the large disparity   will occasionally lead to failure of ParMETIS, even when $ p $ is smaller than $4  $ and $ w_e = n_{sp} $ is used. Specifically, there will be processes which have no elements after parallel mesh partitioning. Therefore, when ParMETIS failure is encountered, we will decrease the weight of each element to $ w_e = p+1 $  and redo the mesh partitioning for the current time step. Note that the computational cost of parallel mesh partitioning is trivial compared to implicit time-stepping. In the following subsection, a simple example is employed to demonstrate the proposed dynamic load balancing strategy.

\subsection{A simple example of dynamic load balancing}
We simulate the  2D isentropic vortex propagation  on a 3D mesh (obtained by extruding a 2D mesh in the $z$ direction for two layers) to demonstrate the dynamic load balancing strategy. The free stream condition is $ (\rho,u,v,w,\text{Ma})^\intercal = (1,1,1,0,0.5)^\intercal$. The fluctuation is defined as~\cite{bassi2015linearly}
\begin{equation}\label{VP_Solution}
\begin{cases}
\delta u=-\frac{\alpha}{2\pi}(y-y_0)e^{\phi(1-r^2)},\\
\delta v=\frac{\alpha}{2\pi}(x-x_0)e^{\phi(1-r^2)},\\
\delta w=0,\\
\delta T=-\frac{\alpha^2(\gamma-1)}{16\phi\gamma\pi^2}e^{2\phi(1-r^2)},\\
\end{cases}	
\end{equation}
where $ \phi = \frac{1}{2}$ and $ \alpha = 5 $ are parameters that define the vortex strength. $ r=(x-x_0)^{2}+(y-y_0)^{2} $ is the distance from any point $ (x,y, z) $ to the center of line the vortex $ (x_0,y_0,z) = (0,0,z) $ at $ t=0 $. 
The domain is within $ [-10,10]\times[-10,10]\times[0,0.8] $. A uniform mesh of $ 50\times50\times2 $ elements is used for the numerical experiments. Periodic boundary conditions are imposed on all boundaries. We only simulate this problem for 60 steps with a time step size $ \Delta t = 0.05 $. In the pseudo transient continuation, $ \Delta \tau_{init} = 0.05 $ and $ \Delta \tau_{max} = 10 $ are used for SER.   For the adaptive solver, the flow field is initialized using uniform $ p_{max} $ discretization.

\begin{figure}		
	\centering
	\begin{subfigure}{0.35\textwidth}
		\includegraphics[width=\textwidth]{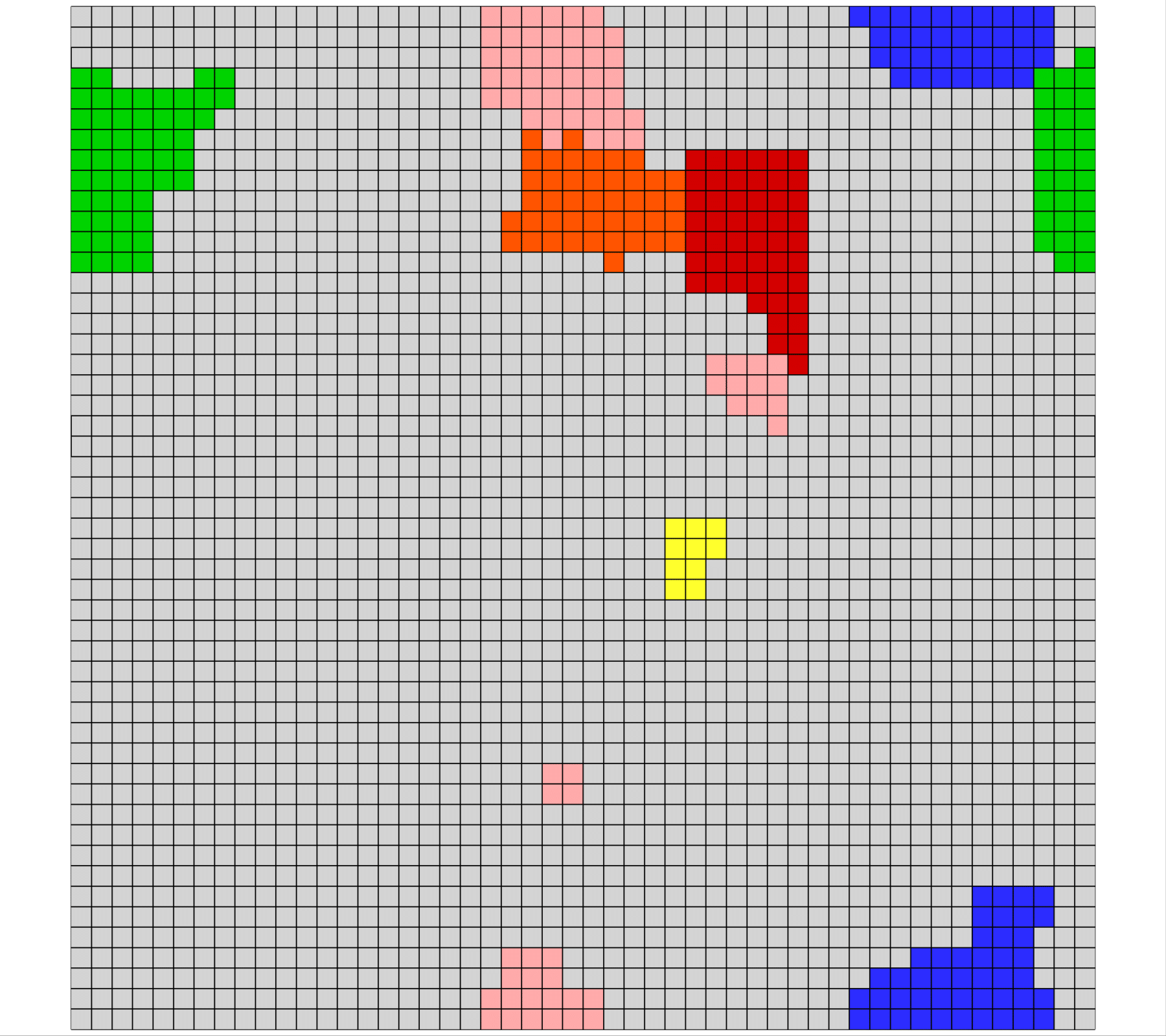} 	
		\subcaption{$ t=0.15 $ }
		\label{part_iter_3}
	\end{subfigure}
	\vspace{5pt}
	\begin{subfigure}{0.35\textwidth}
		\includegraphics[width=\textwidth]{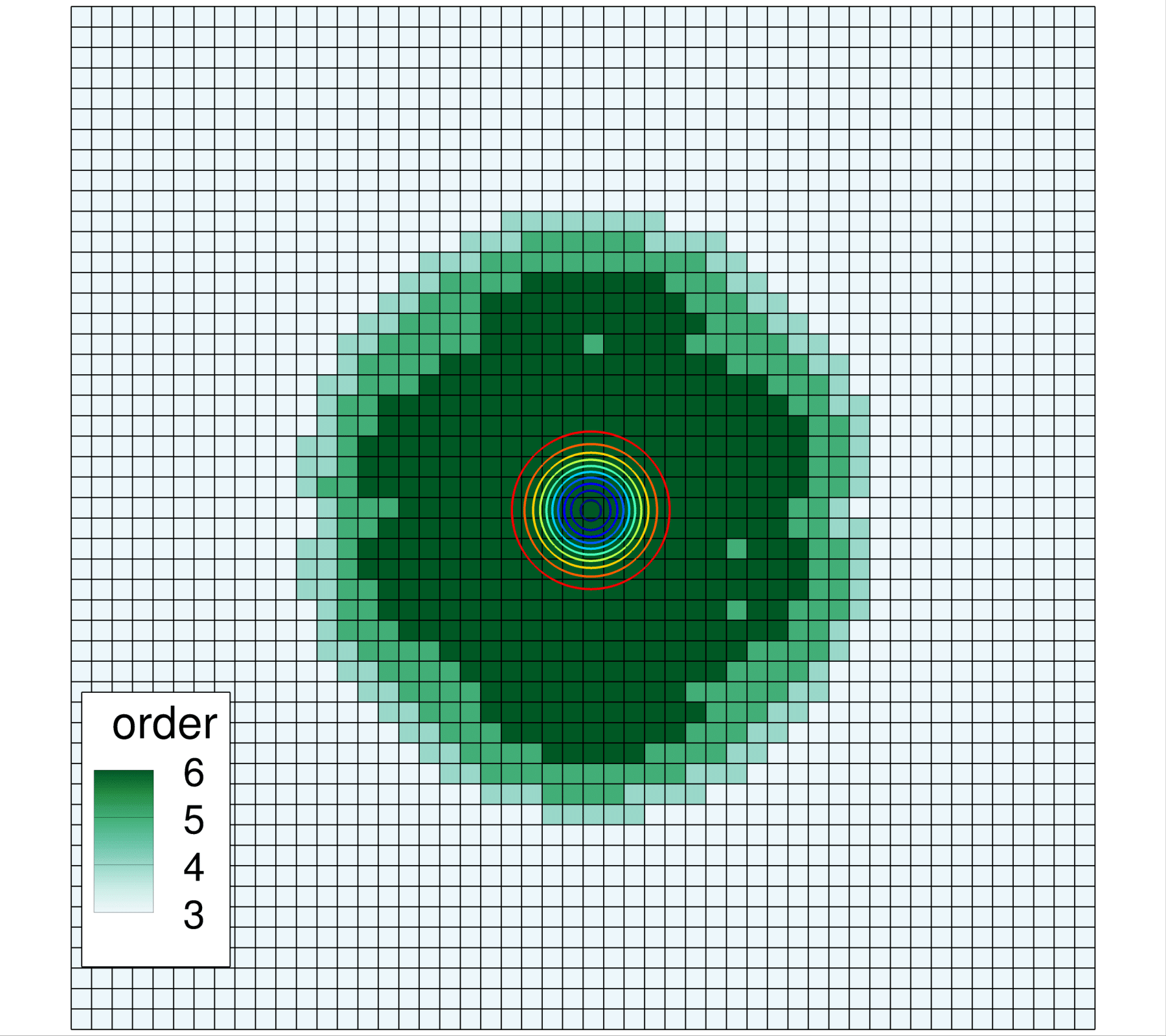}
		\subcaption{$ t=0.15 $ }
		\label{order_iter_3}
	\end{subfigure}
	\vspace{5pt}
	\begin{subfigure}{0.35\textwidth}
		\includegraphics[width=\textwidth]{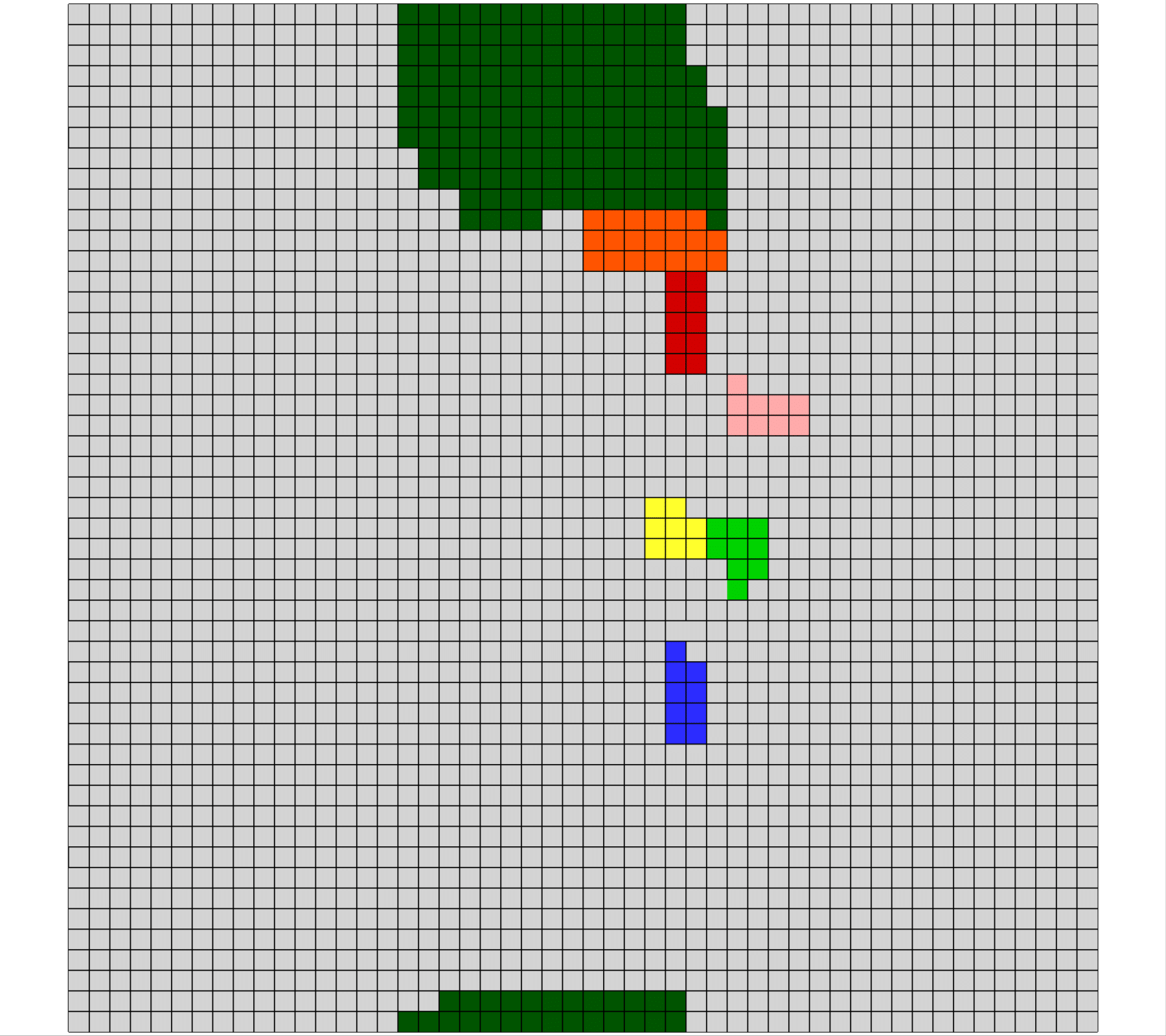} 	
		\subcaption{$ t=0.20 $ }
		\label{part_iter_4}
	\end{subfigure}
	\vspace{5pt}
	\begin{subfigure}{0.35\textwidth}
		\includegraphics[width=\textwidth]{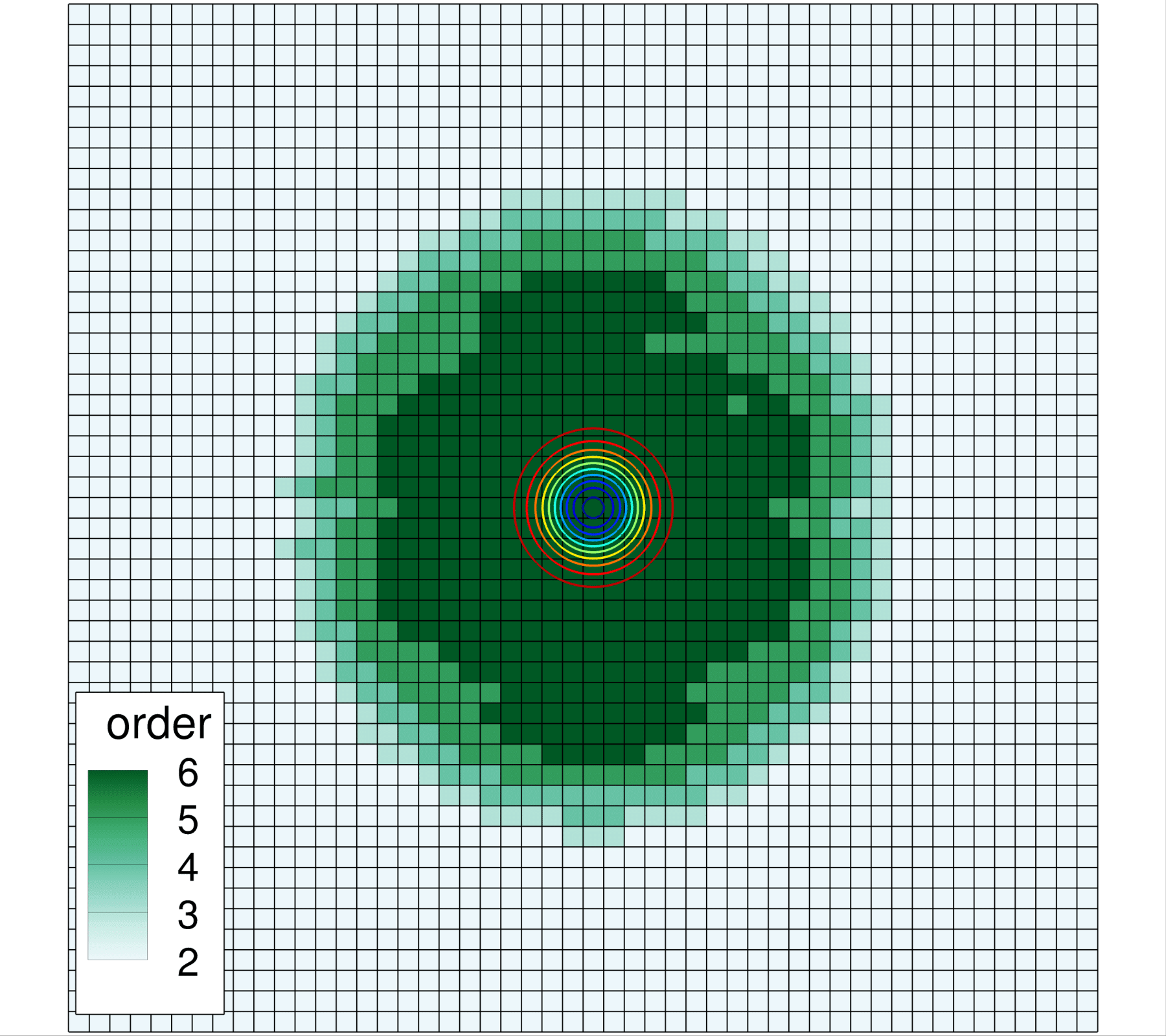} 	
		\subcaption{$ t=0.20 $}
		\label{order_iter_4}
	\end{subfigure}	
	\vspace{5pt}
	\begin{subfigure}{0.35\textwidth}
		\includegraphics[width=\textwidth]{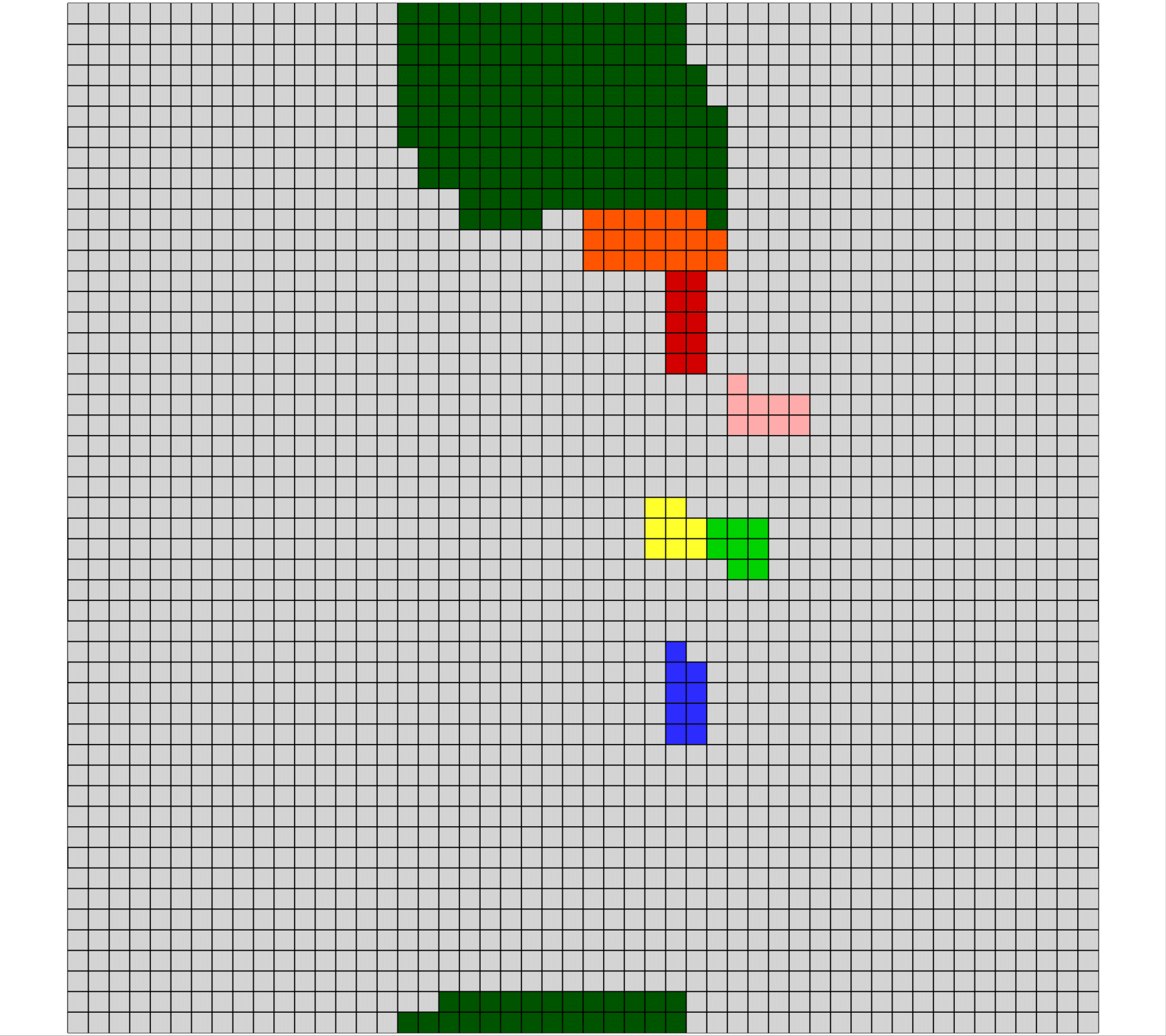}
		\subcaption{$ t=0.25 $  }
		\label{part_iter_5}
	\end{subfigure}
	\vspace{5pt}
	\begin{subfigure}{0.35\textwidth}
		\includegraphics[width=\textwidth]{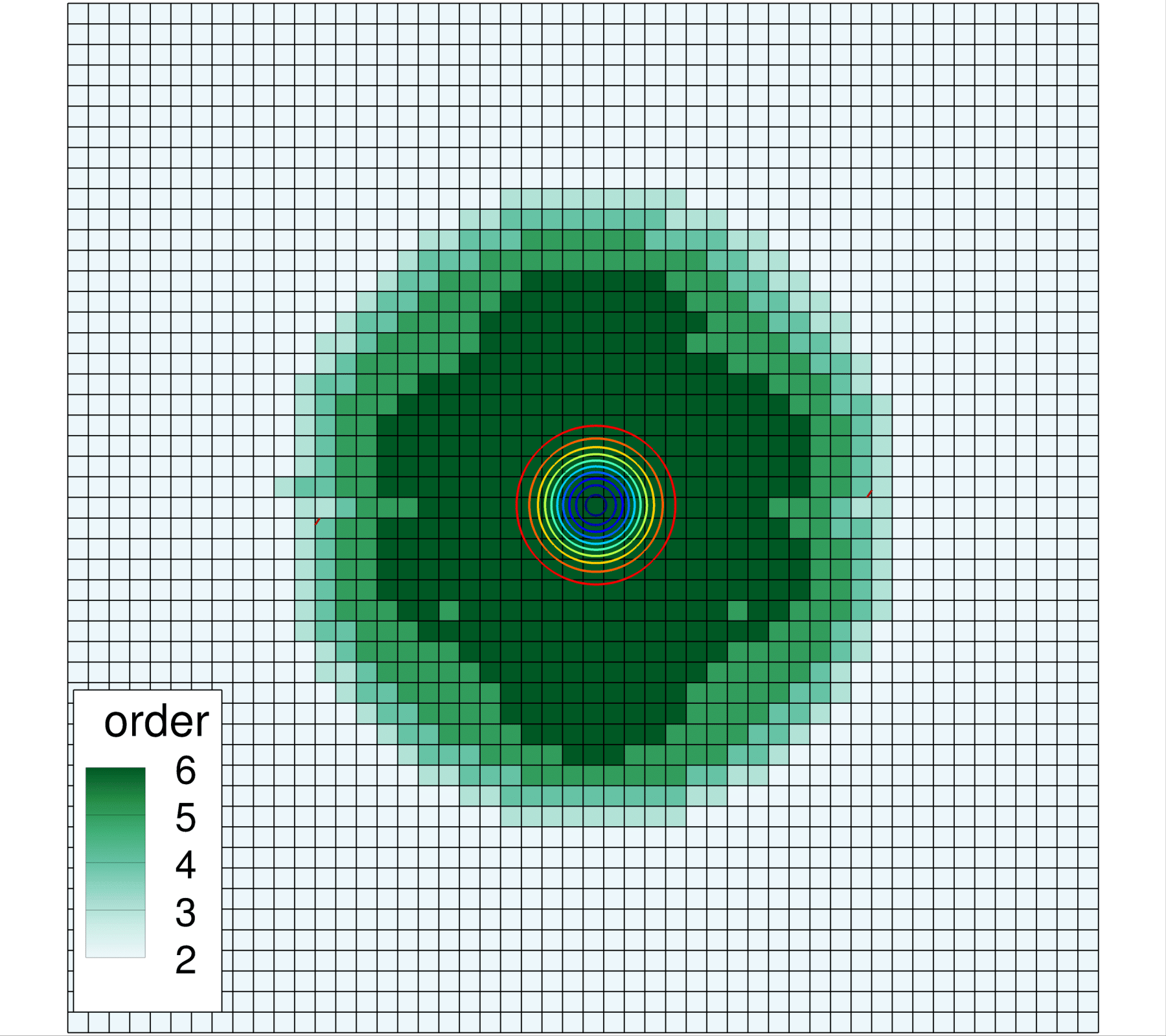} 	
		\subcaption{$ t=0.25 $}
		\label{order_iter_5}
	\end{subfigure}
	\vspace{5pt}
	\begin{subfigure}{0.35\textwidth}
		\includegraphics[width=\textwidth]{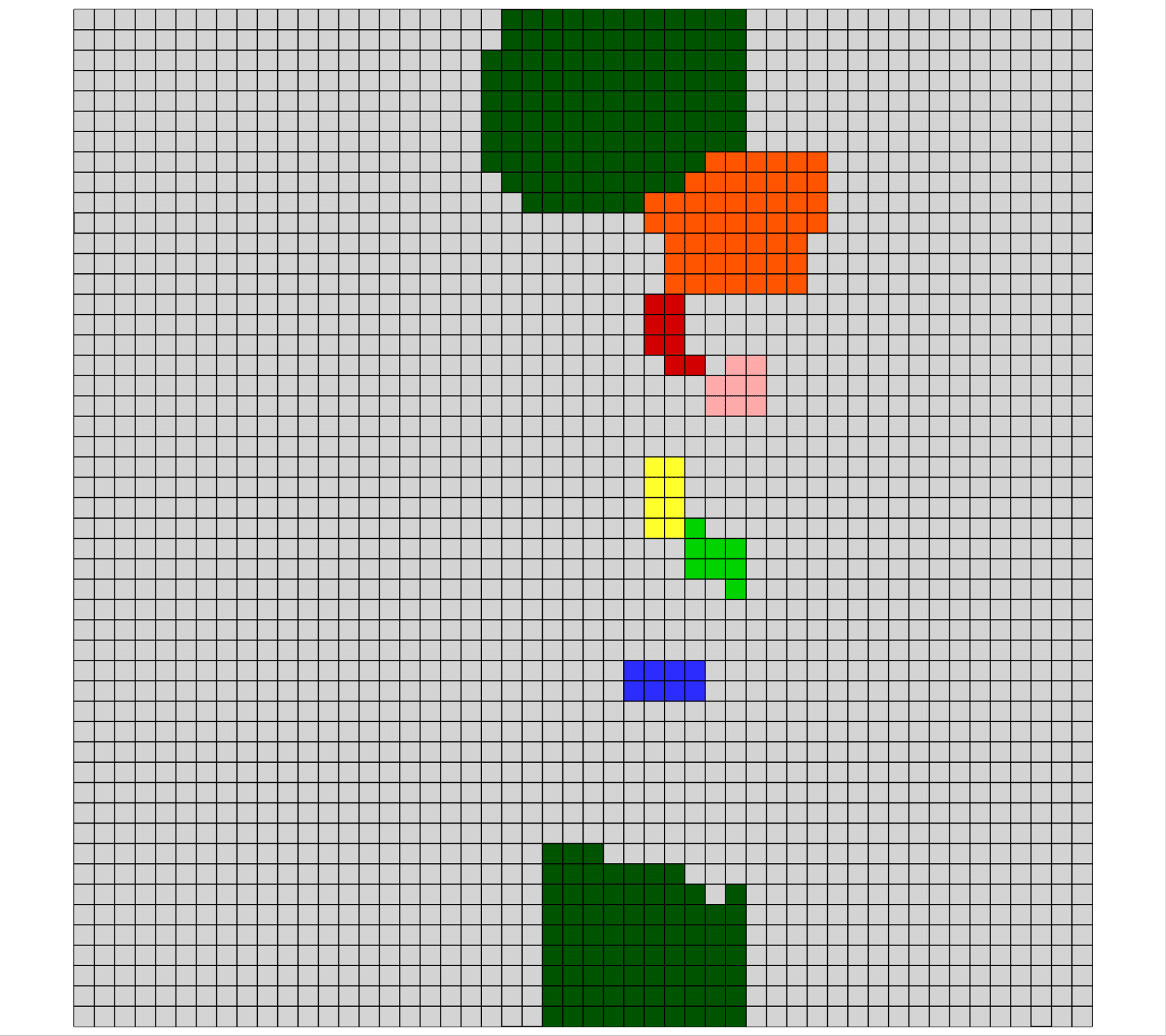} 	
		\subcaption{$ t=0.30 $ }
		\label{part_iter_6}
	\end{subfigure}
	\begin{subfigure}{0.35\textwidth}
		\includegraphics[width=\textwidth]{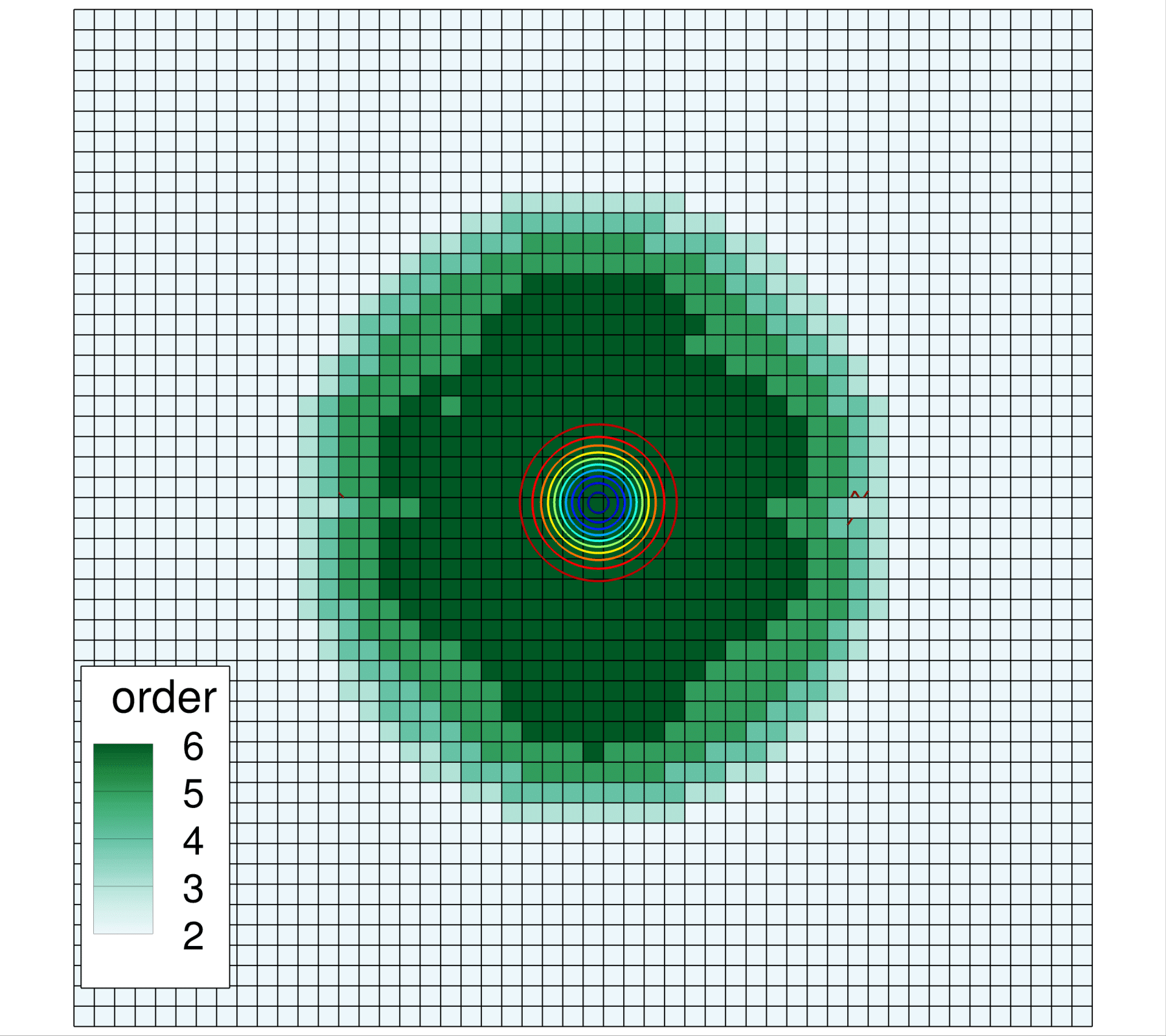} 	
		\subcaption{$ t=0.30 $}
		\label{order_iter_6}
	\end{subfigure}

	\caption{(a), (c), (e), and (g) are elements in Processes 0--6 at four consecutive time steps. Processes 0--6 are colored by red, yellow, green, blue, orange, pink, and dark green, respectively. (b), (d), (f), and (h) are corresponding instantaneous order-of-accuracy distributions.}	
	\label{vp_part_order}	
\end{figure}

72 processes are employed in this section.
Herein, we take $ p^5 $ FR with $ p $-adaptation as an example to show how the elements will be distributed to all processes.
Elements in the first seven processes and the corresponding order-of-accuracy distribution, i.e., $ (p+1) $ distribution, of four consecutive time steps are presented in Figure~\ref{vp_part_order}. From $ t=0.15 $ to $ t=0.20 $, the local polynomial degree at non-critical region will be coarsened to $ p_{min}=1 $ while a circular region surrounding the vortex will maintain high polynomial degrees. Due to this coarsening, the redistributed mesh change drastically. From $ t=0.20 $ to $ t=0.25 $, all the first seven process except the third one do not exchange any elements with other processes. From  $ t=0.25 $ to $ t=0.3 $, even though the order-of-accuracy distribution only changes slightly, almost all the seven processes will send a large portion of local elements to other processes and obtain a significant amount of elements from other processes. For turbulence simulation, we anticipate that the change of the distributed mesh will be more dramatic than this simple problem. 

\begin{figure}		
	\centering
	\begin{subfigure}{0.49\textwidth}
		\includegraphics[width=\textwidth]{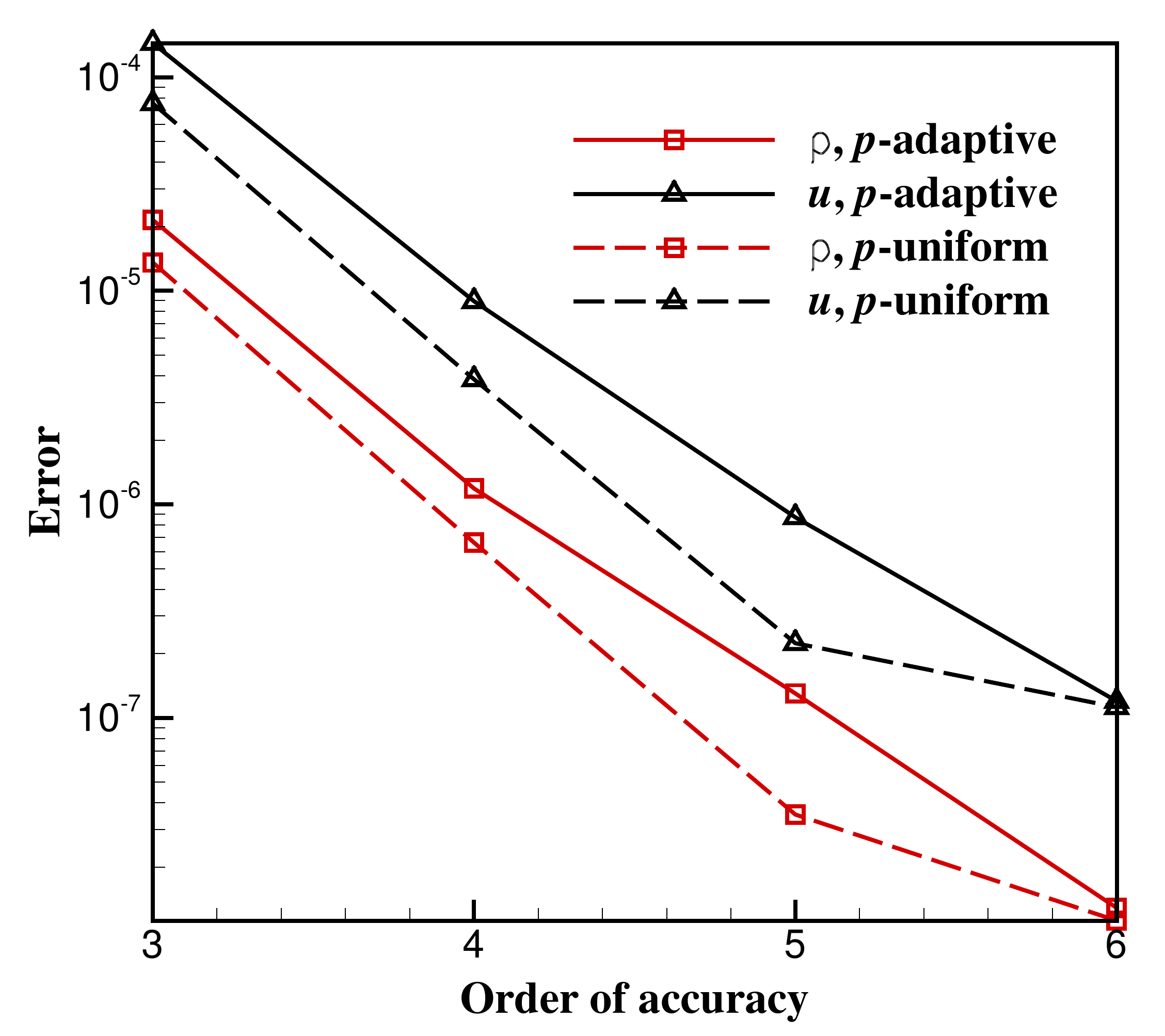}
		\subcaption{}
		\label{vp_prefine_1}
	\end{subfigure}
	\begin{subfigure}{0.49\textwidth}
		\includegraphics[width=\textwidth]{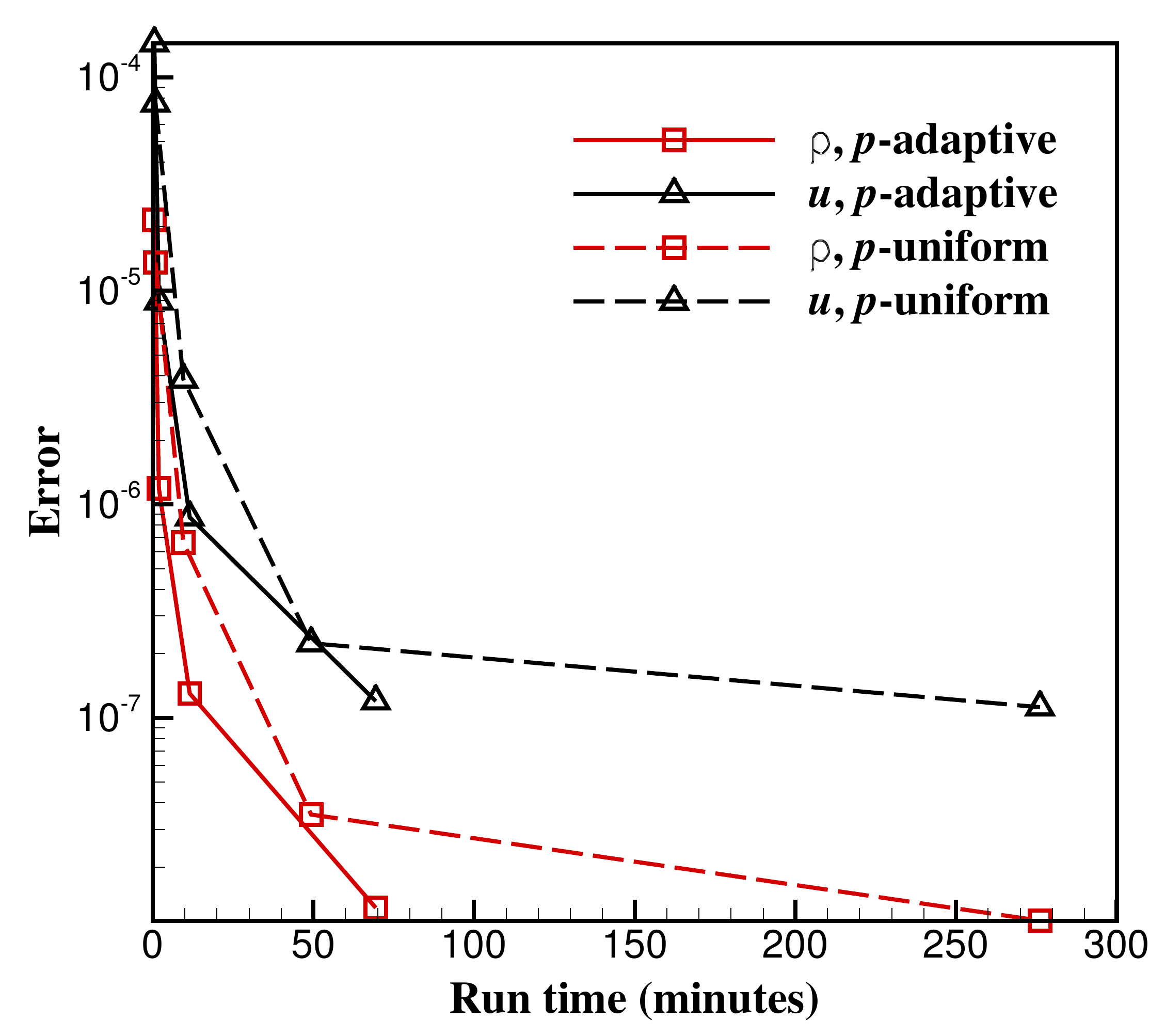} 	
		\subcaption{}
		\label{vp_prefine_2}
	\end{subfigure}
	\begin{subfigure}{0.49\textwidth}
		\includegraphics[width=\textwidth]{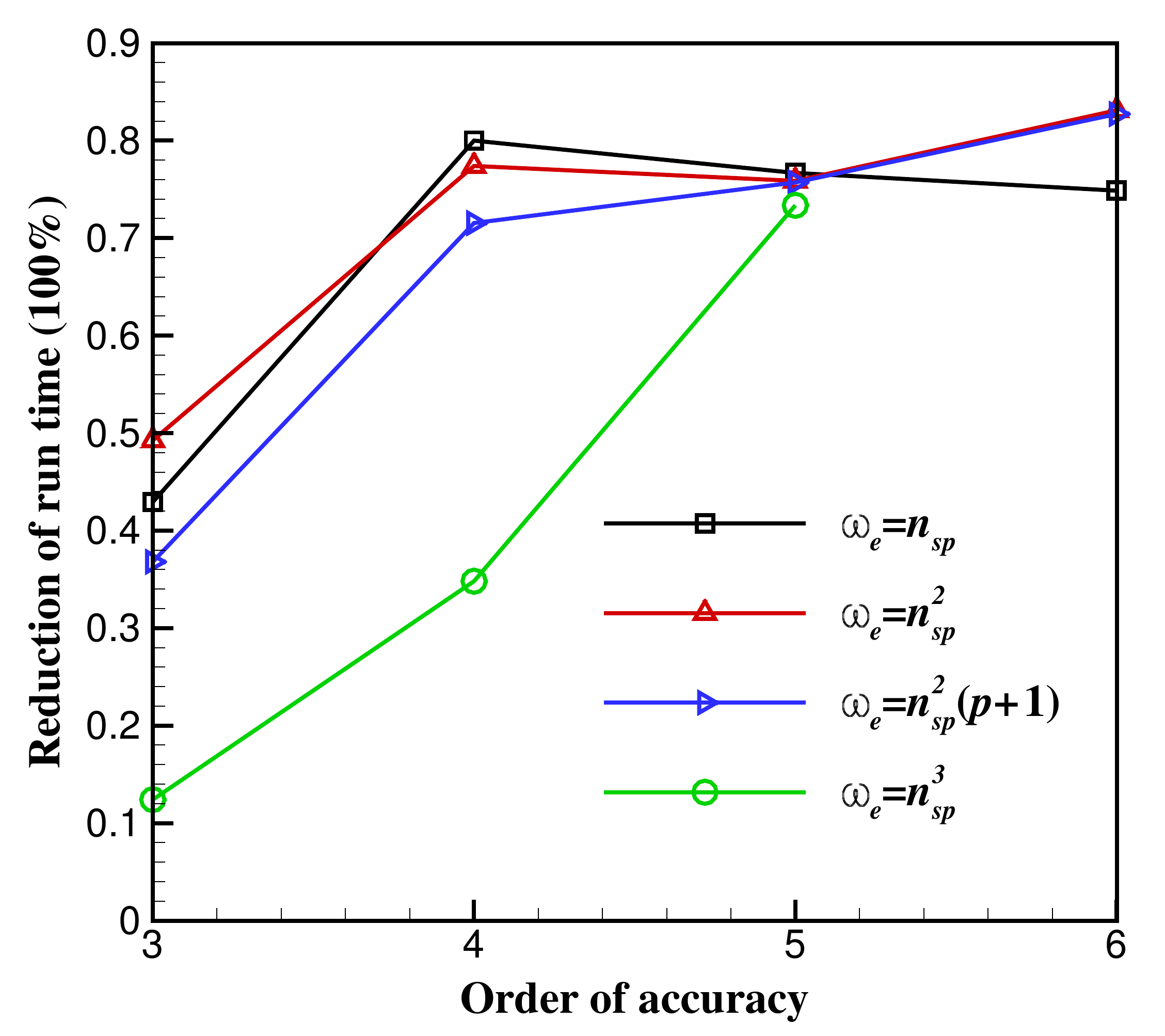} 	
		\subcaption{}
		\label{vp_prefine_3}
	\end{subfigure}
	\caption{$ p/p_{max} $-refinement of the $ p $-uniform and $ p $-adaptive FR solvers for the isentropic  vortex propagation. (a) Error~vs.\ order of accuracy, (b) error~vs.\ run time, and (c) reduction of run time vs.~order of accuracy. For the $ p $-adaptive solver, the order of accuracy indicates the highest order of accuracy, i.e., $ p_{max}+1 $ in the flow field.}	
	\label{vp_p_refinement}	
\end{figure}

 In Figure~\ref{vp_p_refinement}, numerical results from $ p/p_{max} $-refinement studies for the FR solver with/without adaptation are presented. From Figure~\ref{vp_p_refinement}\subref{vp_prefine_1}, it is observed that both the $ p $-uniform solver and $ p $-adaptive solver have spectral convergence. The errors of the $ p $-adaptive and $ p $-uniform solvers are on the same magnitude when the highest polynomial degrees are the same. Run time of different solvers and the run time reduction of the $ p $-adaptive solver using different weight algorithms (with respect to the $ p $-uniform solver) are illustrated in  Figure~\ref{vp_p_refinement}\subref{vp_prefine_2} and  Figure~\ref{vp_p_refinement}\subref{vp_prefine_3}, respectively.
Overall, with proper weights assigned to all the elements, around 80\% run time reduction can be achieved via $ p $-adaptation when $ p_{max}\ge3 $ (highest order of accuracy is no smaller than 4). For $ \omega_e = n_{sp} $, when $ p_{max}>3 $, the parallel efficiency will keep decreasing as $ p_{max} $ increases. This is due to the fact the computational cost of the Jacobian matrix and preconditioner evaluation will grow at  much larger rates than that of the one-time residual evaluation. When $ p_{max} = 5 $, $ \omega_e = n_{sp}^2 $ and  $ \omega_e = n_{sp}^2(p+1) $ have better performance than $ \omega_e = n_{sp} $. However, it is shown that $ \omega_e = n_{sp}^3 $ generally degrades the efficiency than other candidates. Especially, when $ \omega_e = n_{sp}^3 $ and $ p_{max} = 5 $, the $ p $-adaptive solver fails to finish the simulation within the time that the $ p $-uniform solver needs.
In order to achieve optimal performance, both the weight calculation  and the parallel mesh partitioning algorithm should be taken into account.
In the following section, we use $ w_e = n_{sp} $ for the $ p $-adaptive solver when it is applied to under-resolved turbulence simulation since we only consider $ p_{max}\le3 $.

\clearpage
\section{Applications to under-resolved turbulence simulation}\label{numerical_results}
The numerical studies in this section use seven computational nodes in a 
distributed-memory cluster. Each node has two 18-core Intel Xeon Gold 6140 Skylake CPUs (2.3 GHz clock speed, 24.75 MB L3 cache) and 384 GB memory ($ 12 \times 32 $ GB DDR4). The nodes are connected by a network of four 36-port EDR (Enhanced Data Rate) InfiniBand switches (100 Gb/s bandwidth, 90 ns latency). 252 processes are used for every simulation.
The CFD codes and third party libraries, such as ParMETIS and PETSc, are compiled using MPICH 3.2.1 and GCC 7.3.0 compilers. The C++11 standard is used in the compilation of the CFD codes.

\subsection{Under-resolved simulation of the flow over an infinite cylinder}
\begin{figure}		
	\centering
	\begin{subfigure}{0.49\textwidth}
		\includegraphics[width=\textwidth]{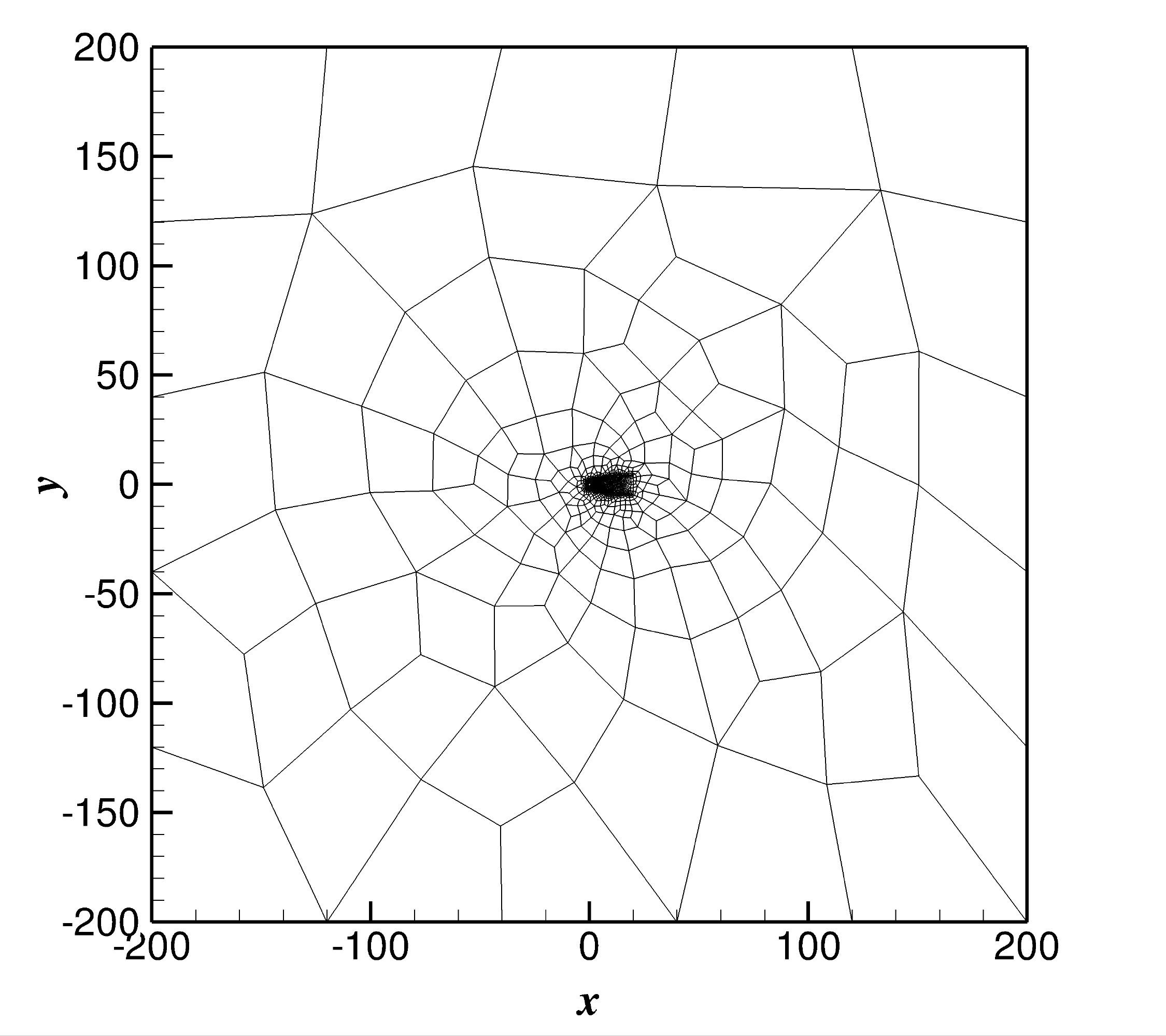}
		\subcaption{Global view}
		\label{cy_global}
	\end{subfigure}
	\begin{subfigure}{0.49\textwidth}
		\includegraphics[width=\textwidth]{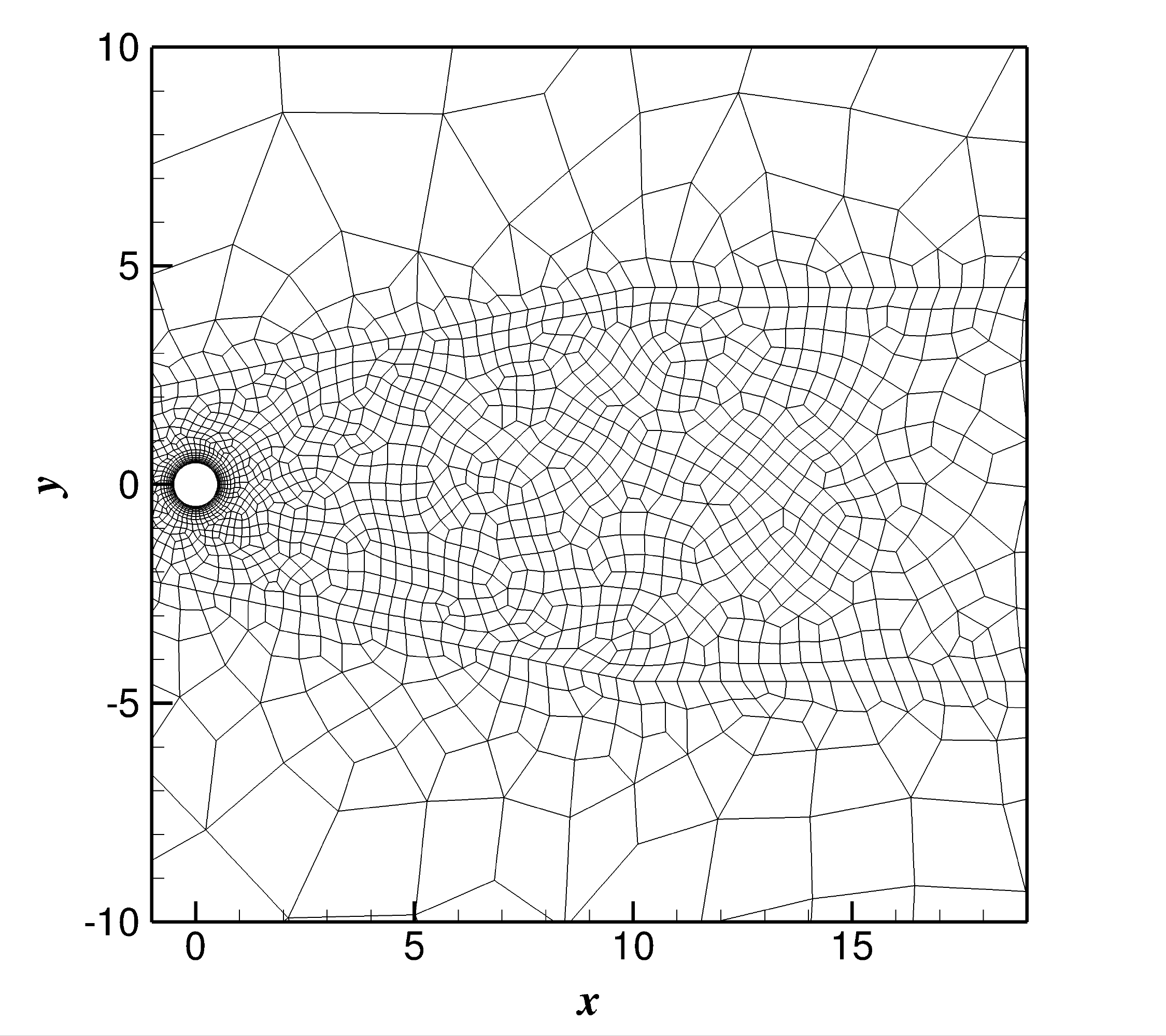} 	
		\subcaption{Close-up view}
		\label{cy_local}
	\end{subfigure}
	\caption{2D views of the unstructured mesh around a circular cylinder. }	
	\label{cylinder_mesh}	
\end{figure}

Long time simulations of the transitional flow over an infinite cylinder are conducted to validate the reliability of the $ p $-adaptive solver in this section. 
The diameter of the cylinder is $ d =1$.
The inflow conditions are set as the vector $ (\rho_\infty, u_\infty, v_\infty, w_\infty, \text{Ma}_\infty)^\intercal=(1,1,0,0,0.1)^\intercal $. 
The Reynolds number of the inflow with respect to the diameter of the cylinder  is $ \text{Re}_d ={\rho_\infty u_\infty d}/{\mu}= 3900 $.  The Prandtl number is $ \text{Pr} = 0.71 $.
A 2D view of the mesh is illustrated in Figure~\ref{cylinder_mesh}. The center of the cylinder sits at the origin. The 3D mesh is obtained by extruding the 2D mesh along the $ z $ direction, i.e., $ (0,0,1)^\intercal $, for eight layers and the thickness of each layer is $ 0.25d $. There are 17694 hexahedral elements and the curved wall boundary is represented by $ p^3 $ elements. 
$ p^2 $ and $ p^3 $ FR are employed with/without $ p $-adaptation. The cylinder surface is treated as a no-slip adiabatic wall. Farfield boundary conditions are applied to outer boundaries. Periodic boundary conditions are imposed at the front and back sides. We employ ESDIRK4   for time integration and $ \Delta t=0.025 $. In the pseudo transient continuation, $ \Delta \tau_{init} = 0.001 $ and $ \Delta \tau_{max} = 0.01 $ are used for SER. The tolerance for the pseudo transient continuation is $ tol_{rel}^{pseudo} = 10^{-4} $ and that of the GMRES solver is $ tol_{rel}^{gmres} = 10^{-1} $.
We run all simulations until $ t_{end} = 800 $. The instantaneous solutions in $ t \in(100,800] $ are used for time averaging.
For $ p^2 $ FR with adaptation,  $ (\nu_{max},\nu_{min}) = (0.1,0.001)$ and  $ (\nu_{max},\nu_{min}) = (0.1,0.01)$ are tested. For $ p^3 $ FR, only $ (\nu_{max},\nu_{min}) = (0.1,0.01)$ is used to carry out the $ p $-adaptation. The flow field is initialized uniformly with the inflow conditions and the $ p $-adaptive solver starts from a uniform $ p^1 $ discretization.

\begin{table}
	\small
	\centering
	\caption{Run time of all simulations of the transitional flow over the infinite cylinder. } 
	\label{cylinder_wall_time}
	\setlength\tabcolsep{4pt}
	\begin{tabular}{rrrrr}
		\hline
		Method&$  (\nu_{max},\nu_{min}) $&Run time & Reduction of& Reduction of \\
		&&(hours)& run time & $ n_{sp}^{tot} $\\
		\hline
		\multirow{3}{1cm}{$ p^2 $ FR}
		& no adaptation     & 27.43 & 0& 0\\
		& $  (0.1,0.001)$ & 22.43 & 18.23\% &49.96\% at $ t=800 $\\
		& $  (0.1, 0.01)$  & 16.55 & 39.66\%&63.43\% at $ t=800 $\\
		\hline	
		\multirow{3}{1cm}{$ p^3 $ FR}
		& no adaptation     & 149.78 & 0&0\\
		& $ (0.1, 0.01)$  & 45.56  & 69.58\% &75.98\% at $ t=800 $\\
		\hline
	\end{tabular}	
\end{table}

\begin{figure}		
	\centering
	\begin{subfigure}{0.9\textwidth}
		\includegraphics[width=\textwidth]{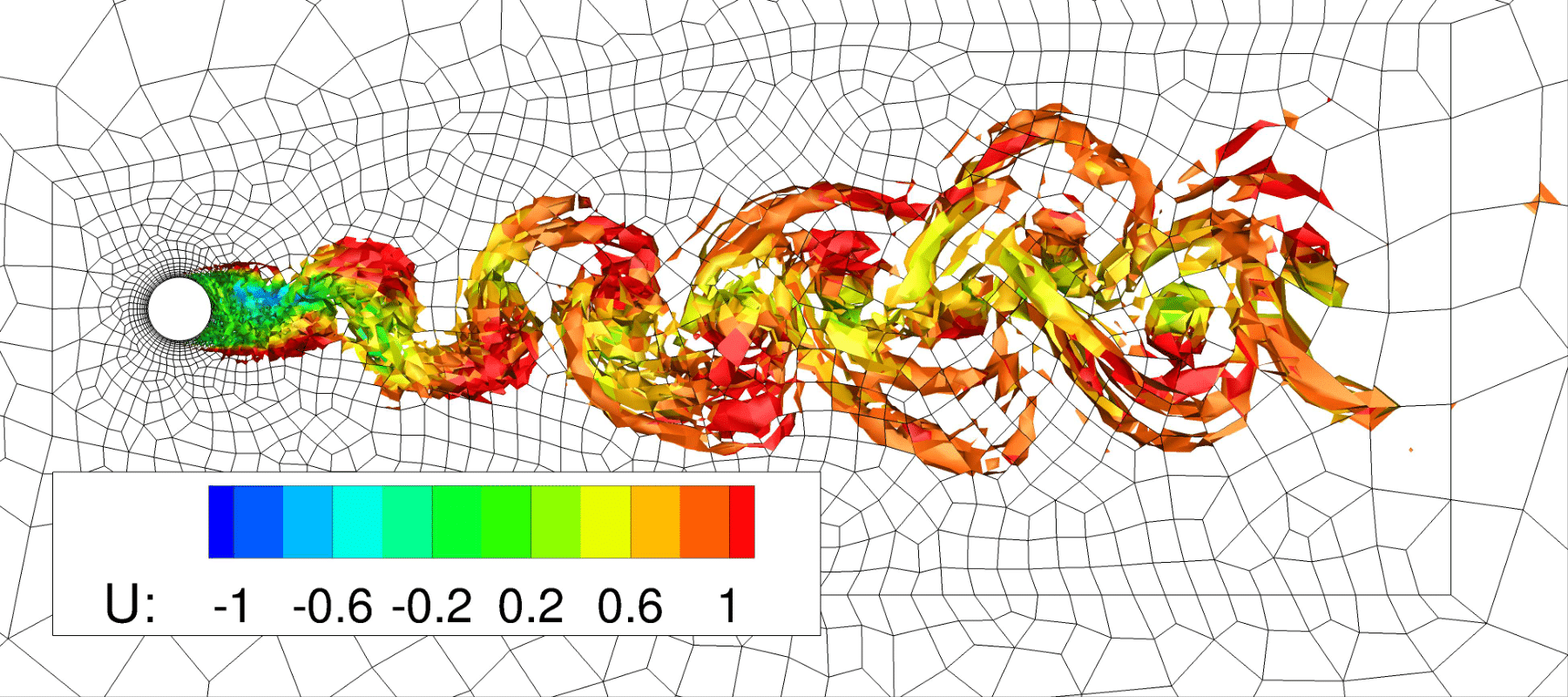}
		\subcaption{$ p^2 $ FR without adaptation}
		\label{cy_qiso_3rd_1}
	\end{subfigure}\\
	\vspace{10pt}
	\begin{subfigure}{0.9\textwidth}
		\includegraphics[width=\textwidth]{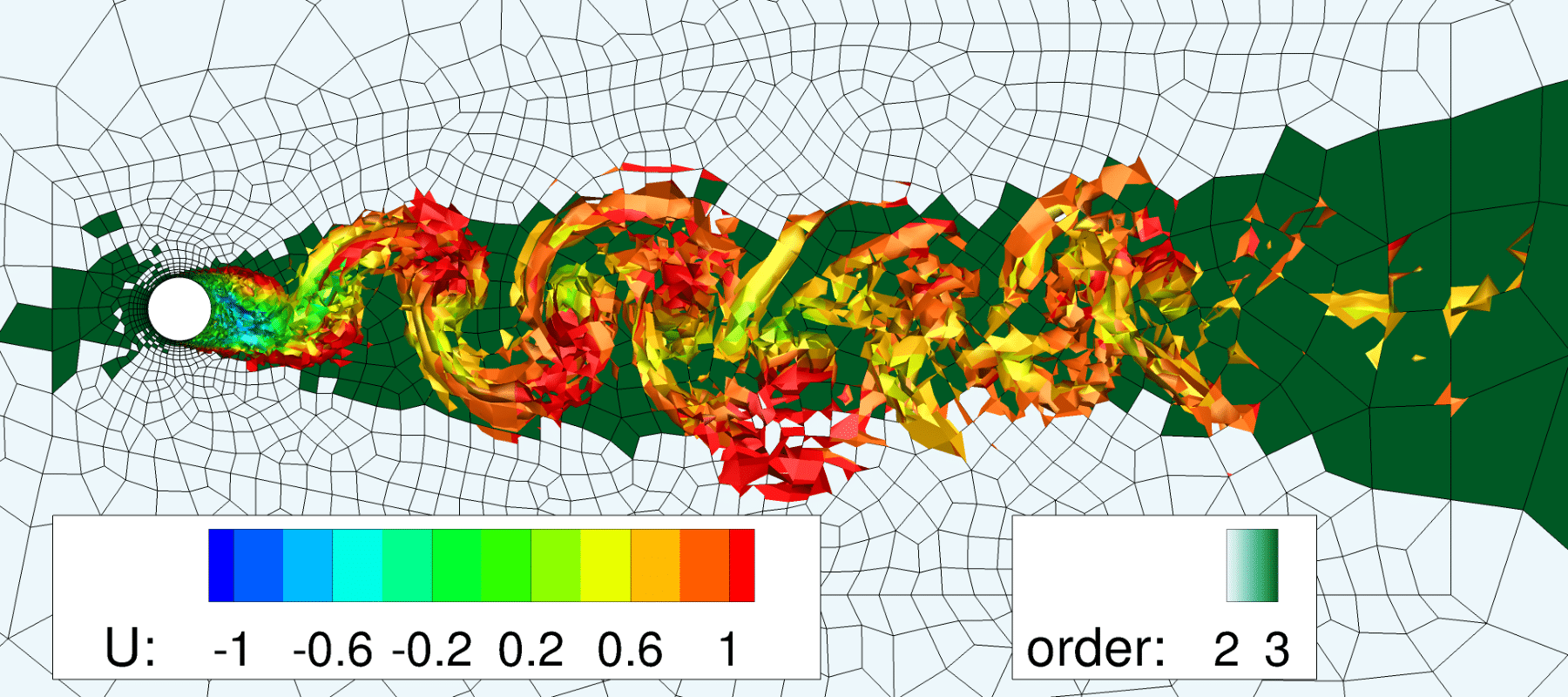} 	
		\subcaption{$ p^2 $ FR with adaptation and $ (\nu_{max},\nu_{min})=(0.1,0.001) $}
		\label{cy_qiso_3rd_2}
	\end{subfigure}\\
	\vspace{10pt}
	\begin{subfigure}{0.9\textwidth}
		\includegraphics[width=\textwidth]{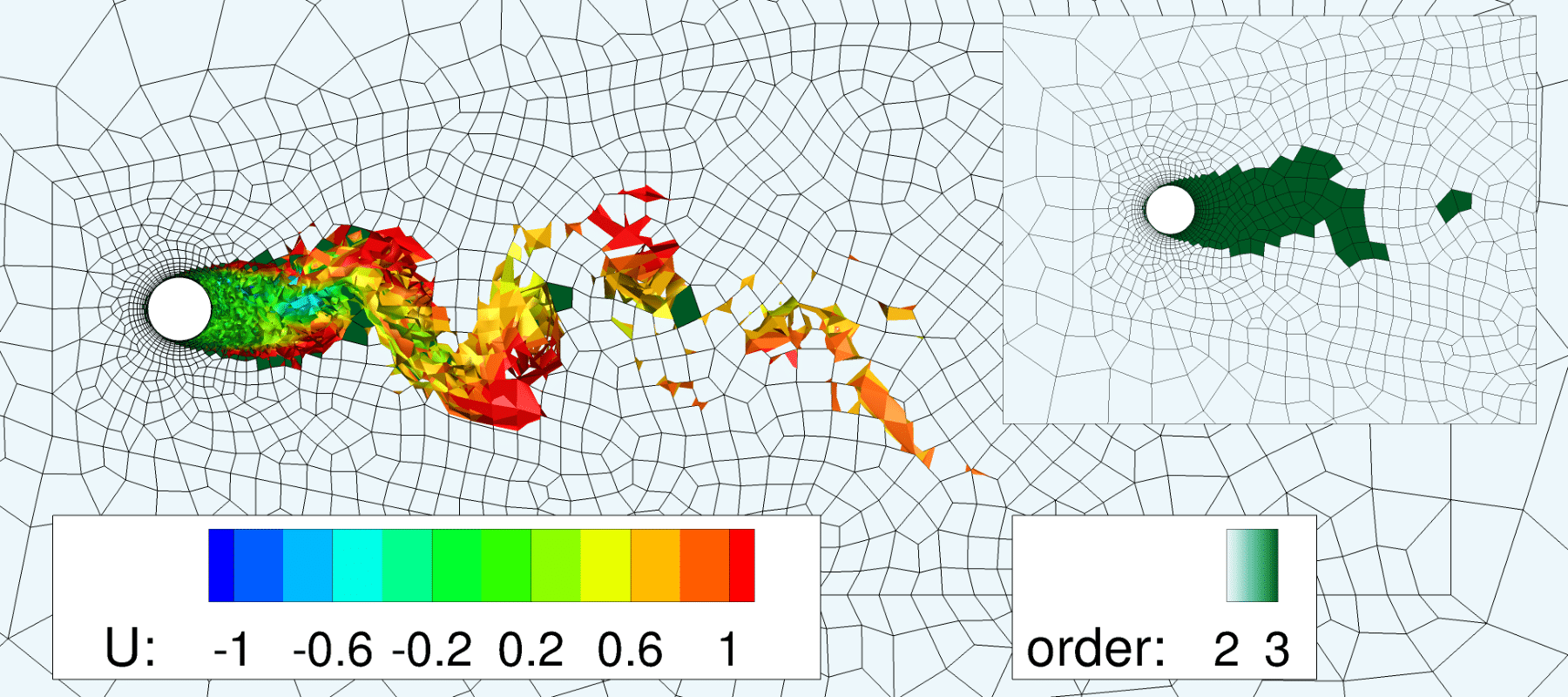} 	
		\subcaption{$ p^2 $ FR with adaptation and $ (\nu_{max},\nu_{min})=(0.1,0.01) $}
		\label{cy_qiso_3rd_3}
	\end{subfigure}
	\caption{Transitional flow over the infinite cylinder at $ \text{Re}=3900 $. Instantaneous isosurfaces of $ Q=0.5 $ colored by velocity component in the $ x $ direction at $ t=800 $. Order-of-accuracy distribution at slice $ z=0 $ is turned on in (b) and (c). A close-up view of the near wall region is also presented in (c).}	
	\label{cylinder_qiso_3rd}	
\end{figure}

\begin{figure}		
	\centering
	\begin{subfigure}{0.49\textwidth}
		\includegraphics[width=\textwidth]{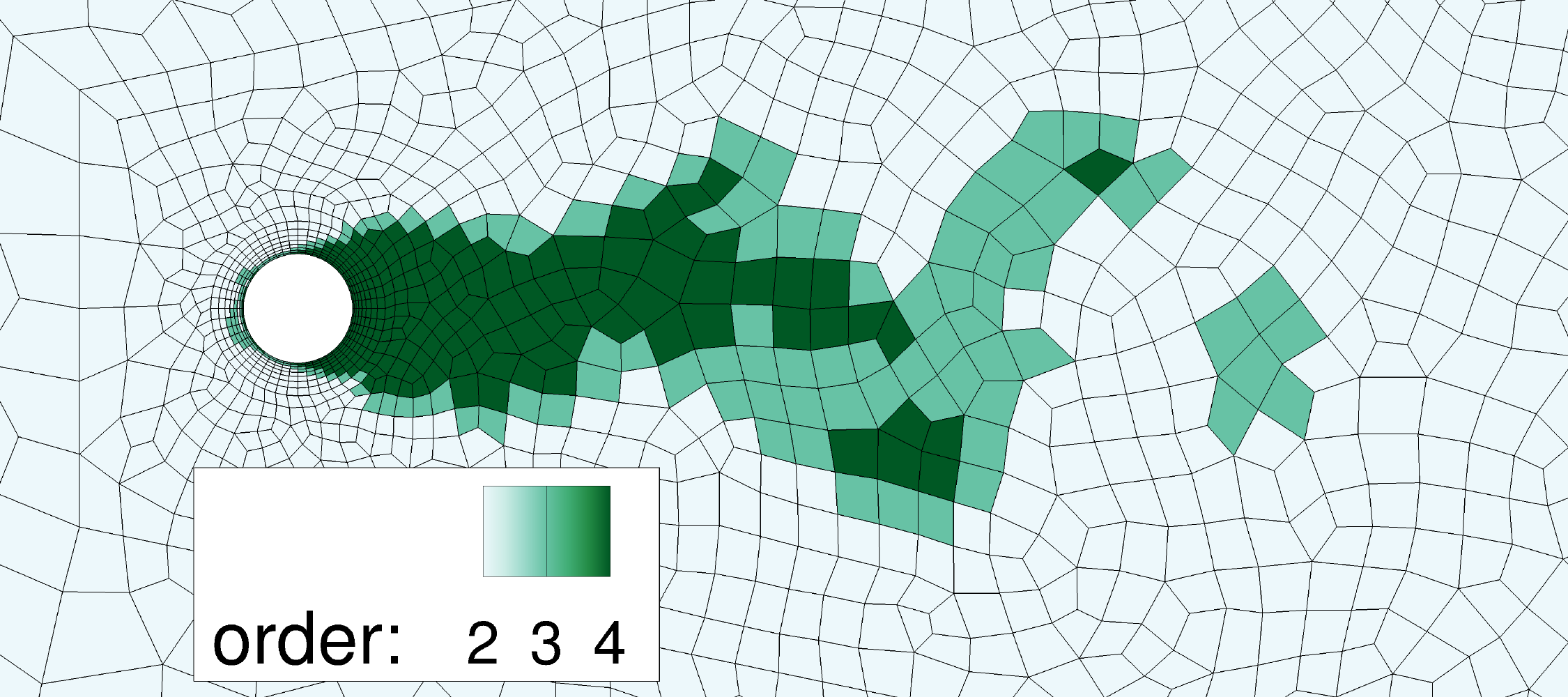}
		\subcaption{$ z=0.5 $}
		\label{4th_slice_1}
	\end{subfigure}
	\begin{subfigure}{0.49\textwidth}
		\includegraphics[width=\textwidth]{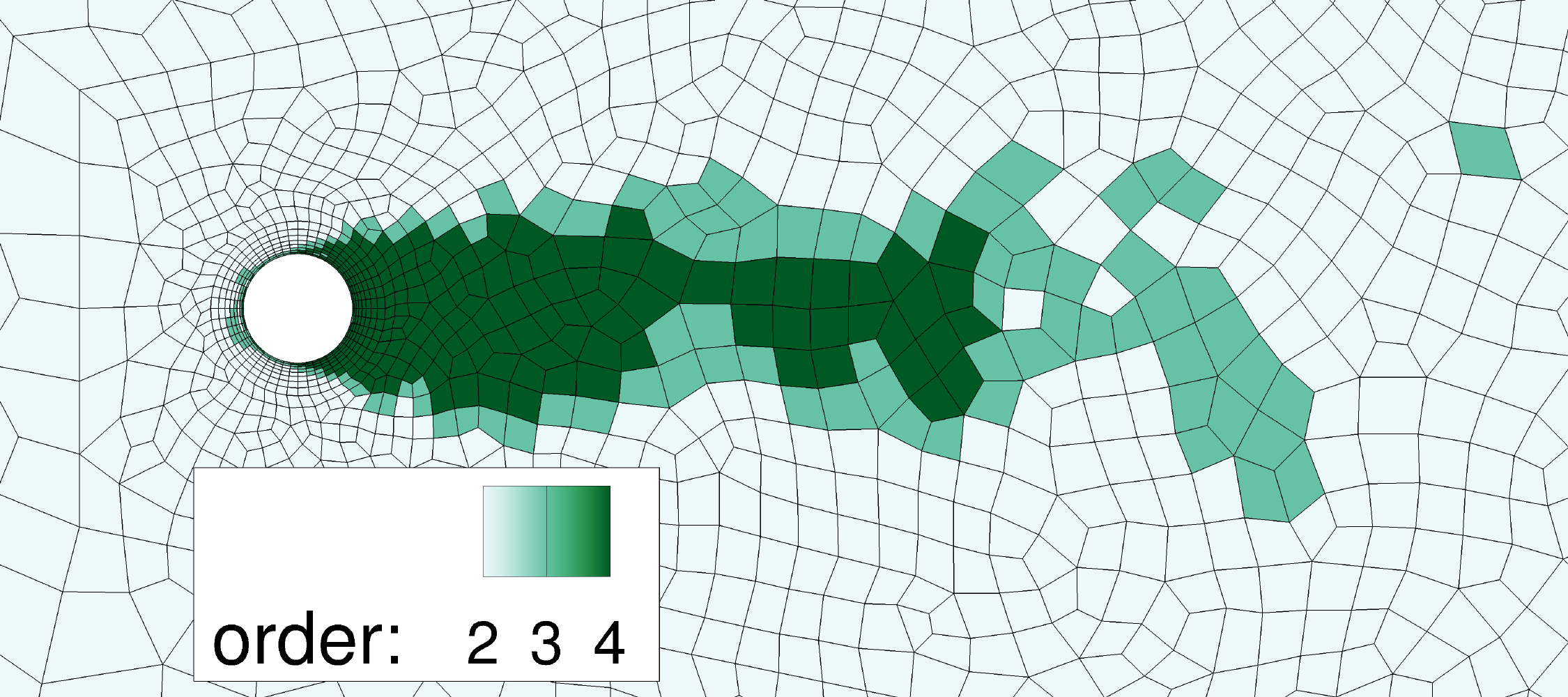} 	
		\subcaption{$ z=1 $}
		\label{4th_slice_2}
	\end{subfigure}\\
	\vspace{10pt}
	\begin{subfigure}{0.49\textwidth}
		\includegraphics[width=\textwidth]{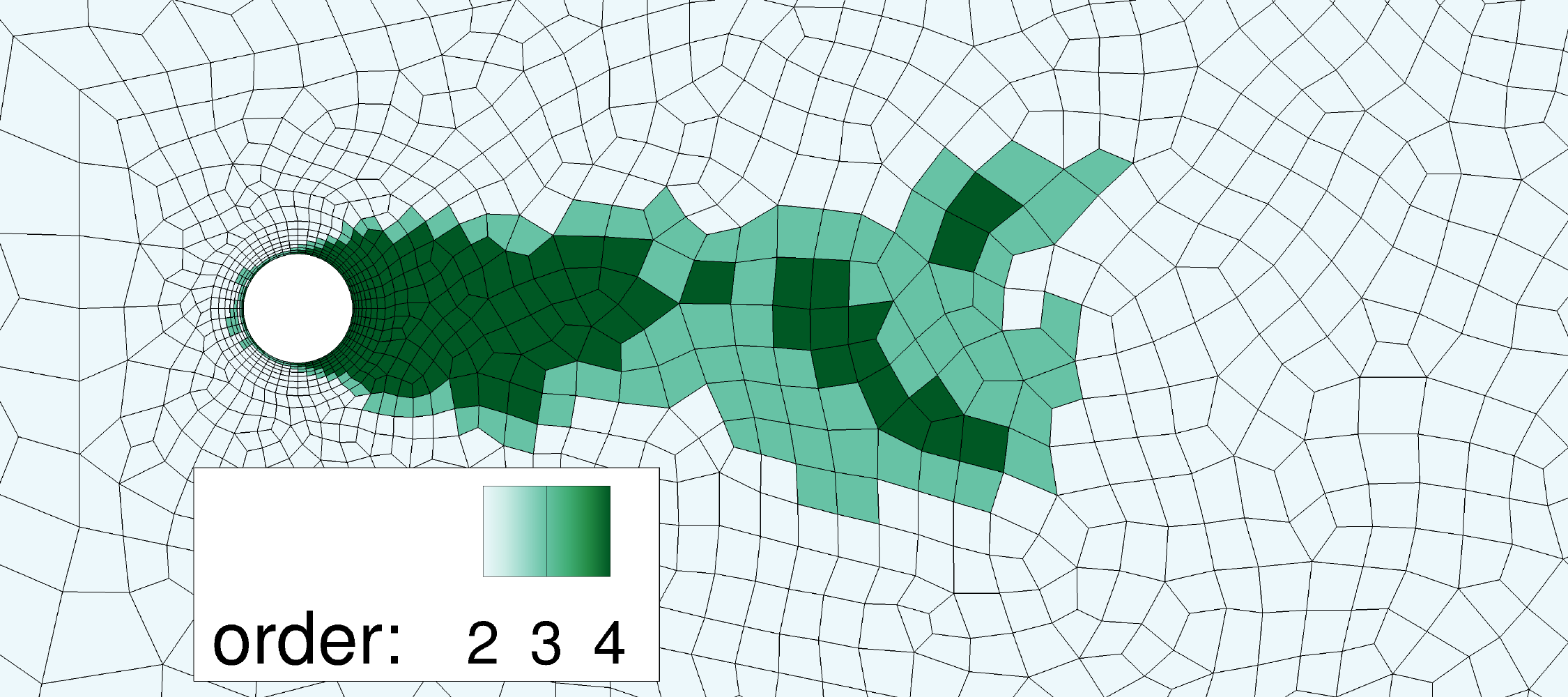} 	
		\subcaption{$ z=1.5 $}
		\label{4th_slice_3}
	\end{subfigure}
	\begin{subfigure}{0.49\textwidth}
		\includegraphics[width=\textwidth]{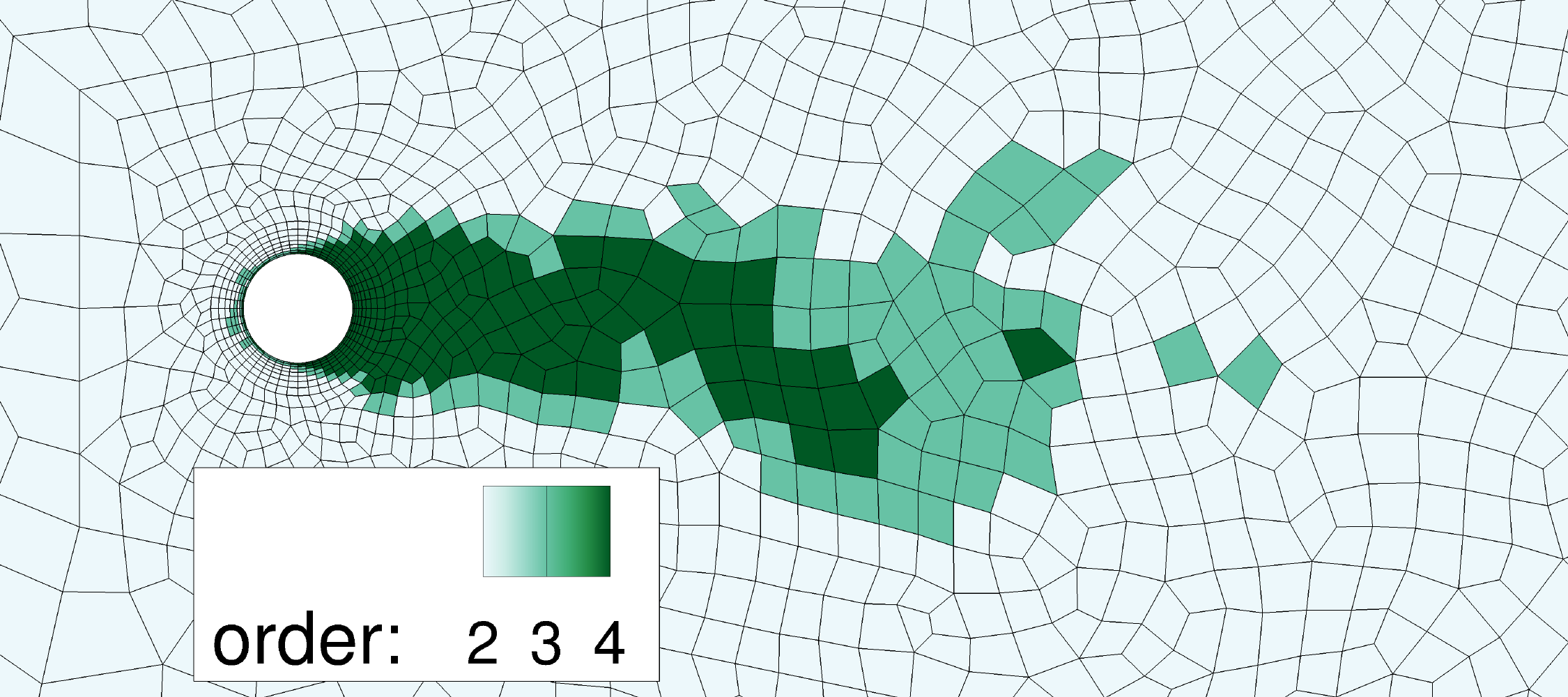} 	
		\subcaption{$ z=2 $}
		\label{4th_slice_4}
	\end{subfigure}
	\caption{Instantaneous order-of-accuracy distribution of $p^3 $ FR with adaptation at different slices when $ t=800 $.}	
	\label{cylinder_4th_slices}	
\end{figure}

\begin{figure}		
	\centering
	\begin{subfigure}{0.23\textwidth}
		\includegraphics[width=\textwidth]{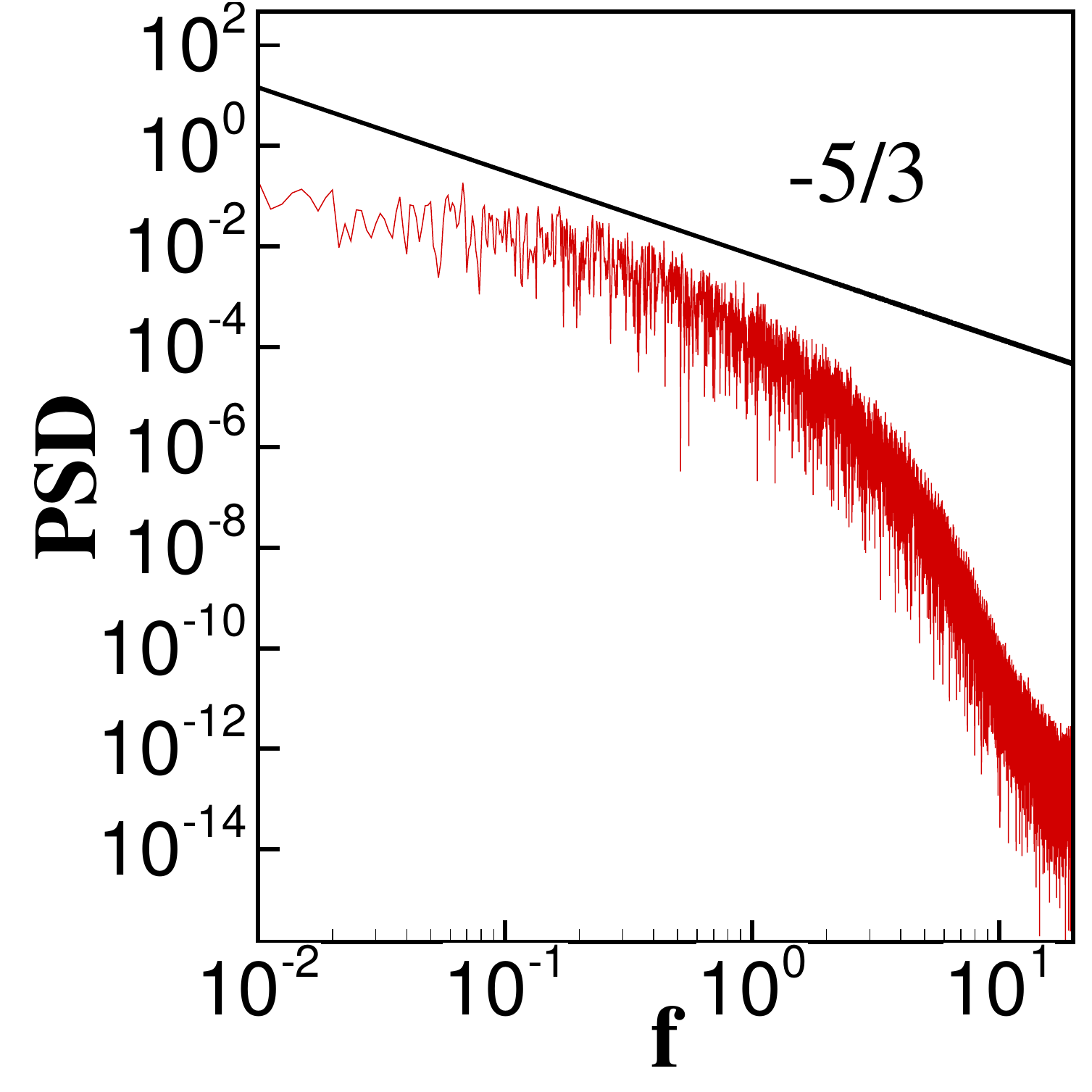}
		\subcaption{$ (0.58,0,1)^\intercal $ }
		\label{cylinder_psd_3rd_1}
	\end{subfigure}
	\begin{subfigure}{0.23\textwidth}
		\includegraphics[width=\textwidth]{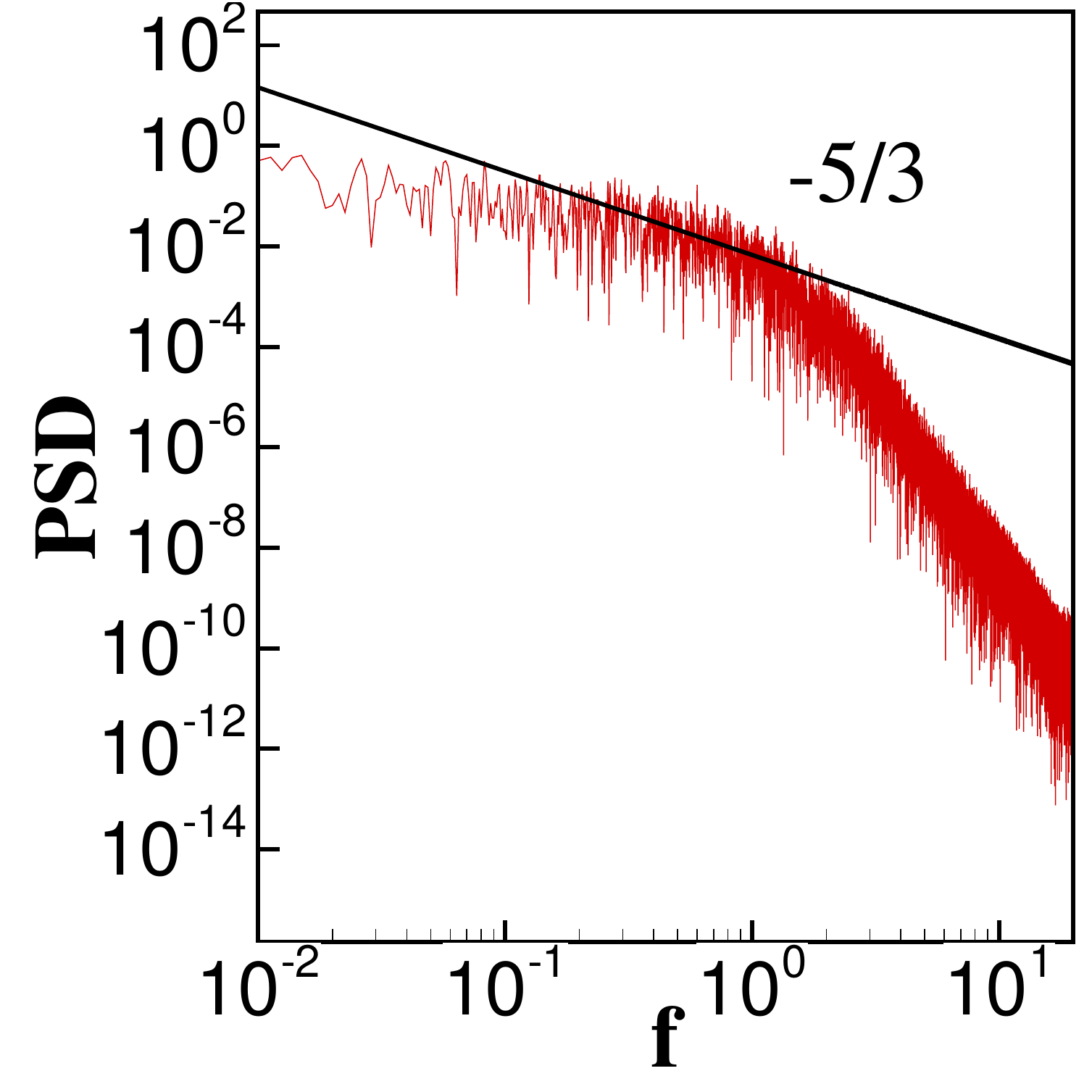} 	
		\subcaption{$ (1.54,0,1)^\intercal $}
		\label{cylinder_psd_3rd_2}
	\end{subfigure}
	\begin{subfigure}{0.23\textwidth}
		\includegraphics[width=\textwidth]{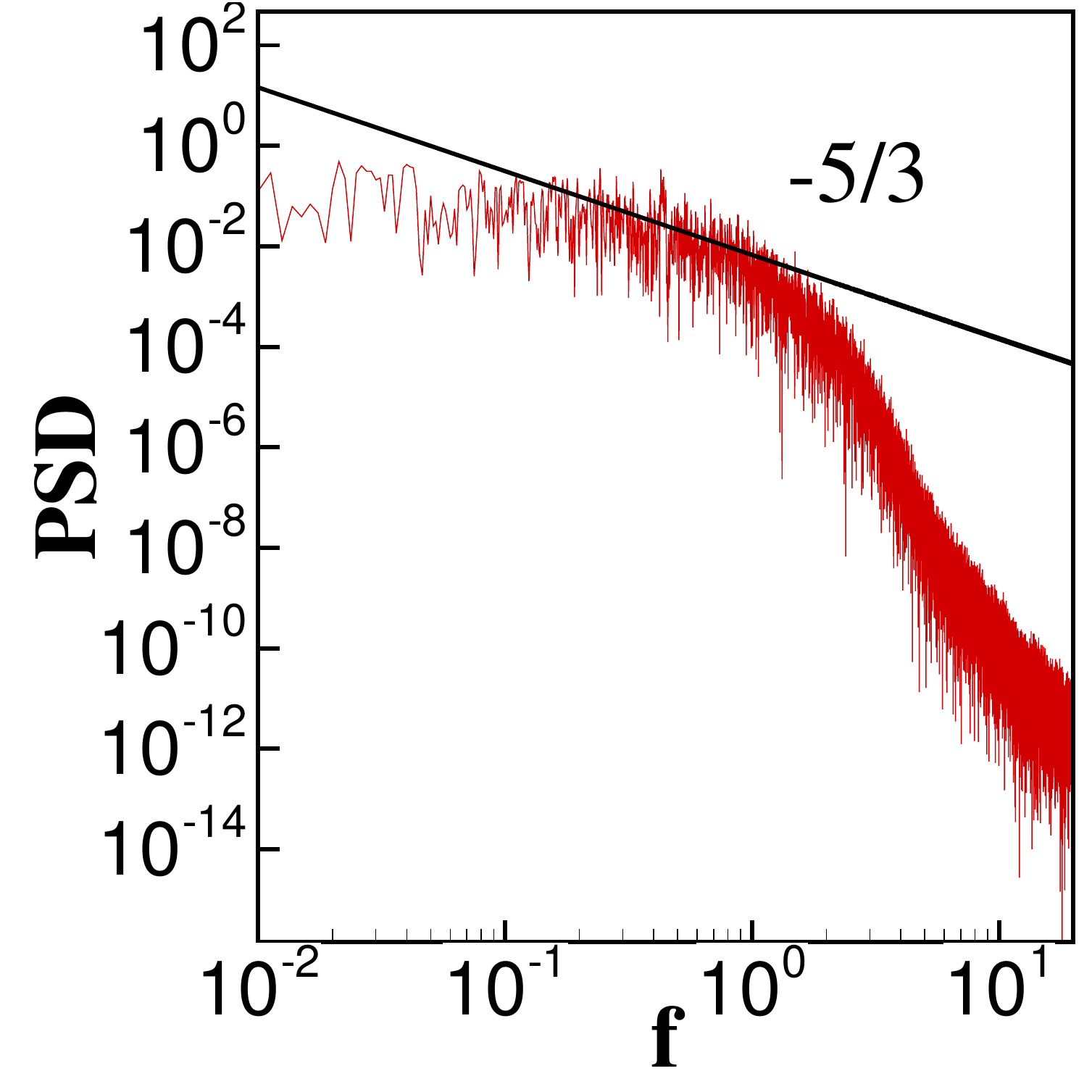} 	
		\subcaption{$ (6,0,1)^\intercal $}
		\label{cylinder_psd_3rd_3}
	\end{subfigure}
	\begin{subfigure}{0.23\textwidth}
		\includegraphics[width=\textwidth]{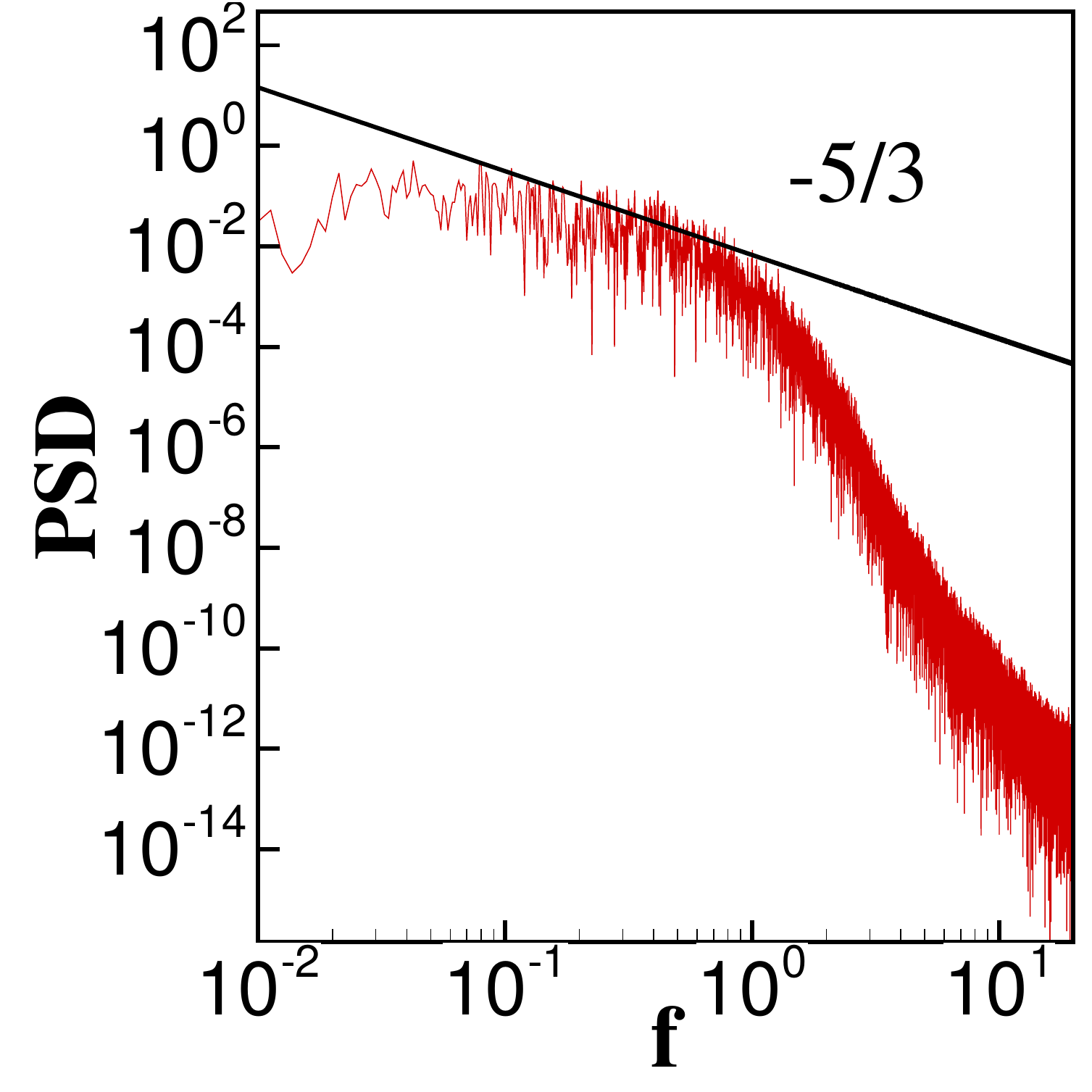} 	
		\subcaption{$ (10,0,1)^\intercal $}
		\label{cylinder_psd_3rd_4}
	\end{subfigure}\\
	\vspace{5pt}
	\begin{subfigure}{0.23\textwidth}
		\includegraphics[width=\textwidth]{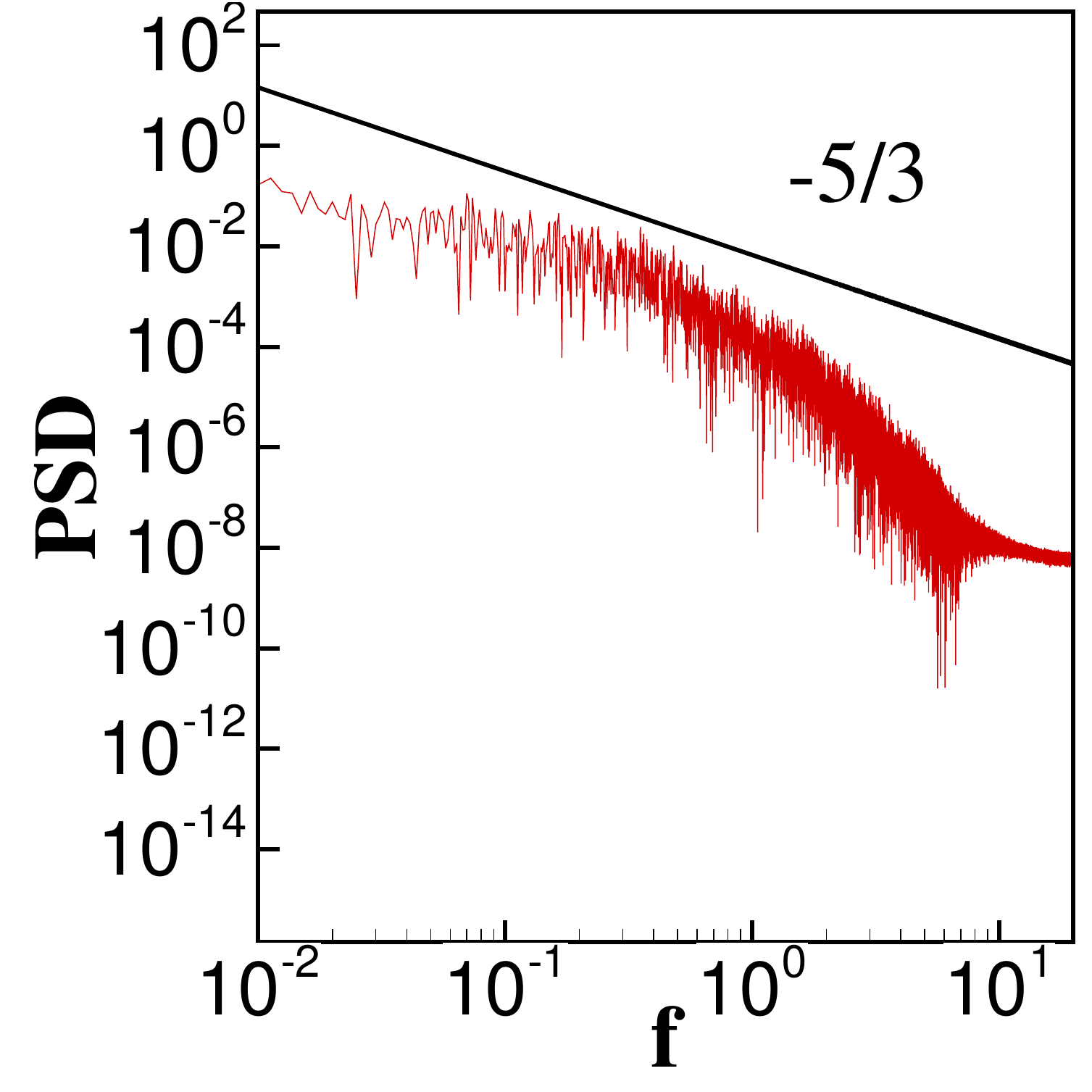}
		\subcaption{$ (0.58,0,1)^\intercal $ }
		\label{cylinder_psd_3rd_1_p1}
	\end{subfigure}
	\begin{subfigure}{0.23\textwidth}
		\includegraphics[width=\textwidth]{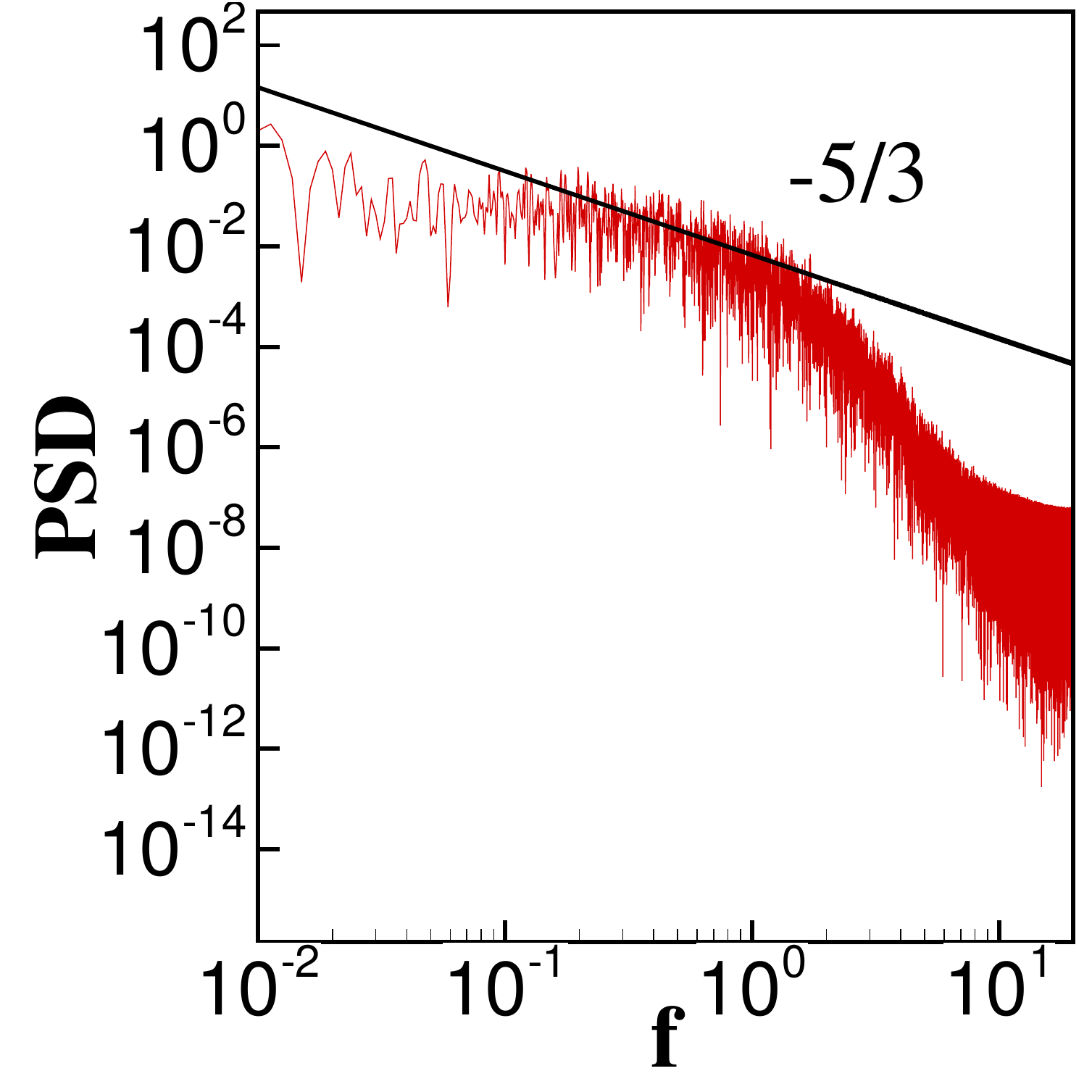} 	
		\subcaption{$ (1.54,0,1)^\intercal $}
		\label{cylinder_psd_3rd_2_p1}
	\end{subfigure}
	\begin{subfigure}{0.23\textwidth}
		\includegraphics[width=\textwidth]{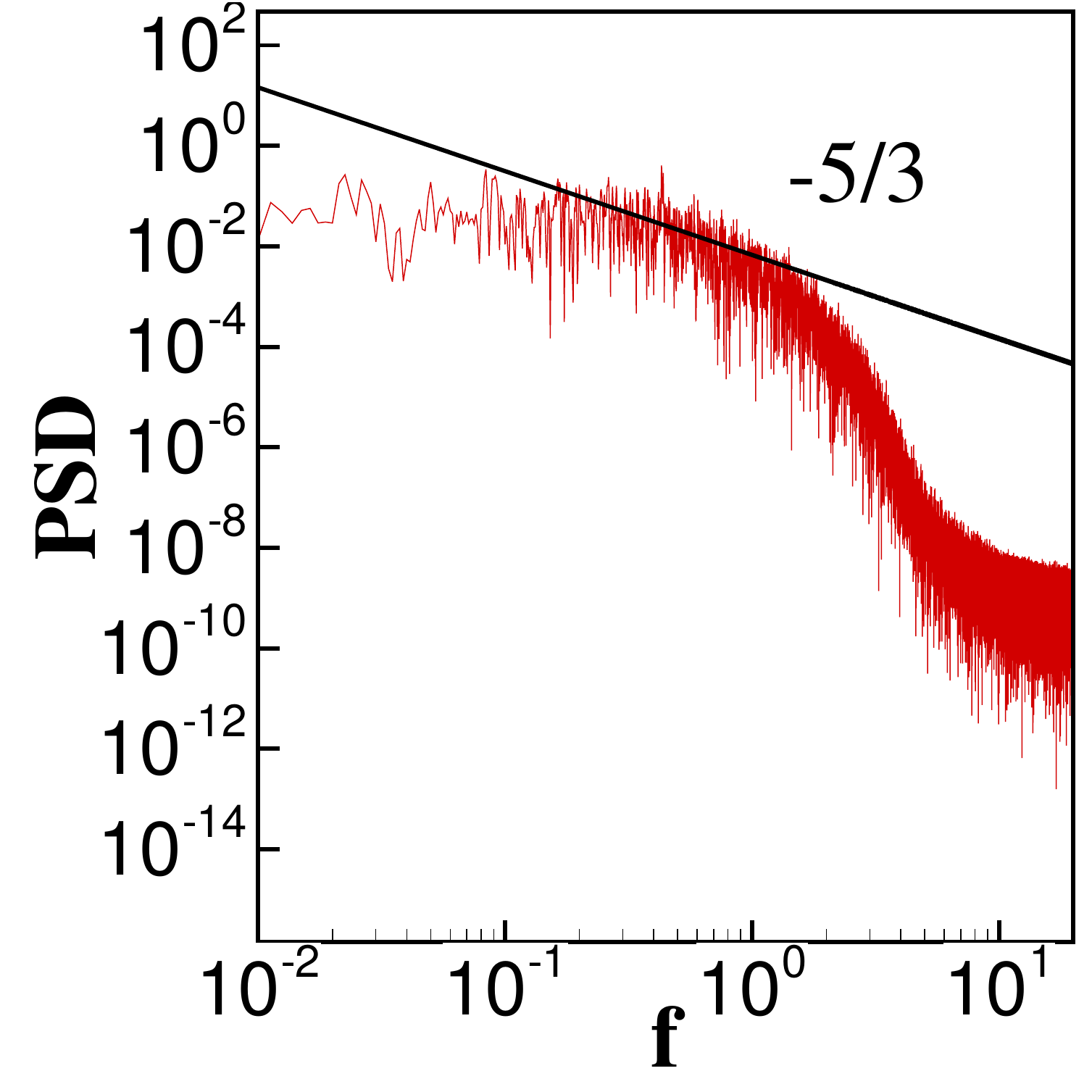} 	
		\subcaption{$ (6,0,1)^\intercal $}
		\label{cylinder_psd_3rd_3_p1}
	\end{subfigure}
	\begin{subfigure}{0.23\textwidth}
		\includegraphics[width=\textwidth]{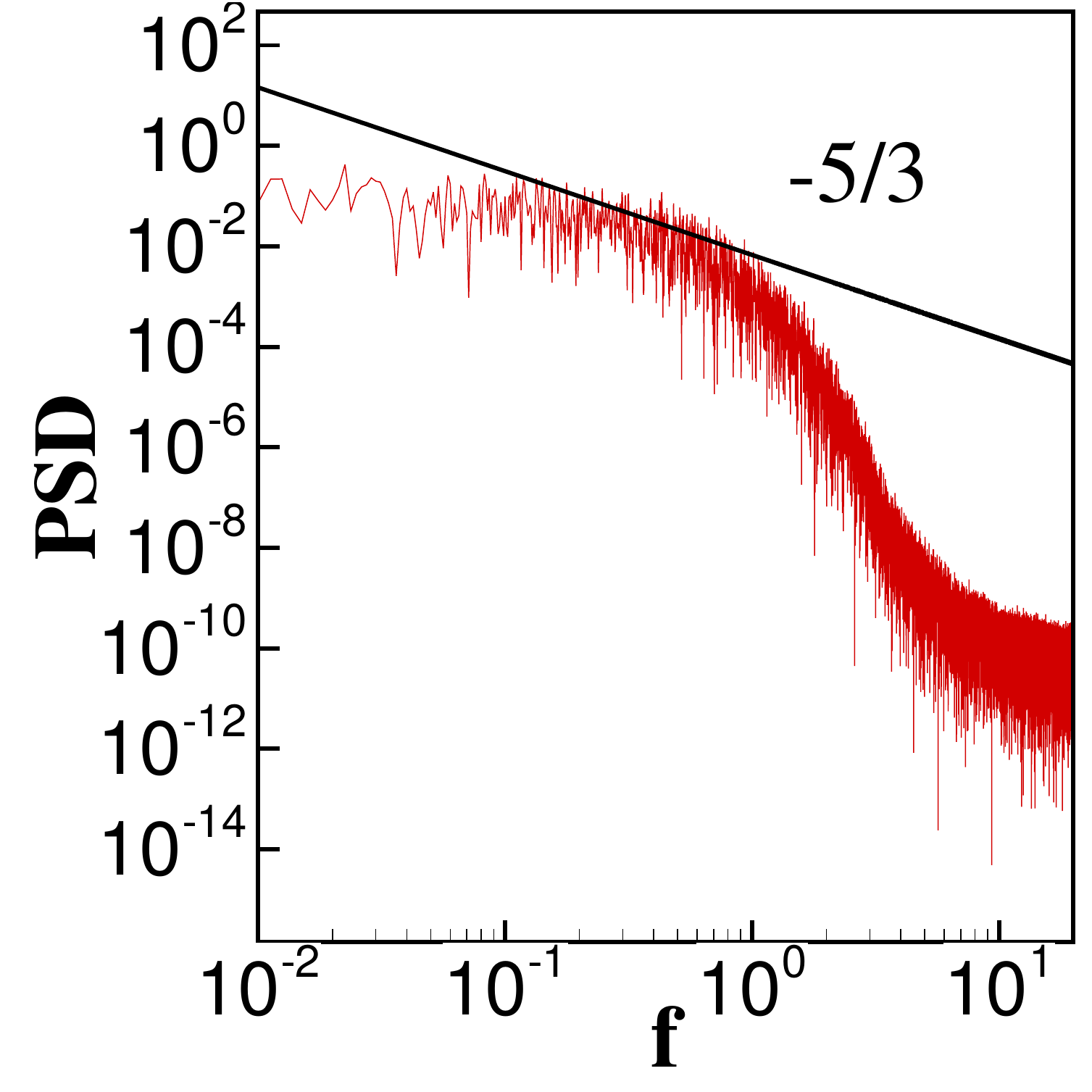} 	
		\subcaption{$ (10,0,1)^\intercal $}
		\label{cylinder_psd_3rd_4_p1}
	\end{subfigure}\\
	\vspace{5pt}
	\begin{subfigure}{0.23\textwidth}
		\includegraphics[width=\textwidth]{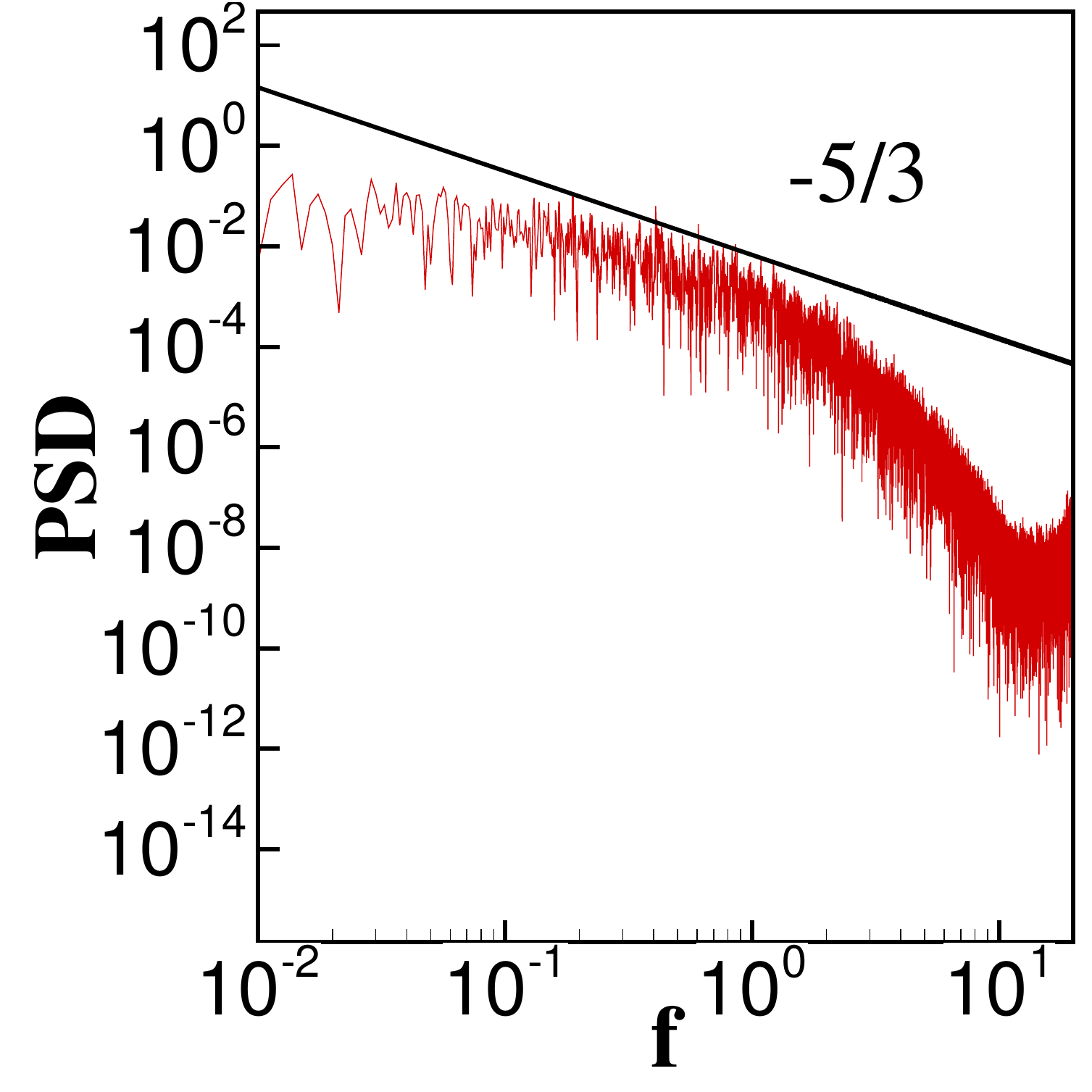}
		\subcaption{$ (0.58,0,1)^\intercal $ }
		\label{cylinder_psd_3rd_1_p2}
	\end{subfigure}
	\begin{subfigure}{0.23\textwidth}
		\includegraphics[width=\textwidth]{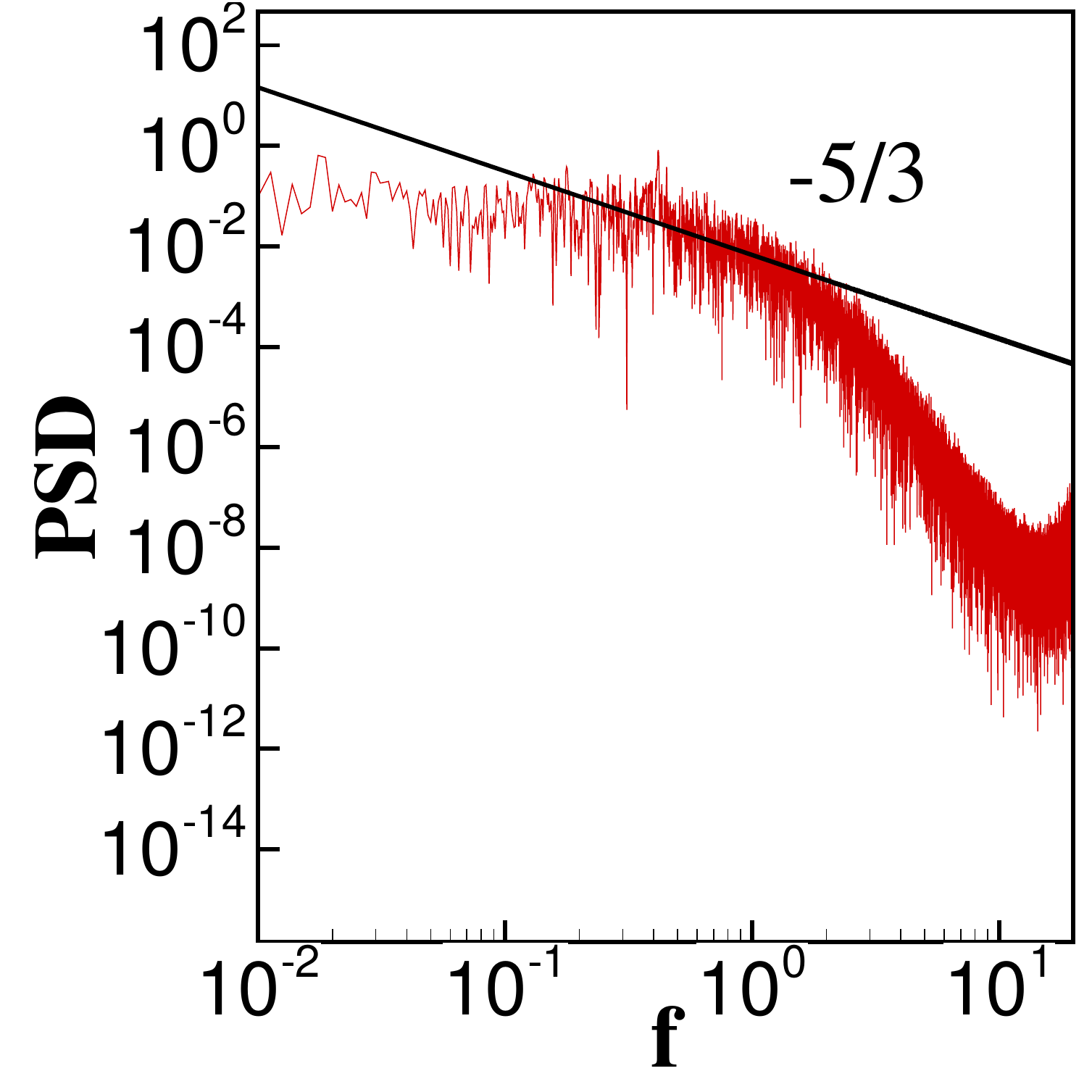} 	
		\subcaption{$ (1.54,0,1)^\intercal $}
		\label{cylinder_psd_3rd_2_p2}
	\end{subfigure}
	\begin{subfigure}{0.23\textwidth}
		\includegraphics[width=\textwidth]{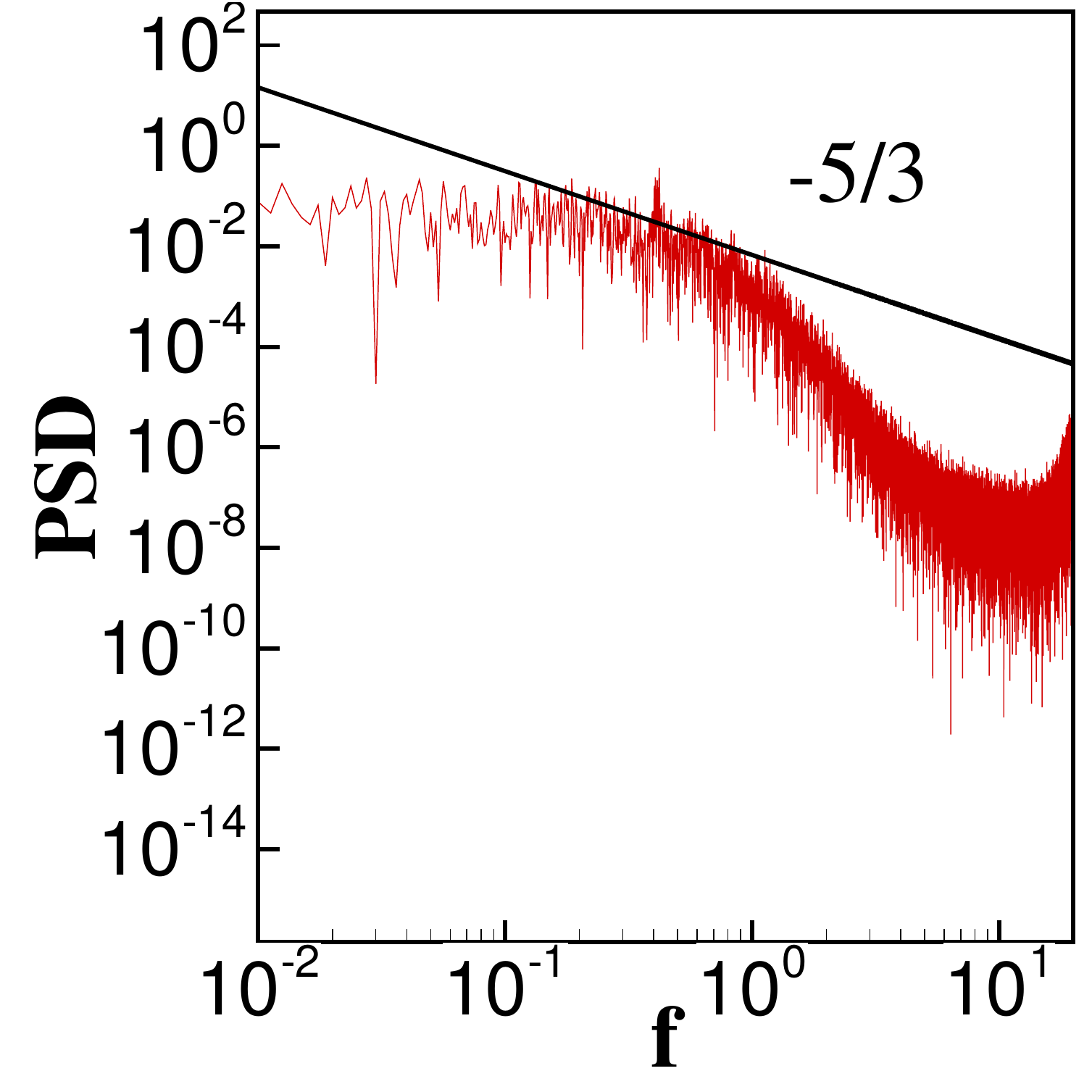} 	
		\subcaption{$ (6,0,1)^\intercal $}
		\label{cylinder_psd_3rd_3_p2}
	\end{subfigure}
	\begin{subfigure}{0.23\textwidth}
		\includegraphics[width=\textwidth]{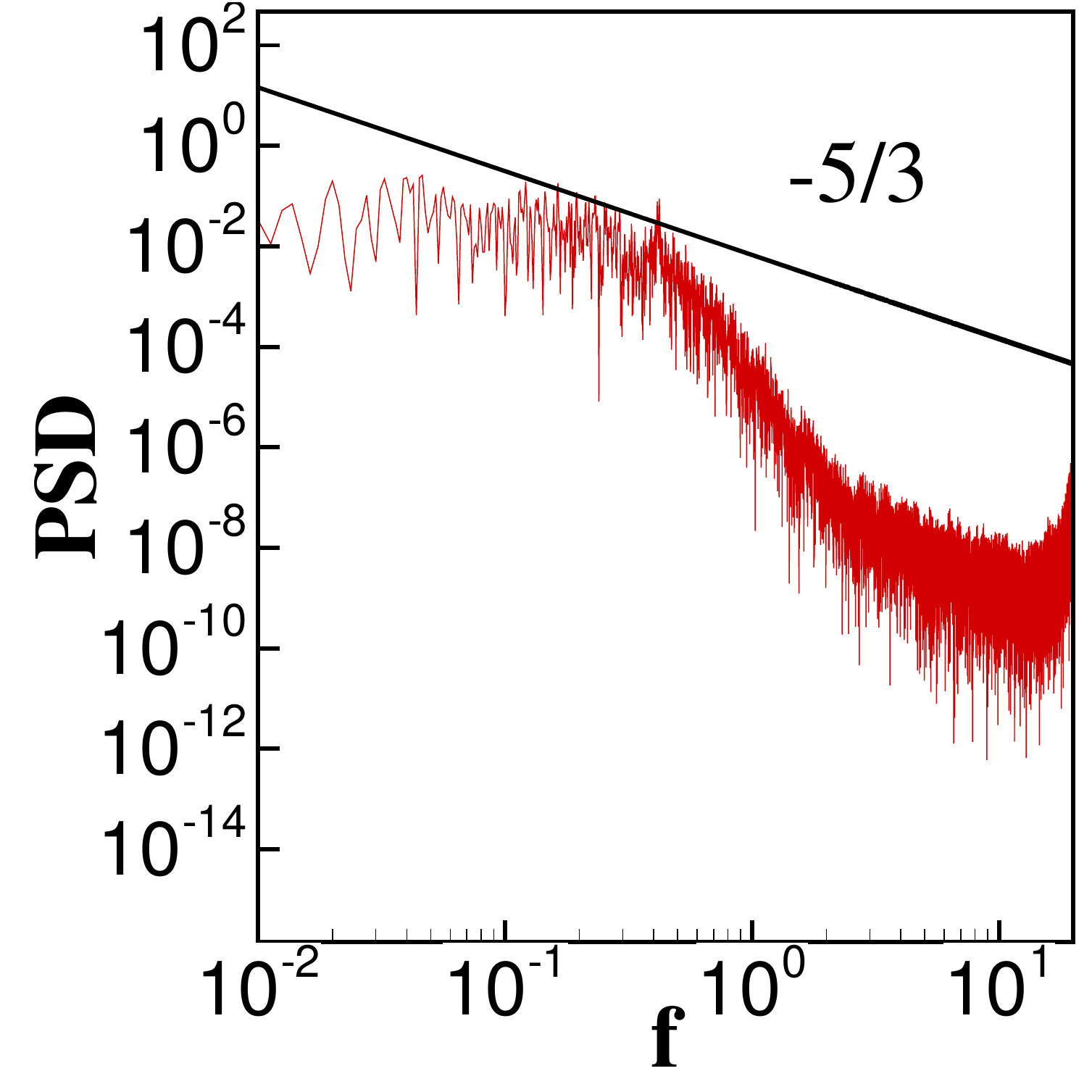} 	
		\subcaption{$ (10,0,1)^\intercal $}
		\label{cylinder_psd_3rd_4_p2}
	\end{subfigure}\\
	\vspace{5pt}
	\begin{subfigure}{0.23\textwidth}
		\includegraphics[width=\textwidth]{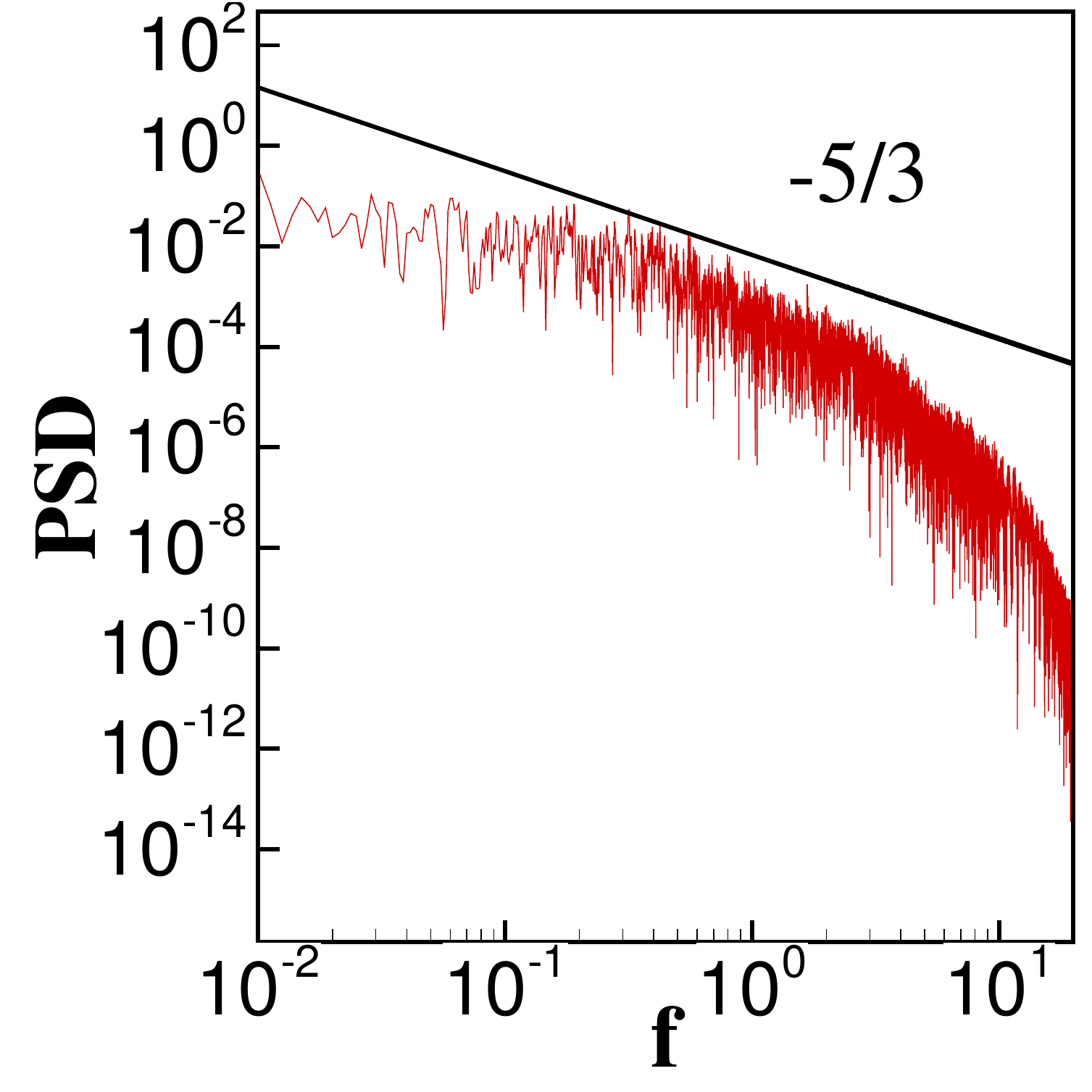}
		\subcaption{$ (0.58,0,1)^\intercal $ }
		\label{cylinder_psd_4th_1}
	\end{subfigure}
	\begin{subfigure}{0.23\textwidth}
		\includegraphics[width=\textwidth]{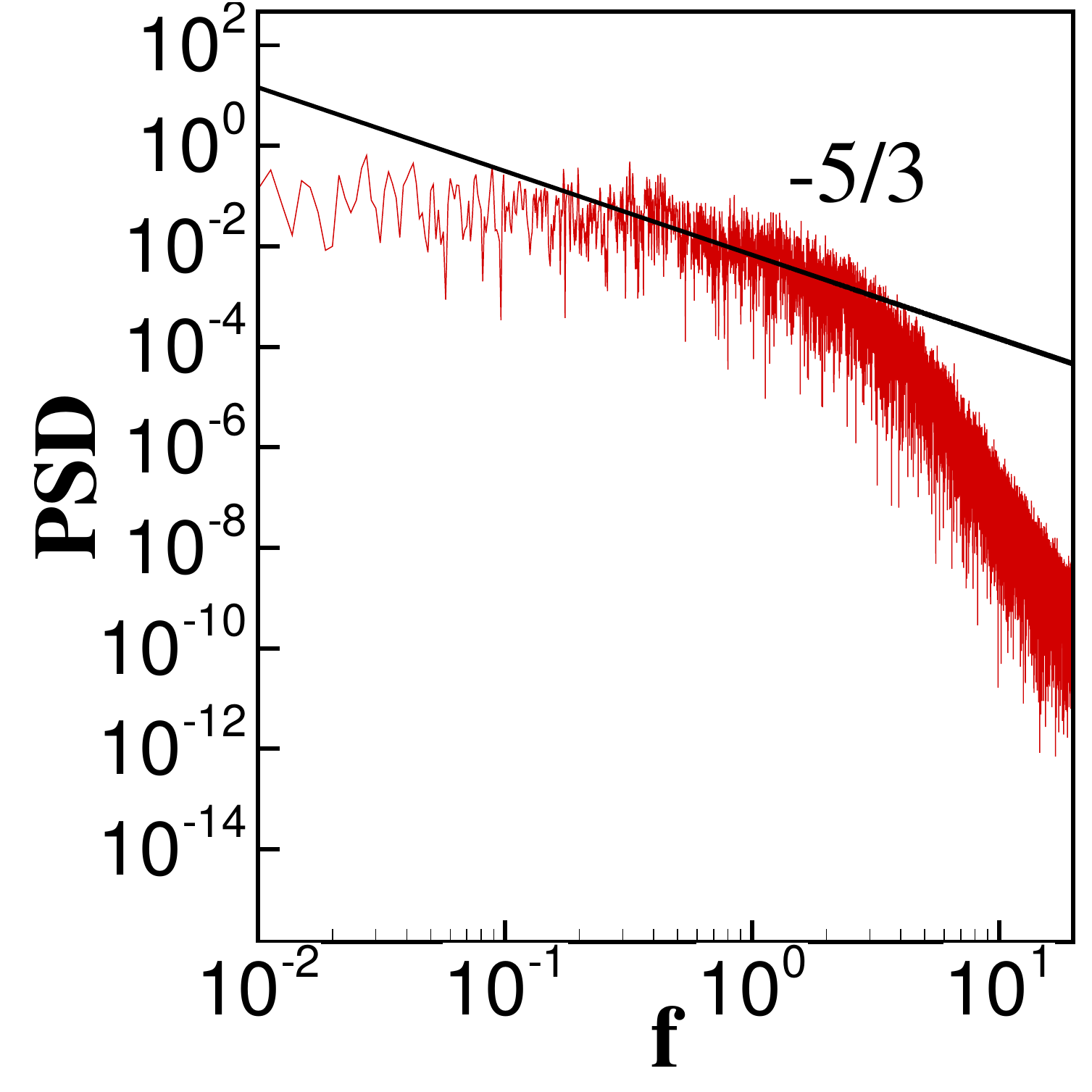} 	
		\subcaption{$ (1.54,0,1)^\intercal $ }
		\label{cylinder_psd_4th_2}
	\end{subfigure}
	\begin{subfigure}{0.23\textwidth}
		\includegraphics[width=\textwidth]{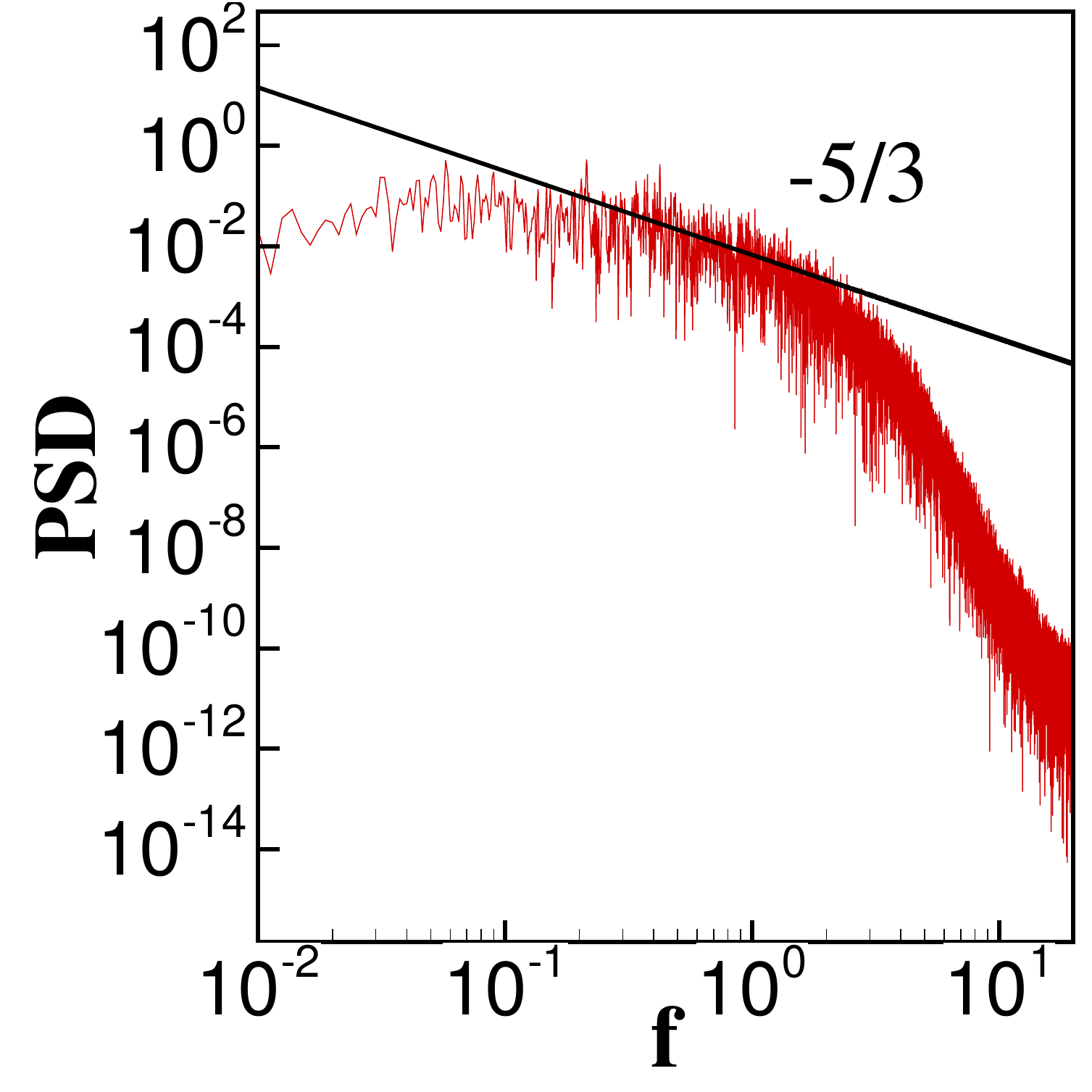} 	
		\subcaption{$ (6,0,1)^\intercal $ }
		\label{cylinder_psd_4th_3}
	\end{subfigure}
	\begin{subfigure}{0.23\textwidth}
		\includegraphics[width=\textwidth]{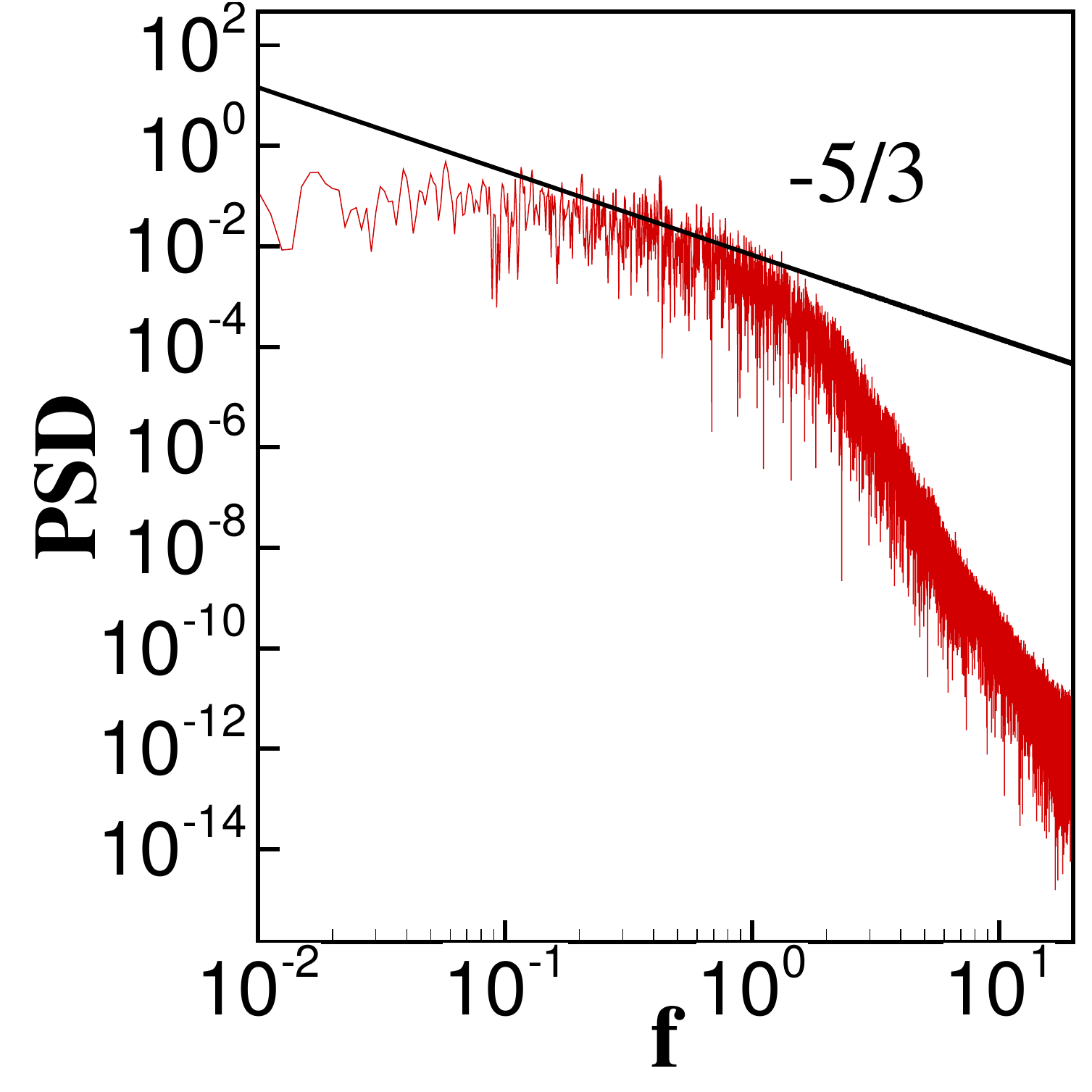} 	
		\subcaption{$ (10,0,1)^\intercal $ }
		\label{cylinder_psd_4th_4}
	\end{subfigure}\\
	\vspace{5pt}
	\begin{subfigure}{0.23\textwidth}
		\includegraphics[width=\textwidth]{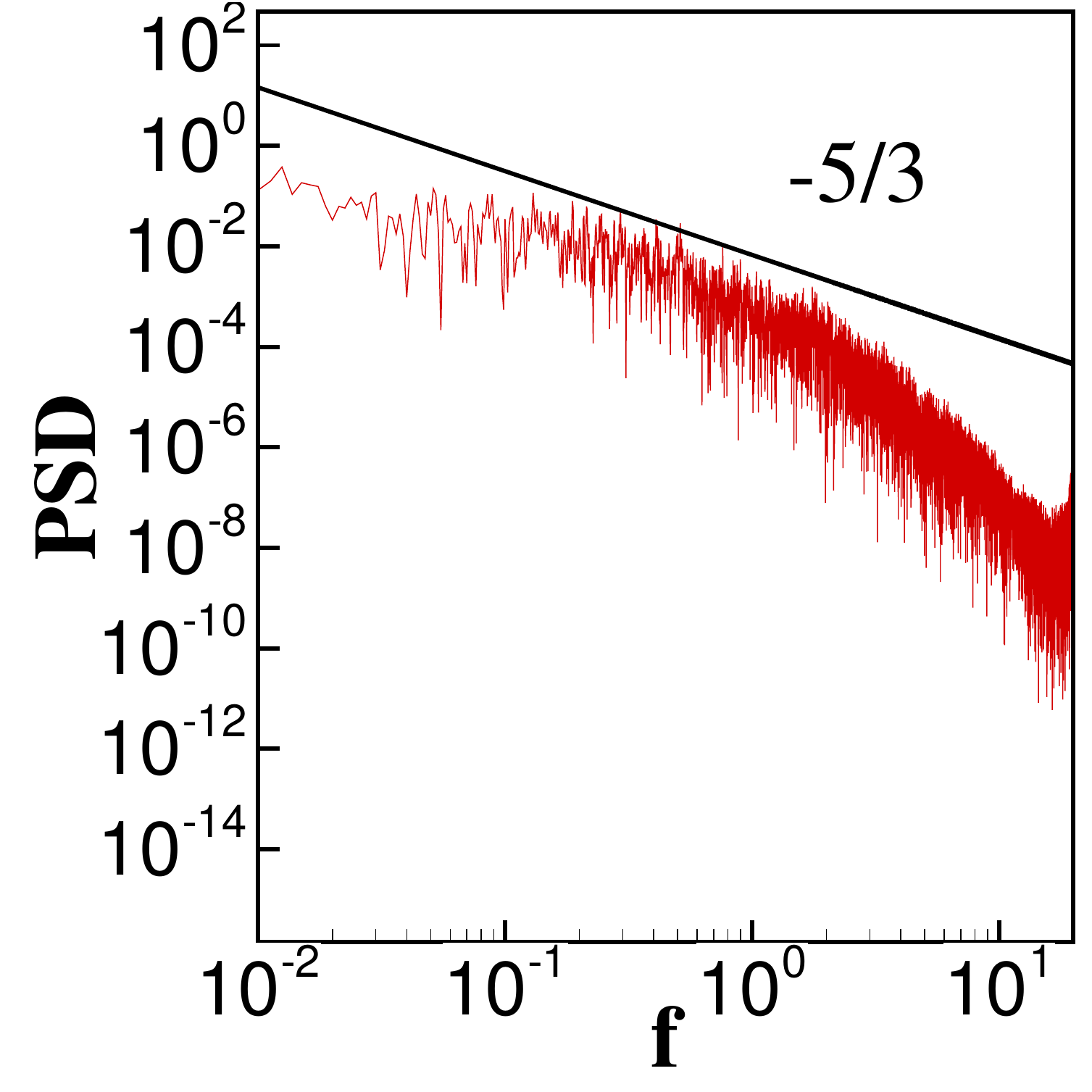}
		\subcaption{$ (0.58,0,1)^\intercal $ }
		\label{cylinder_psd_4th_1_p1}
	\end{subfigure}
	\begin{subfigure}{0.23\textwidth}
		\includegraphics[width=\textwidth]{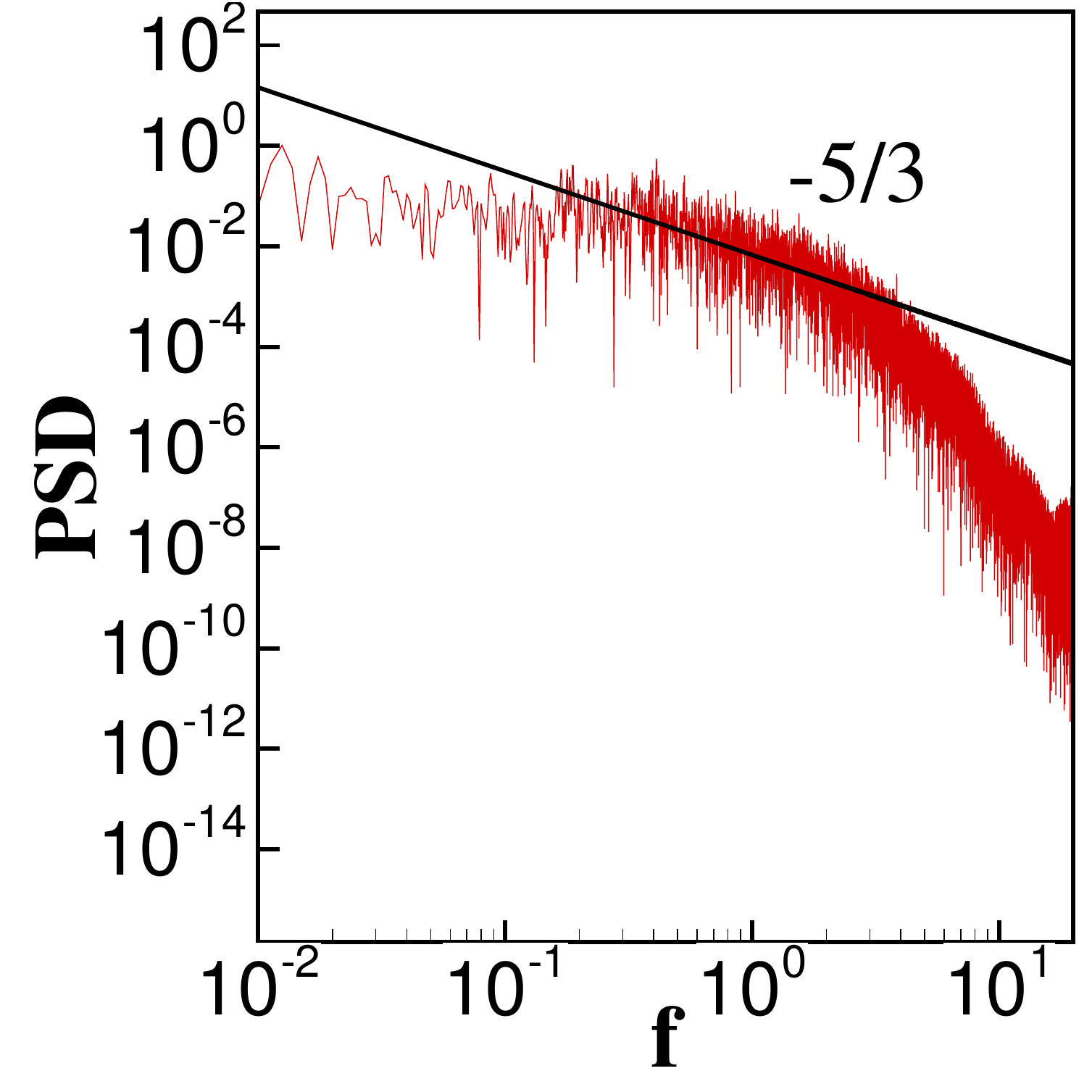} 	
		\subcaption{$ (1.54,0,1)^\intercal $ }
		\label{cylinder_psd_4th_2_p1}
	\end{subfigure}
	\begin{subfigure}{0.23\textwidth}
		\includegraphics[width=\textwidth]{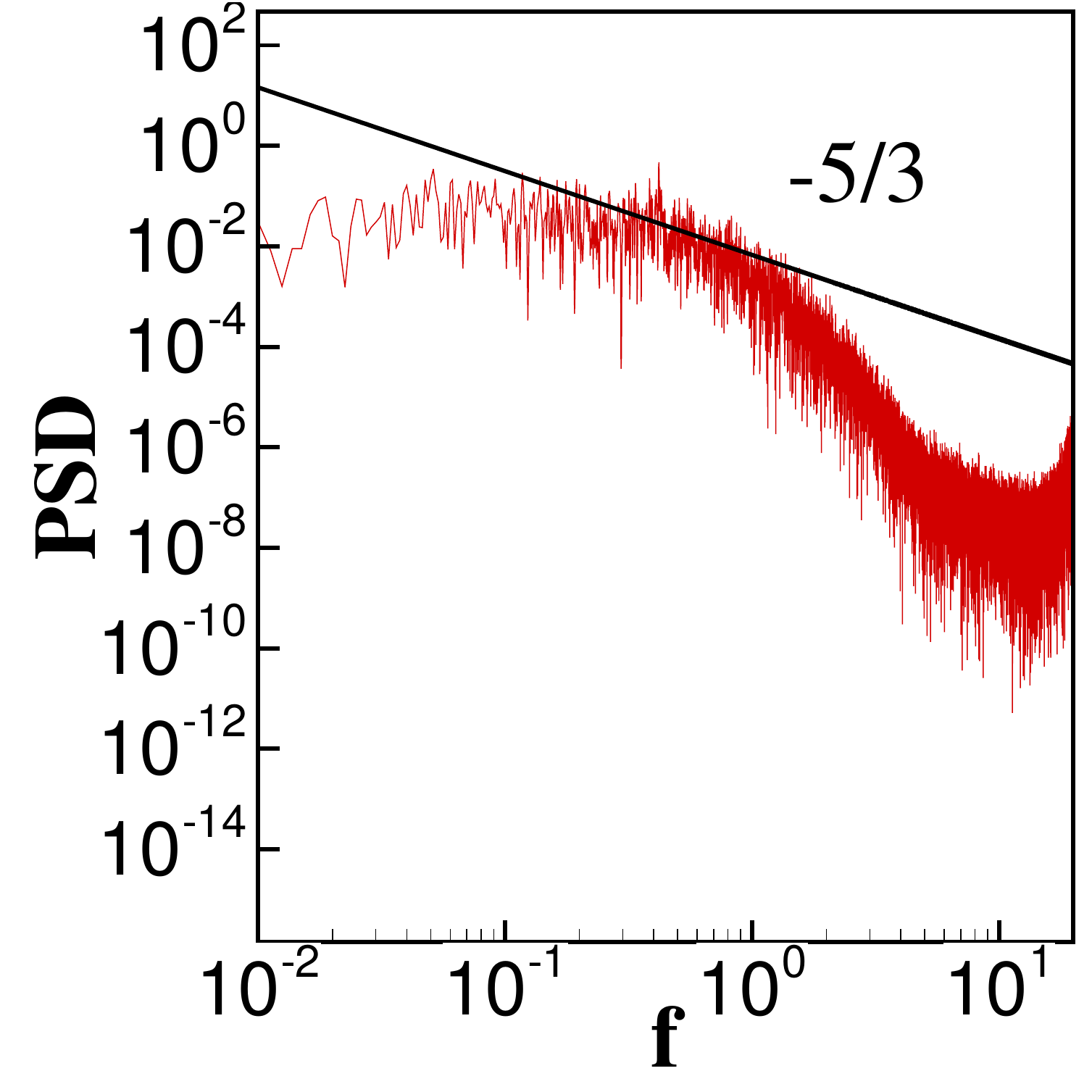} 	
		\subcaption{$ (6,0,1)^\intercal $ }
		\label{cylinder_psd_4th_3_p1}
	\end{subfigure}
	\begin{subfigure}{0.23\textwidth}
		\includegraphics[width=\textwidth]{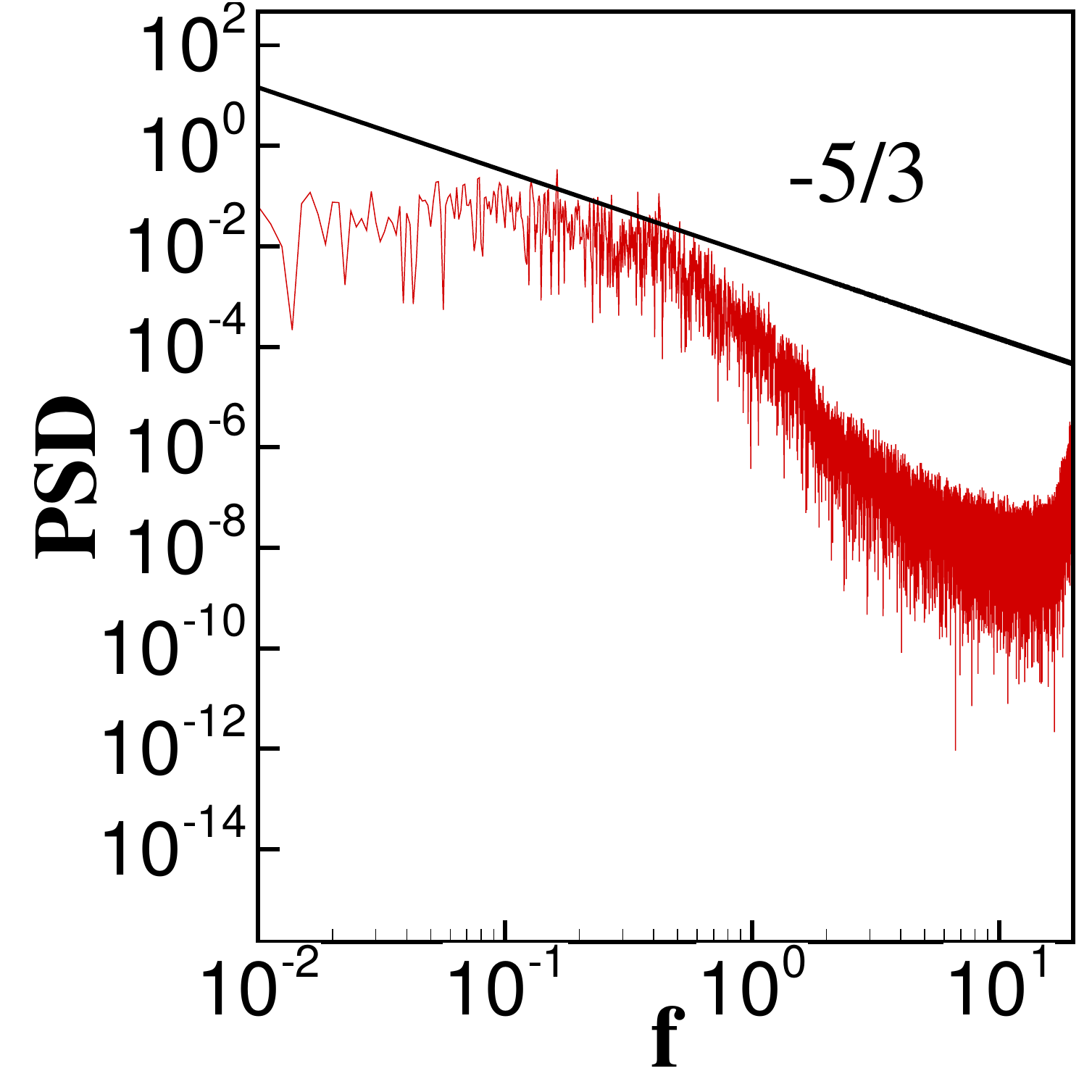} 	
		\subcaption{$ (10,0,1)^\intercal $ }
		\label{cylinder_psd_4th_4_p1}
	\end{subfigure}
	\caption{Power spectral density of the total velocity at different locations in the wake region. (a)--(d) adaptive $ p^2 $ FR without adaptation, (e)--(h) adaptive $ p^2 $ FR with $ (\nu_{max},\nu_{min})=(0.1,0.001) $, (i)--(l) adaptive $ p^2 $ FR with $ (\nu_{max},\nu_{min})=(0.1,0.01) $, (m)--(p) $ p^3 $ FR without adaptation, and (q)--(t) adaptive $ p^3 $ FR with $(\nu_{max},\nu_{min})=(0.1,0.01) $. A line of slope $ -5/3 $ is  added to every graph as a reference.}	
	\label{cylinder_psd_3rd}	
\end{figure}

The run time of all simulations is documented in Table~\ref{cylinder_wall_time}. Overall, the $ p $-adaptive solver can reduce a significant amount of run time. Particularly, for $ p^3 $ FR, the adaptive solver can reduce the run time by 69.58\% when $ (\nu_{max},\nu_{min}) = (0.1,0.01)$.  At $ t=t_{end} $, the adaptive $ p^2 $ FR solver has 239,073 solution points when $ (\nu_{max},\nu_{min}) = (0.1,0.001)$ and 17,4701 solution points when $ (\nu_{max},\nu_{min}) = (0.1,0.01)$; the adaptive $ p^3 $ FR solver has 271,958 solution points. When the turbulence is fully developed, the total number of $ p $-refined elements will be similar at different time steps. In general, the reduction of run time and that of $ n_{sp}^{tot} $ are consistent with each other and $ p $-adaptation with larger $ p_{max} $ is encouraged as shown in~Table~\ref{cylinder_wall_time}. 
From the isosurfaces of $ Q $-criterion, where $ Q=0.5 $, illustrated in Figure~\ref{cylinder_qiso_3rd}, it is intuitive that the $ p $-adaptive solver is more dissipative than $ p $-uniform solver in the wake region  away from the cylinder. The order-of-accuracy   distributions of the adaptive solver with different adaptation parameters at slice $ z=0 $ are also presented. When $ \nu_{min} $ is decreased from $ 0.01 $ to $ 0.001 $, the $ p^2 $ region will substantially extend into the wake region away from the cylinder. Thus, the reduction in run time will decrease. The order-of-accuracy distributions of the adaptive $ p^3 $ FR solver at different slices are shown in Figure~\ref{cylinder_4th_slices} to give a better presentation of the local  $ p $-adaptation. 

\begin{figure}		
	\centering
	\begin{subfigure}{0.49\textwidth}
		\includegraphics[width=\textwidth]{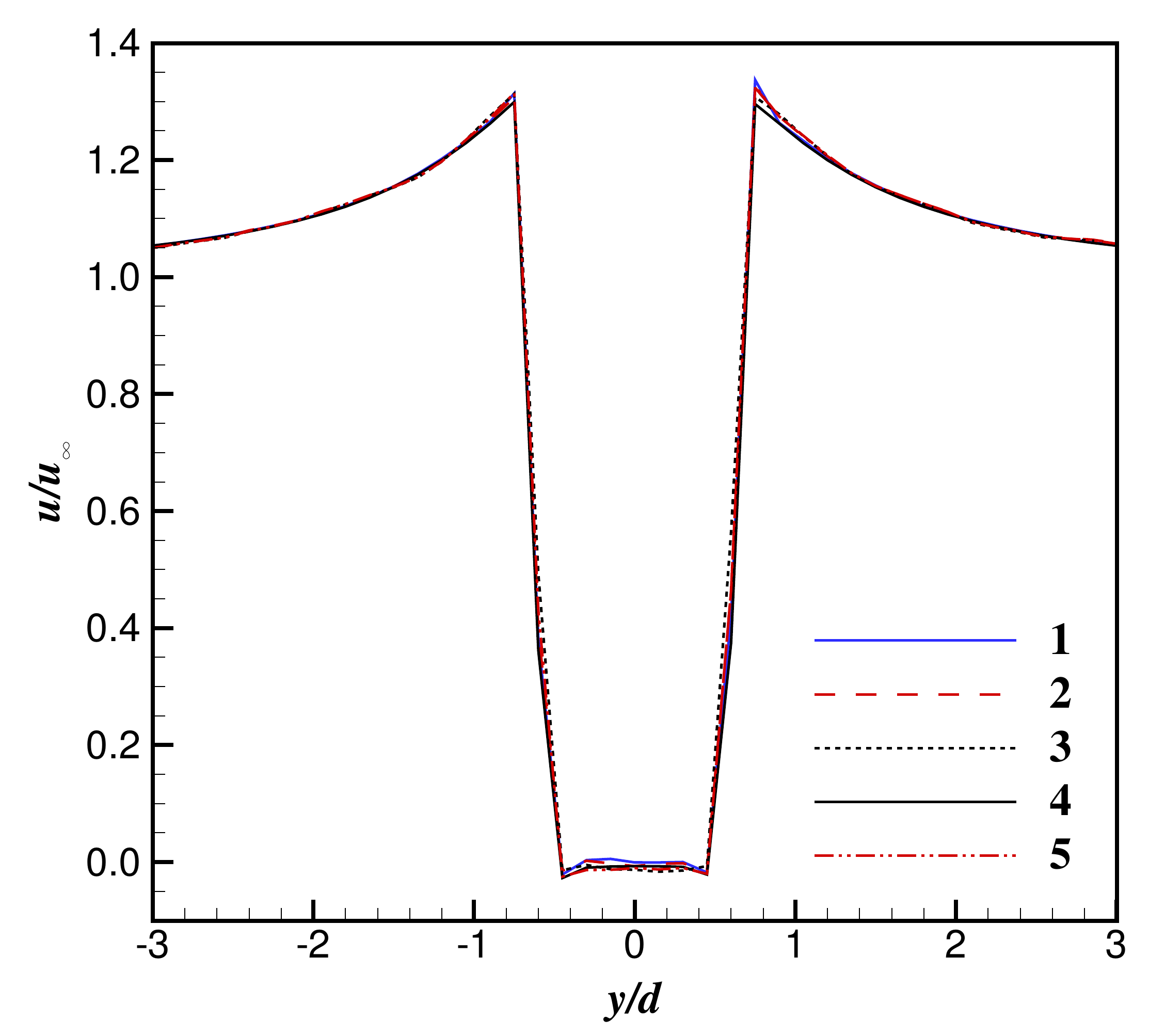}
		\subcaption{$ x = 0.58$ }
		\label{cy_u1}
	\end{subfigure}
	\begin{subfigure}{0.49\textwidth}
		\includegraphics[width=\textwidth]{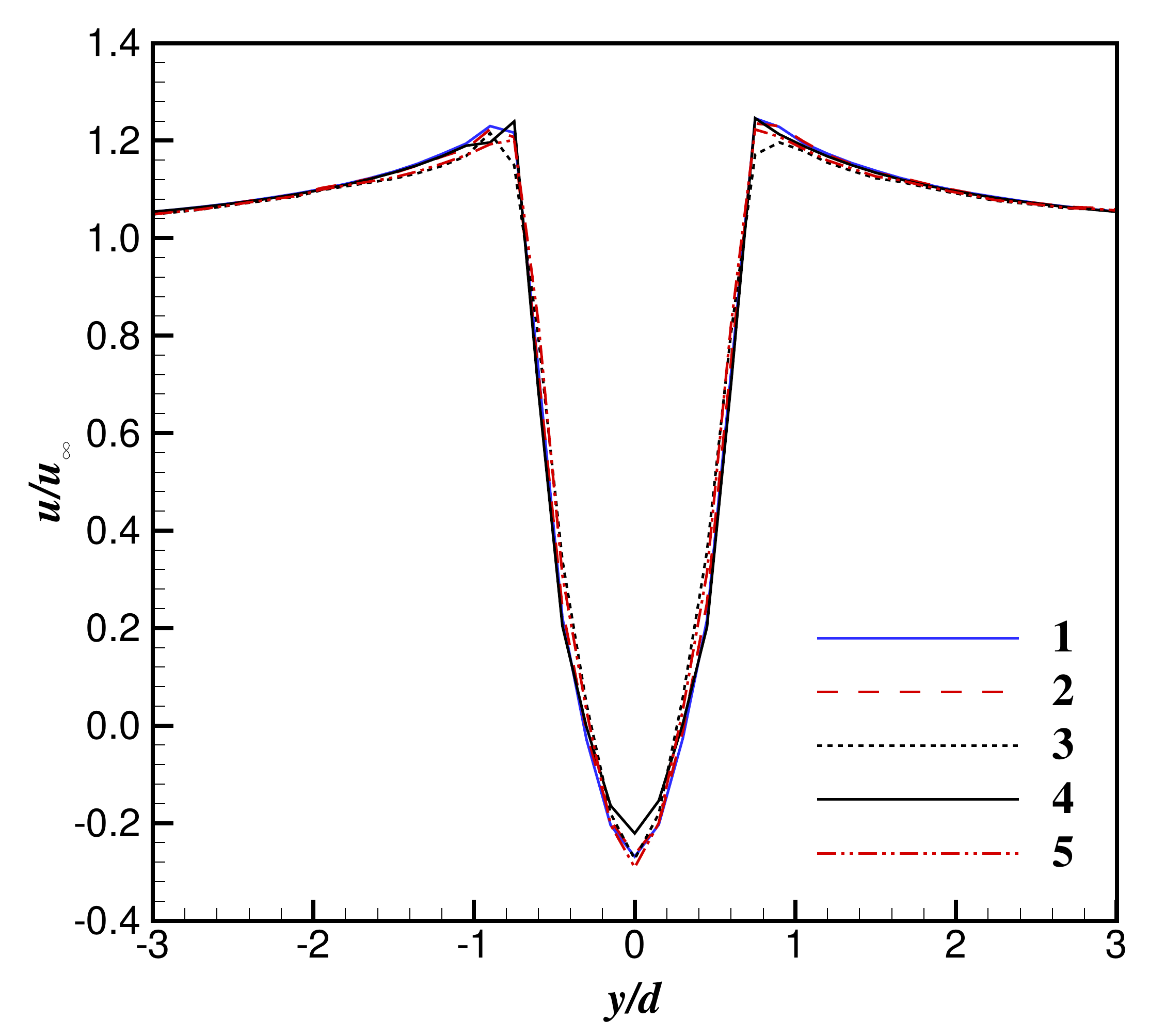} 	
		\subcaption{$ x = 1.54 $ }
		\label{cy_u2}
	\end{subfigure}\\
	\vspace{10pt}
	\begin{subfigure}{0.49\textwidth}
		\includegraphics[width=\textwidth]{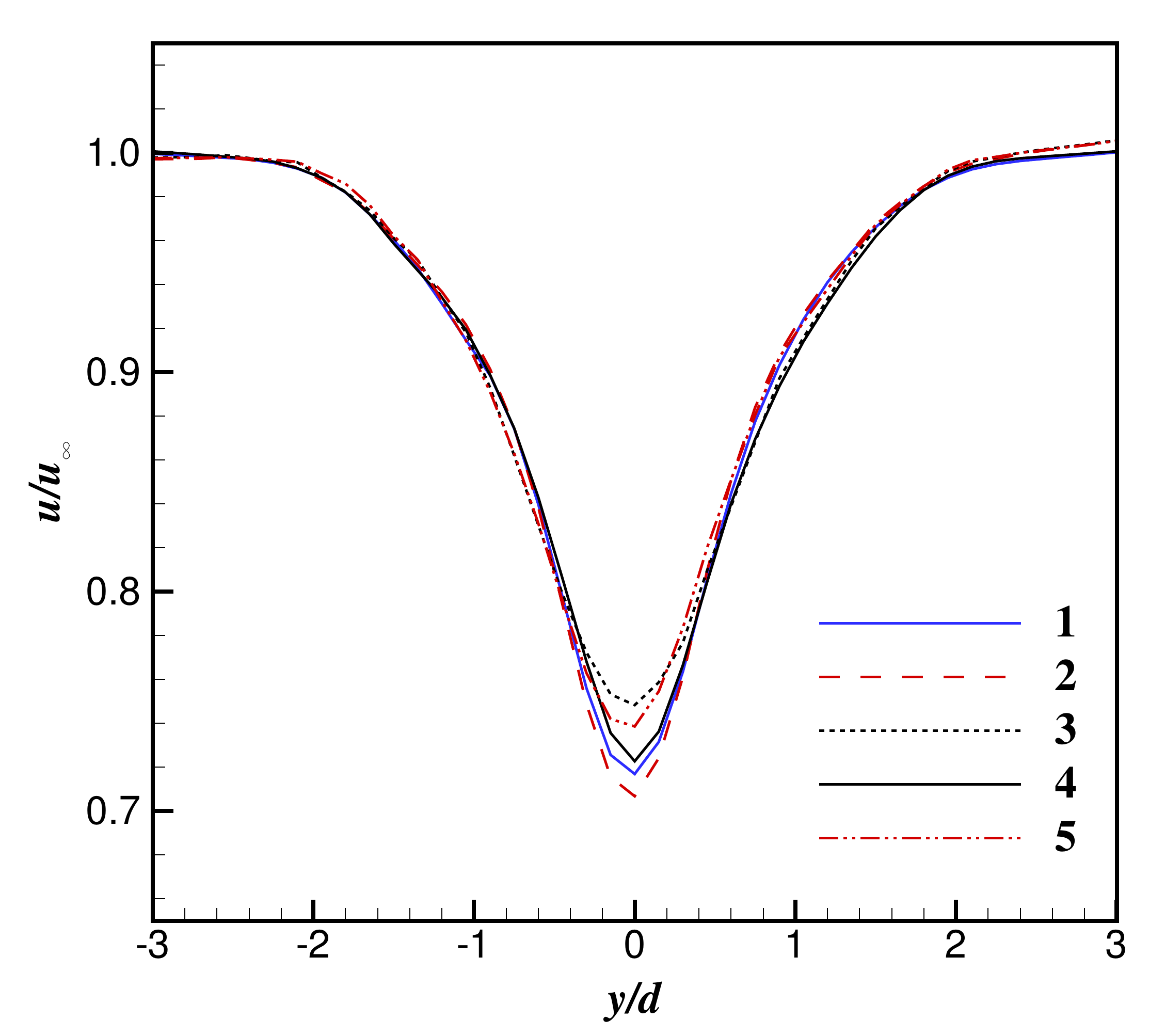}
		\subcaption{$ x = 6$ }
		\label{cy_u3}
	\end{subfigure}
	\begin{subfigure}{0.49\textwidth}
		\includegraphics[width=\textwidth]{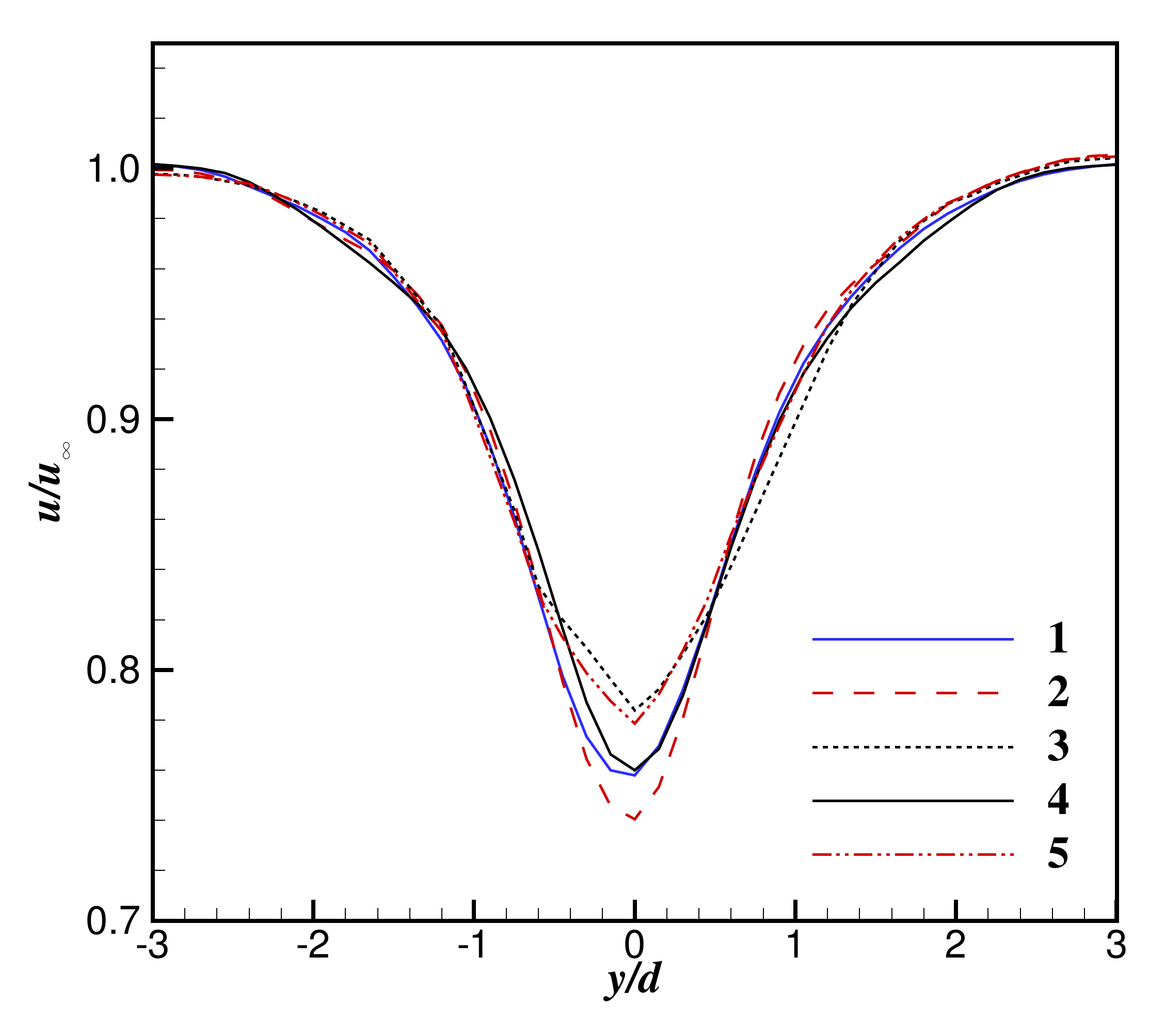} 	
		\subcaption{$ x = 10 $ }
		\label{cy_u4}
	\end{subfigure}
	\caption{Profiles of velocity component $ u $ in $ x $-direction at different locations. Legend 1, $ p $-uniform $ p^2 $ FR; Legend 2, $ p $-adaptive $ p^2 $ FR with  $ (\nu_{max},\nu_{min}) = (0.1,0.001)$; Legend 3, $ p $-adaptive $ p^2 $ FR with  $ (\nu_{max},\nu_{min}) = (0.1,0.01)$; Legend 4, $ p $-uniform $ p^3 $ FR; Legend 5, $ p $-adaptive $ p^3 $ FR with $ (\nu_{max},\nu_{min}) = (0.1,0.01)$;}	
	\label{cylinder_velocity}	
\end{figure}

We further examine the power spectral density (PSD) of the total velocity at four locations in the wake region, namely $ (0.58,0,1)^\intercal $, $ (1.54,0,1)^\intercal $, $ (6,0,1)^\intercal$, and $ (10,0,1)^\intercal $ as presented in Figure~\ref{cylinder_psd_3rd}. A reference line of slope $ -5/3 $ is included in every graph. Compared to the DNS results in~\cite{lehmkuhl2013low}, a large portion of the inertial range can be resolved at the first two points.  At the last two points, $ (6,0,1)^\intercal$, and $ (10,0,1)^\intercal$,  the adaptive solver gets more dissipative as the parameter $ \nu_{min} $ increases. The velocity profiles  at different positions on the $ x $-axis in Figure~\ref{cylinder_velocity}, where $ y/d\in[-3,3] $, further demonstrates this observation. At $ x=0.58 $ and $ x=1.54 $, the velocity profiles of the $ p $-adaptive FR methods are close to those of the $ p $-uniform FR methods. At $ x=6 $ and $ x=10 $, the local extrema of the $ p $-adaptive FR are largely dissipated, even when $ (\nu_{min},\nu_{max}) = (0.1,0.001) $.  

\begin{figure}	
	\centering
	\begin{subfigure}{0.49\textwidth}
		\includegraphics[width=\textwidth]{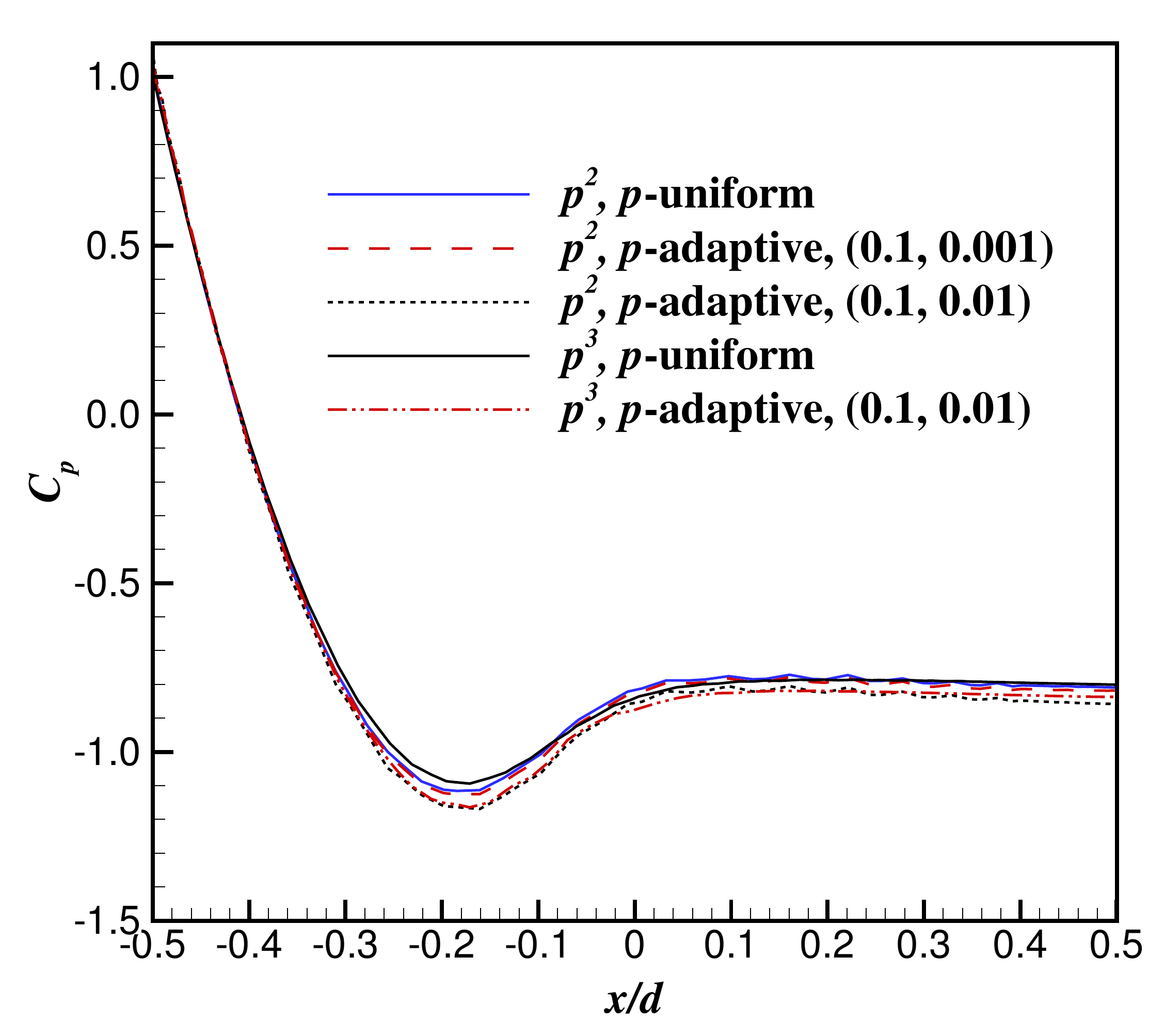}
		\subcaption{Surface pressure coefficient $ C_p $}
		\label{cy_cp}
	\end{subfigure}
	\begin{subfigure}{0.49\textwidth}
		\includegraphics[width=\textwidth]{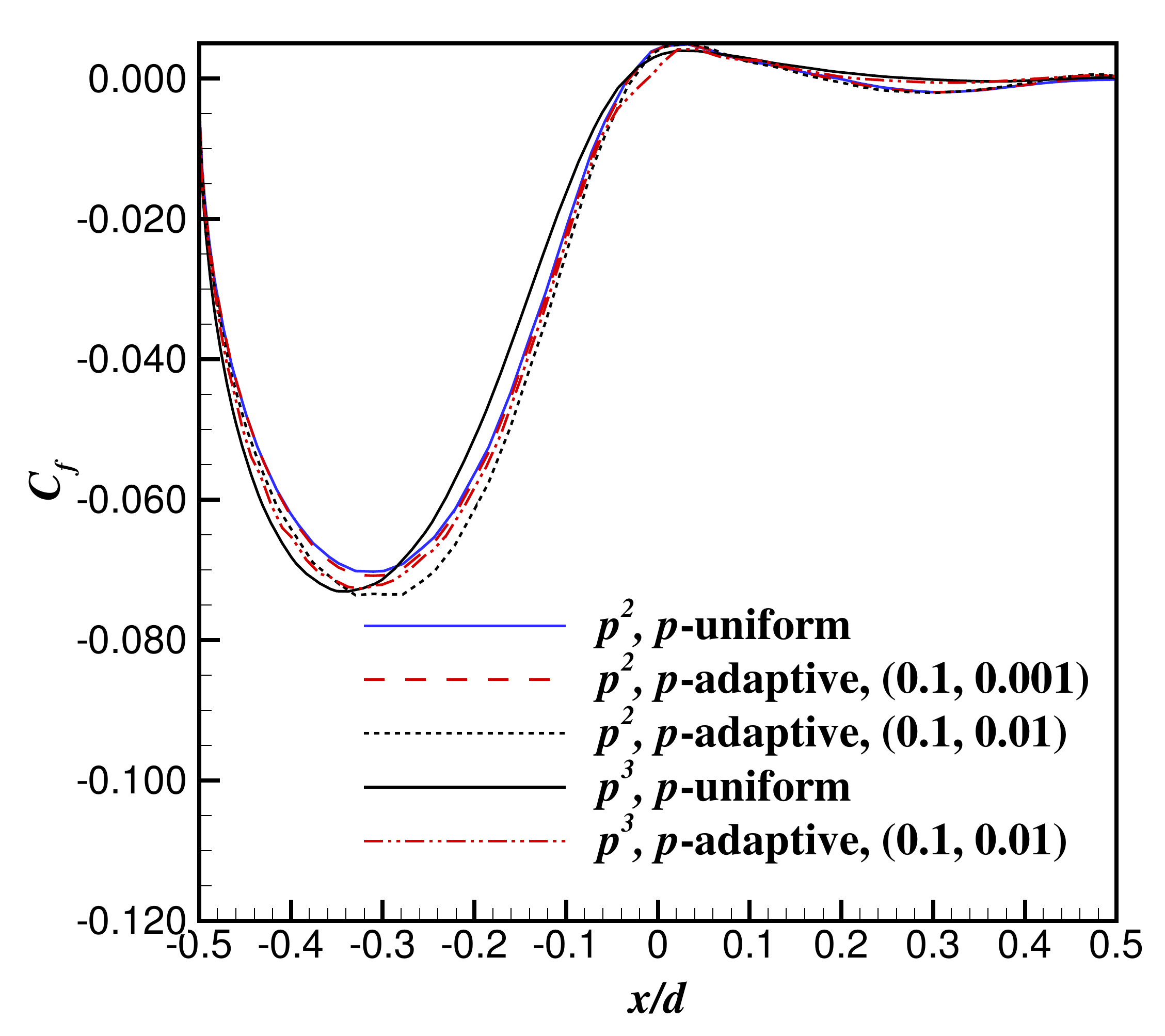} 	
		\subcaption{Surface friction coefficient $ C_f $}
		\label{cy_cf}
	\end{subfigure}
	\caption{ Surface $ C_p $ and $ C_f $ in the averaged field of the transitional flow over the infinite cylinder at $ \text{Re}=3900 $.}	
	\label{cylinder_surface_force}	
\end{figure}

The surface pressure coefficient $ C_p $ and surface friction coefficient $ C_f $ on the $ y>0 $ side of the cylinder   are presented in Figure~\ref{cylinder_surface_force}. When $ (\nu_{max},\nu_{min}) = (0.1,0.001)$, the results of the adaptive $ p^2 $ FR method is 
close to that of the $ p $-uniform $ p^2 $ FR method. When $ (\nu_{max},\nu_{min}) = (0.1,0.01)$, the results of adaptive solver are still comparable to those of the $ p $-uniform solver even though a large portion of the wake region uses $ p^1 $ polynomials only.

\clearpage
\subsection{Under-resolved simulation of the transitional flow over the SD7003 wing}
\begin{figure}		
	\centering
	\begin{subfigure}{0.49\textwidth}
		\includegraphics[width=\textwidth]{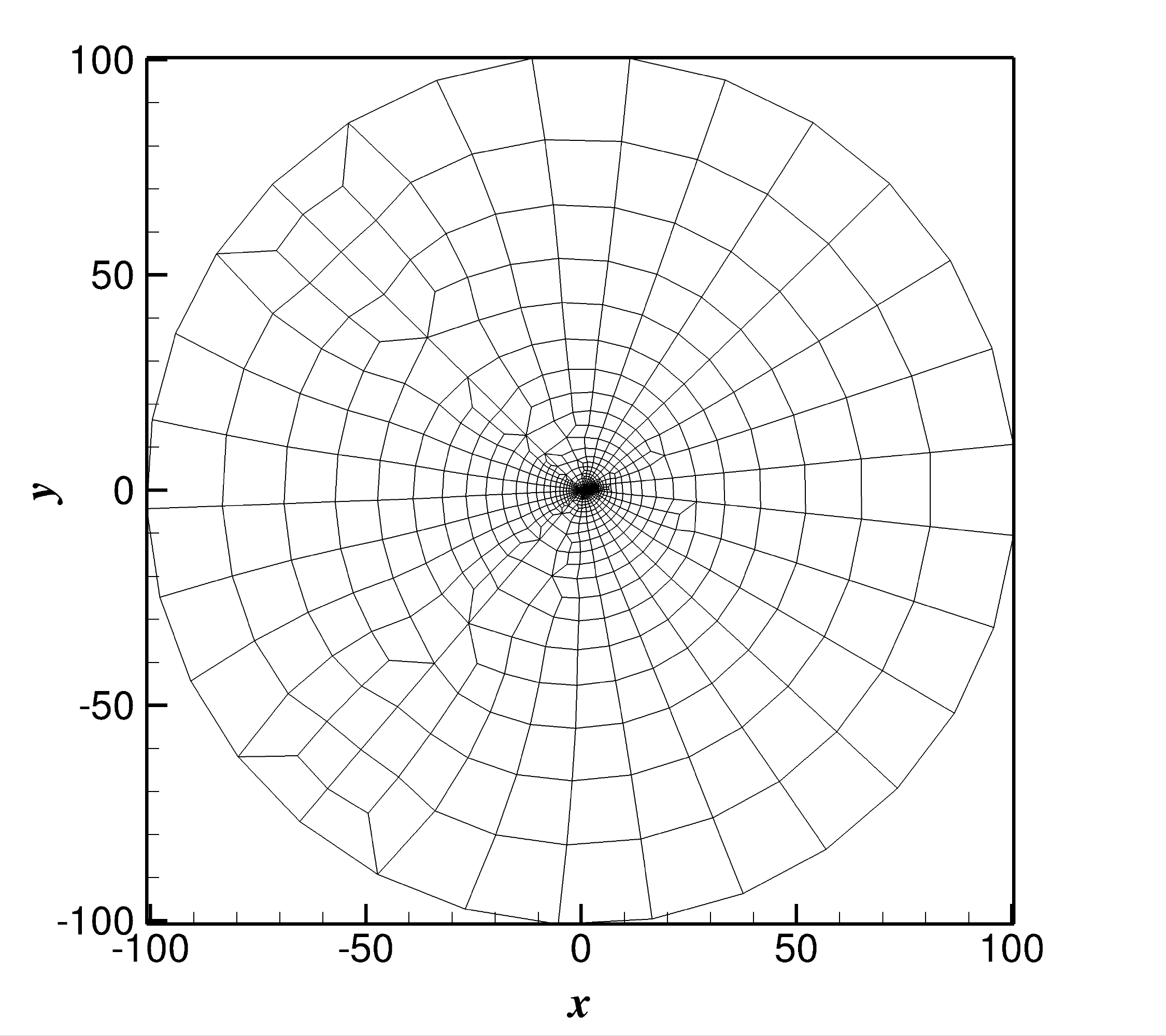}
		\subcaption{Global view}
		\label{sd_global}
	\end{subfigure}
	\begin{subfigure}{0.49\textwidth}
		\includegraphics[width=\textwidth]{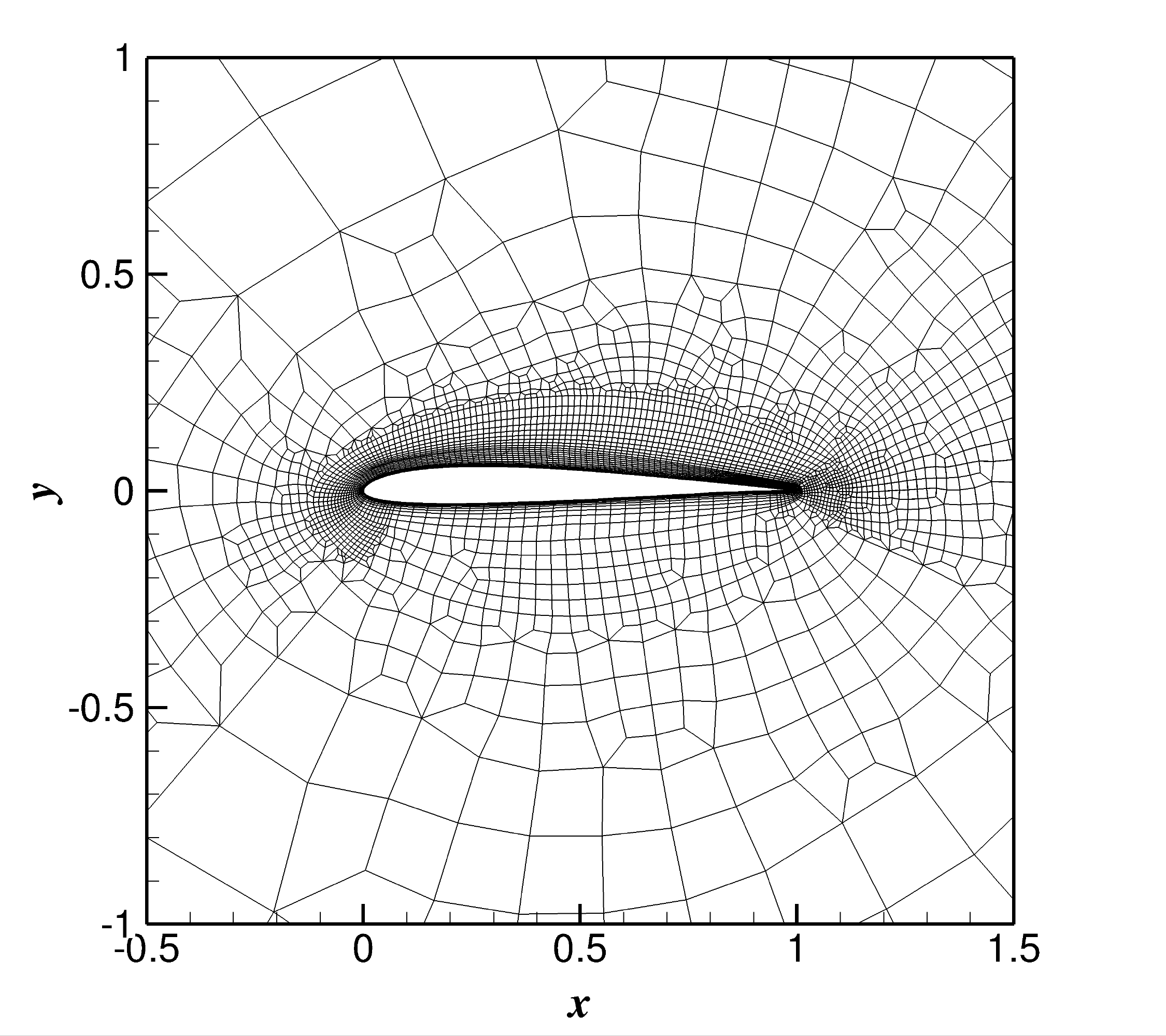} 	
		\subcaption{Close-up view}
		\label{sd_local}
	\end{subfigure}
	\caption{2D views of the unstructured mesh around the SD7003 wing.}	
	\label{sd7003_mesh}	
\end{figure}

In this section, we apply the $ p $-adaptive solver to the simulation of the transitional flow over the SD7003 wing. The geometry of the wing is obtained from the $1^{\text{st}}$ International Workshop on High-Order CFD Methods. The chord length of the wing is $ c = 1 $ with the sharp trailing edge rounded by an arc of radius $ r\approx 0.0004 $. The inflow conditions are $ (\rho_\infty, u_\infty, v_\infty, w_\infty, \text{Ma}_\infty)^\intercal=(1,1,0,0,0.1)^\intercal $. The angle of attack of the inflow is $ 8^\circ $. The Reynolds number of the inflow with respect to the chord length of the wing is $ \text{Re}_c ={\rho_\infty u_\infty c}/{\mu}= 60000 $. The Prandtl number is $ \text{Pr} = 0.72 $.  2D views of the unstructured mesh used for under-resolved simulation are illustrated in Figure~\ref{sd7003_mesh}. The height of the first layer of elements in the normal direction of the wing is $ 0.0003c $. We extrude the 2D mesh along the $ z $ direction to obtain the 3D mesh. The first 3D mesh has 20 layers in the $ z $ direction and 109,540 hexahedral elements in total. The thickness of each layer is $ 0.01c $. 
The second one has ten layers in the spanwise direction and each layer has a thickness of  $ 0.02c $. Taking advantage of $ p $-adaptation, only the $ p $-adaptive  $ p^2 $ and $ p^3 $ FR methods  are used to simulate this problem. We will provide the reduction in the total number of solution points $ n_{sp}^{tot} $ to demonstrate the efficiency of $ p $-adaptation. 
The   $ p $-adaptation parameters are set as $ (\nu_{max},\nu_{min})=(0.1,0.001)$. 

The time step $ \Delta t =0.002 $ and tolerance for the pseudo transient continuation $ tol_{rel}^{pseudo} = 10^{-4}$ are employed for all numerical experiments.  For the simulation on the 20-layer mesh, ESDIRK2 is employed for time integration with $ tol_{rel}^{gmres} = 10^{-1}$ for the GMRES solver. In the pseudo transient continuation,  $ \Delta \tau_{min} = 0.0002 $ and $ \Delta \tau_{max} = 0.004 $ are used for SER. We first run the simulation with adaptive $ p^2 $ FR until $ t_1 = 26 $. The instantaneous values of conservative variables in $ t \in (20,26] $ are used for averaging. Then we increase the $ p_{max} $ to $ p_{max} = 3 $ and resume the simulation until  $ t_2 = 42 $. Then, averaging is done for  $ t\in(36,42] $.
When $ p^3 $  FR with $ p $-adaptation is used to simulate this problem on the 10-layer mesh, aliasing errors will lead to failure. Therefore, we employ a simple nodal polynomial filtering method proposed by Fisher and Mullen~\cite{fischer2001filter} for every element whose polynomial degree exceeds two. The $ p^2 $ polynomial is employed as a basis to perform a cut-off as  $ \widetilde{\boldsymbol{q}} _{p^3}= (1-\alpha)\boldsymbol{q}_{p^3} + \alpha \mathbb{P}_{p^2}^{p^3} \boldsymbol{q}_{p^3} $, where $ \mathbb{P}_{p^2}^{p^3} $ is the projection operation from $ p^3 $ to $ p^2 $ and $ \alpha = 0.2 $. With this nodal polynomial  filtering, a small amount of dissipation is introduced to stabilize the numerical methods. ESDIRK4 serves as the time integrator for the simulation on the 10-layer mesh with adaptive $ p^3 $ FR. To increase the robustness of the pseudo transient continuation, $ tol_{rel}^{gmres} = 10^{-2}$, $ \Delta \tau_{min} = 0.0002 $, and $ \Delta \tau_{max} = 0.002 $ are used. We run the simulation until $ t_1=26 $ only. Solutions in $ t \in (20,26] $ are averaged for statistics.

\begin{figure}		
	\centering
	\begin{subfigure}{0.49\textwidth}
		\includegraphics[width=\textwidth]{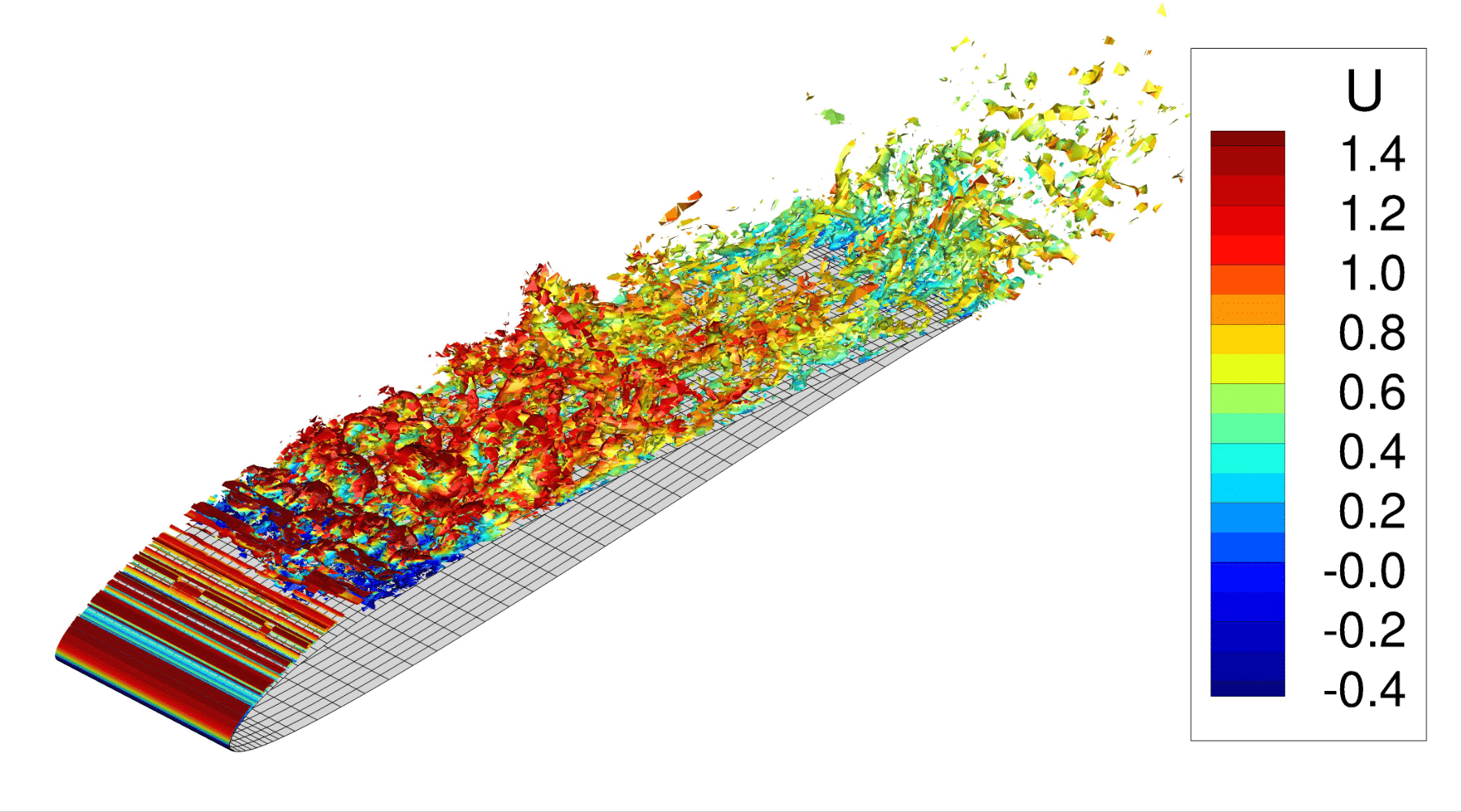}
		\subcaption{Adaptive $ p^2 $ FR, $ t=26 $}
		\label{sd_u}
	\end{subfigure}
	\begin{subfigure}{0.49\textwidth}
		\includegraphics[width=\textwidth]{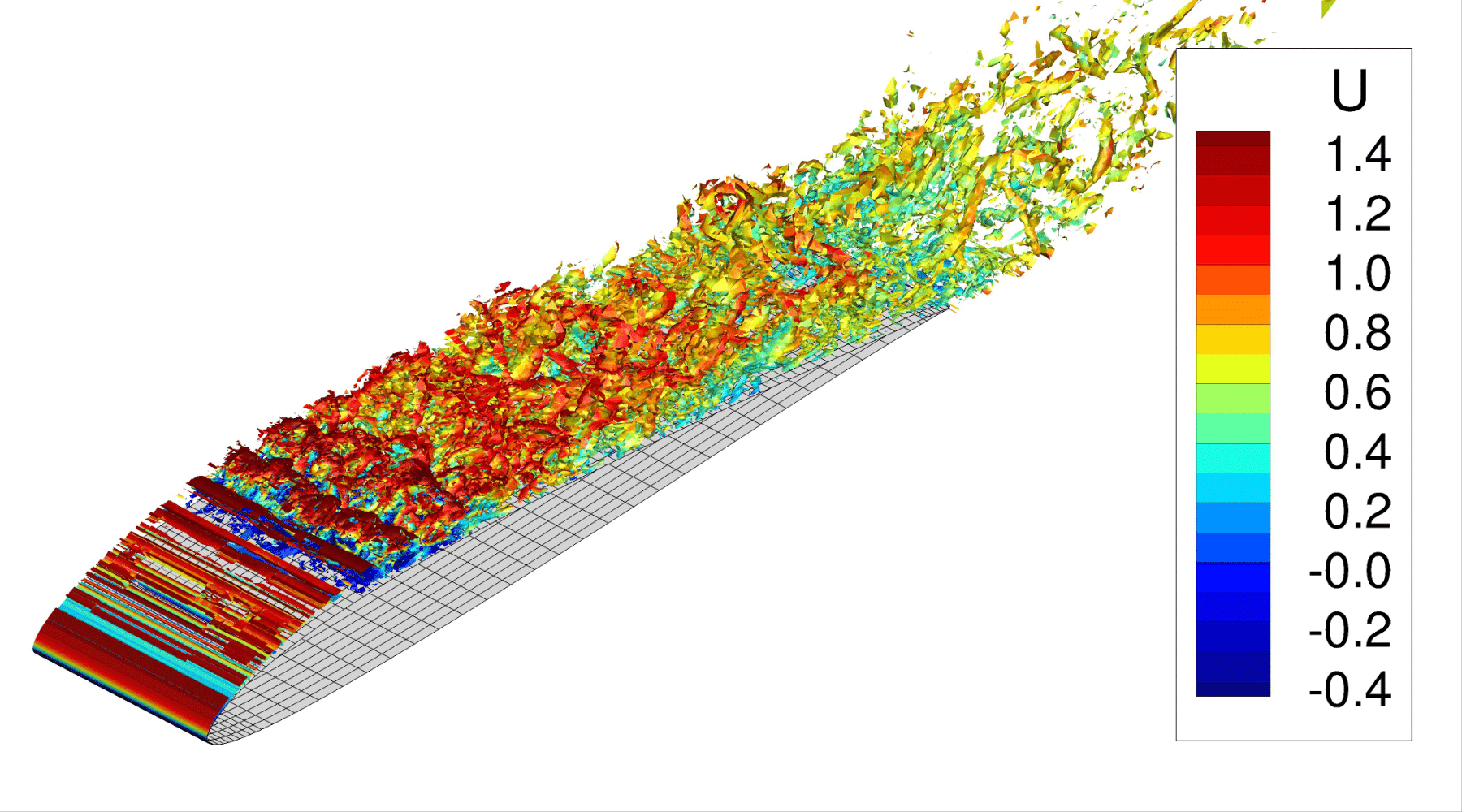} 	
		\subcaption{Adaptive $ p^3 $ FR, $ t=40 $}
		\label{sd_order}
	\end{subfigure}
	\caption{Instantaneous $ Q $-isosurfaces colored by  velocity component $ u  $ in the $ x $ direction when simulating the transitional flow on the 20-layer mesh.}	
	\label{qiso_sd_20_layer}	
\end{figure}

\begin{figure}		
	\centering
	\begin{subfigure}{0.49\textwidth}
		\includegraphics[width=\textwidth]{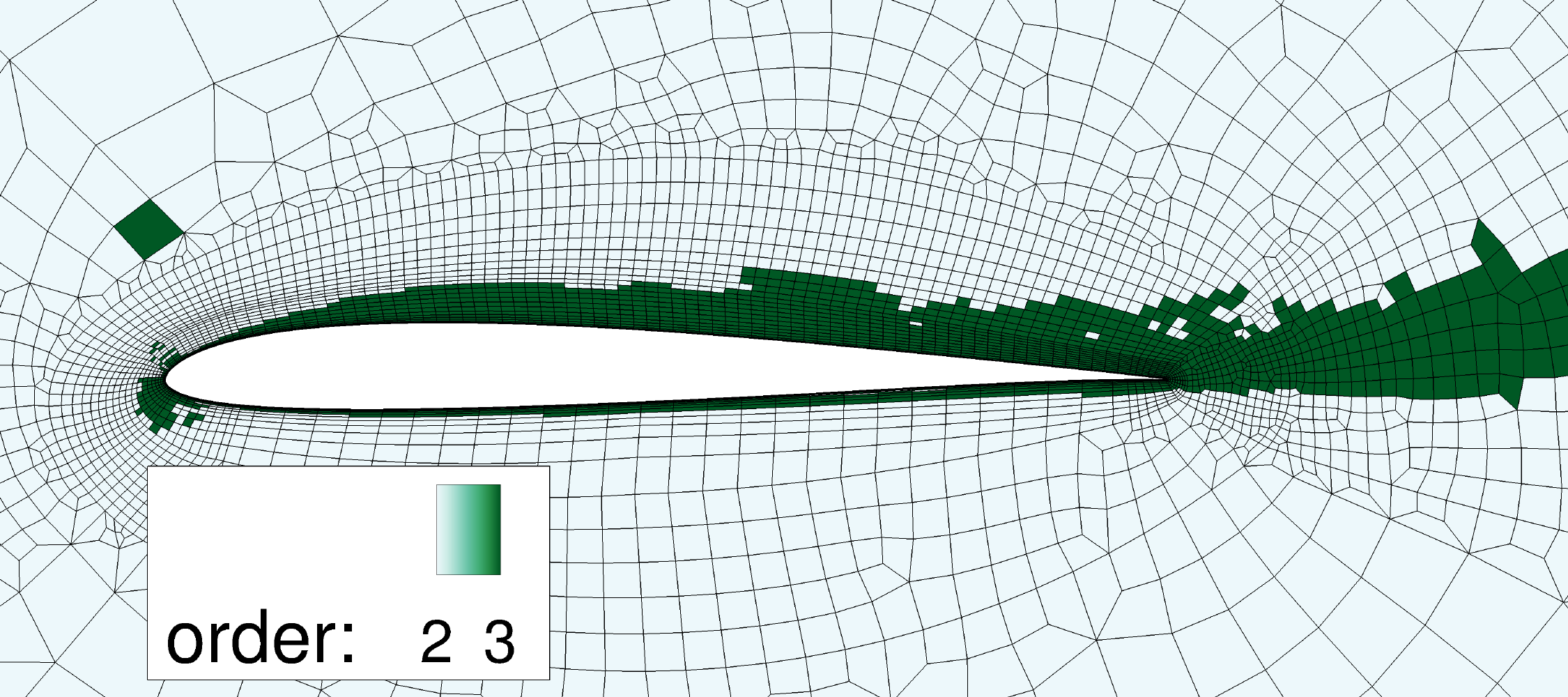}
		\subcaption{$ z=0.05 $}
		\label{sd_20_p2_slice_1}
	\end{subfigure}
	\begin{subfigure}{0.49\textwidth}
		\includegraphics[width=\textwidth]{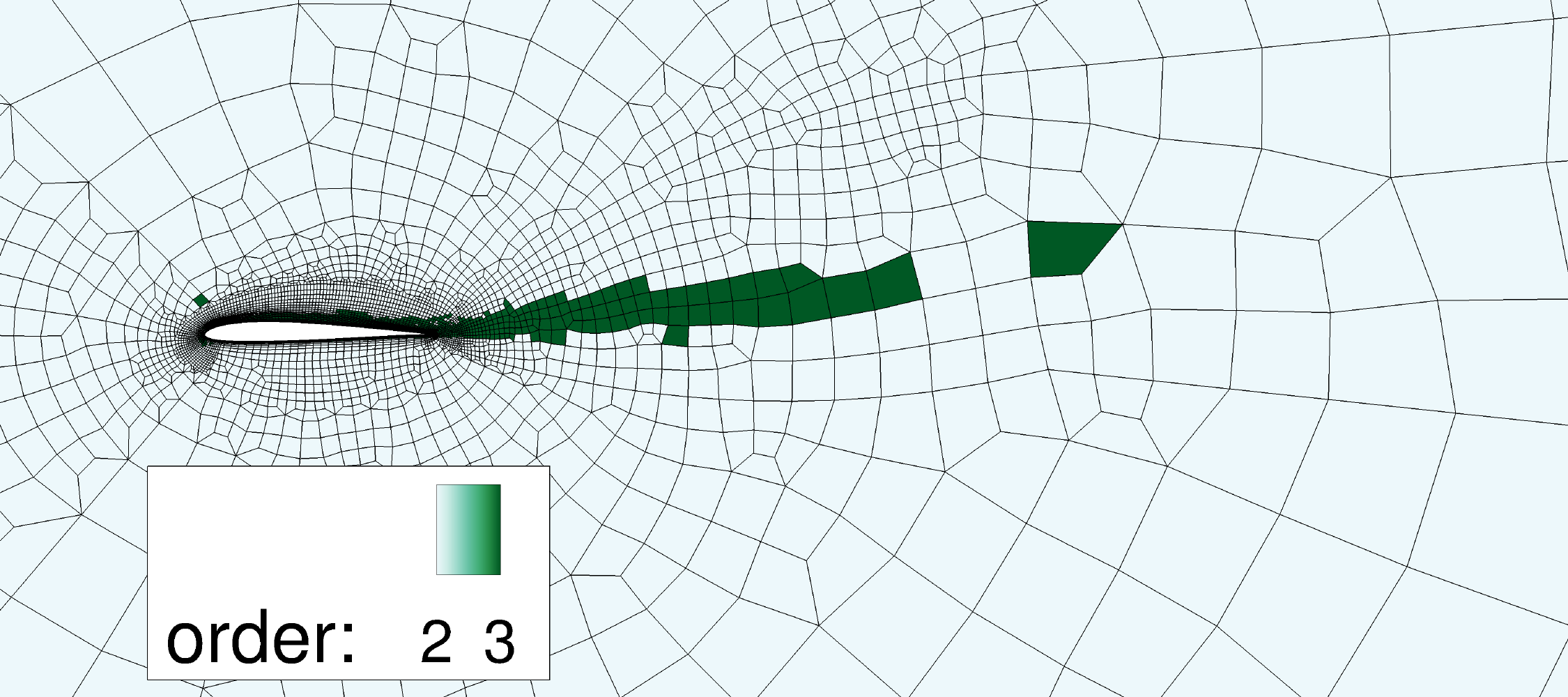} 	
		\subcaption{$ z=0.05 $}
		\label{sd_20_p2_slice_4}
	\end{subfigure}	\\
	\vspace{10pt}
	\begin{subfigure}{0.49\textwidth}
		\includegraphics[width=\textwidth]{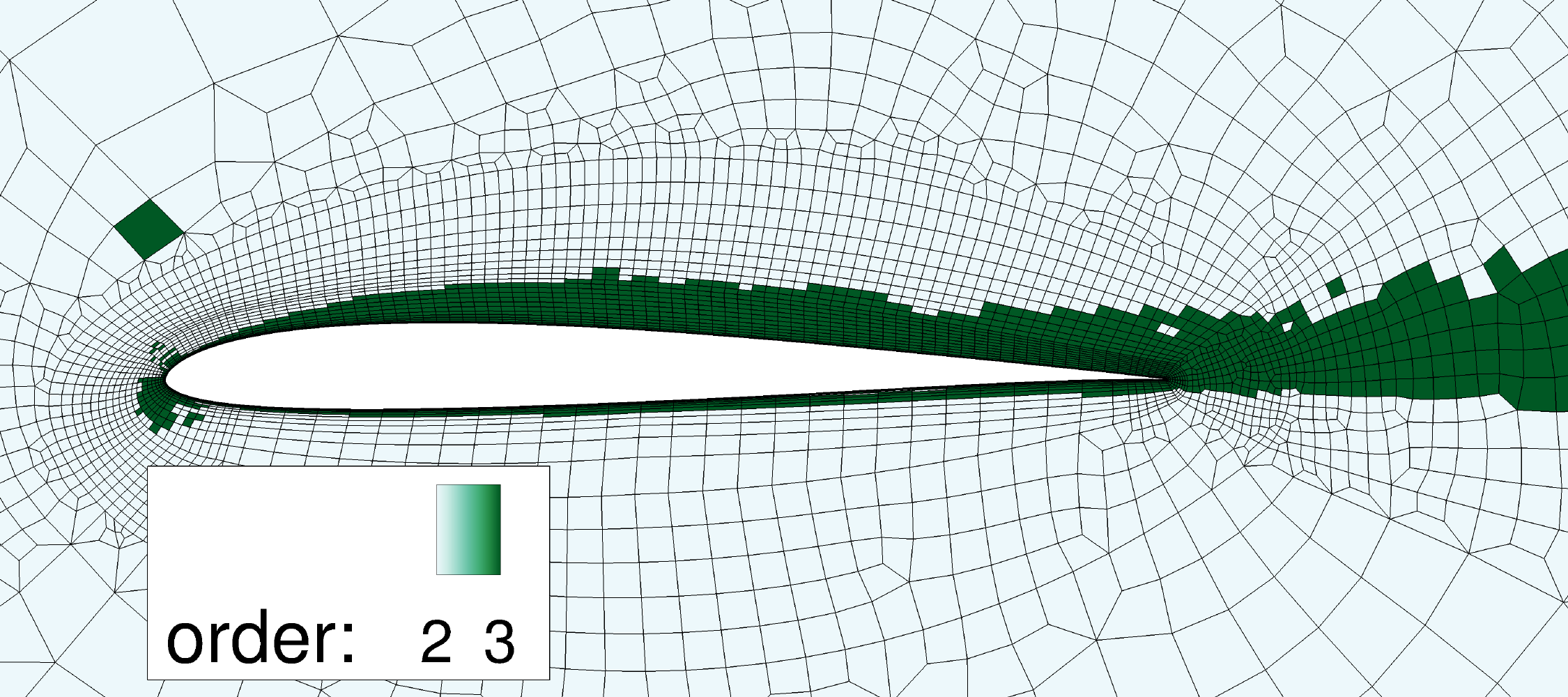} 	
		\subcaption{$ z=0.1 $}
		\label{sd_20_p2_slice_2}
	\end{subfigure}
	\begin{subfigure}{0.49\textwidth}
		\includegraphics[width=\textwidth]{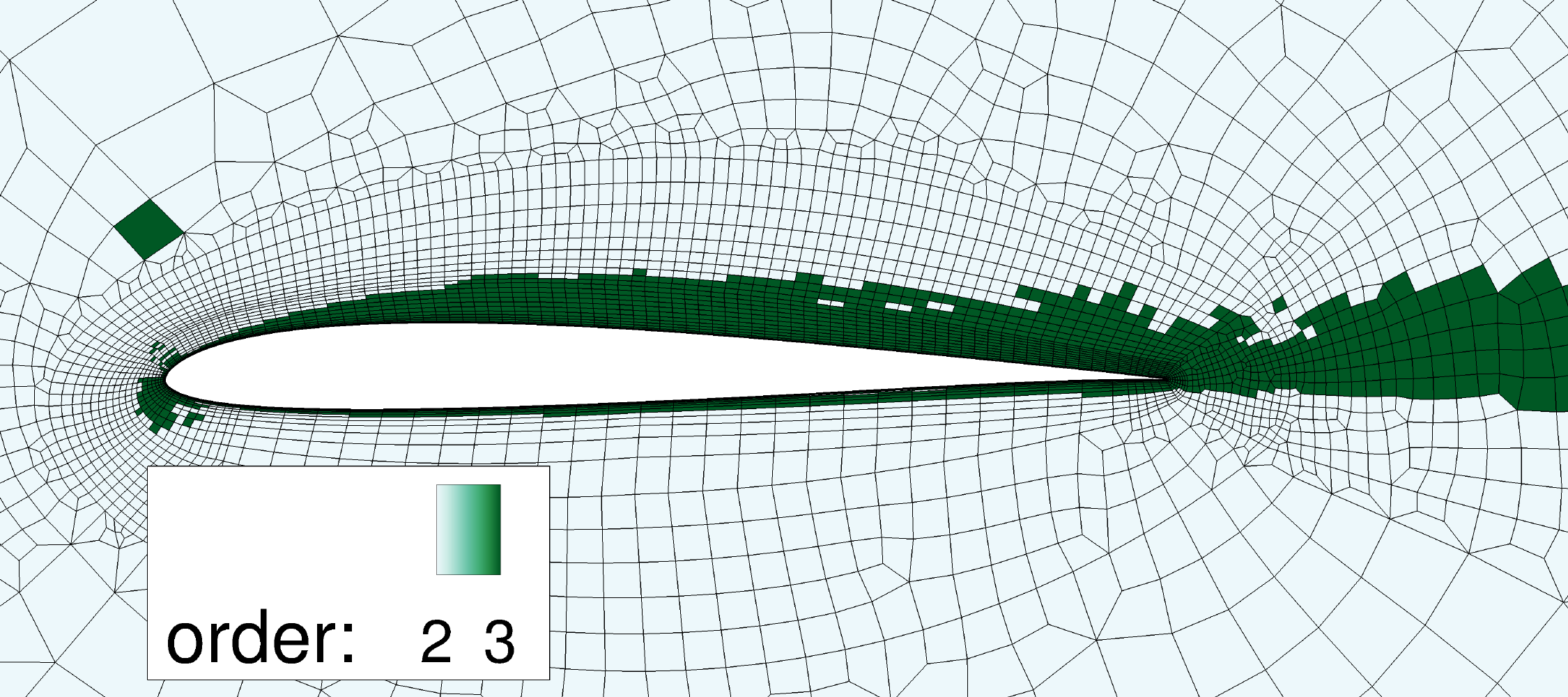} 	
		\subcaption{$ z=0.15 $}
		\label{sd_20_p2_slice_3}
	\end{subfigure}
	\caption{Instantaneous order-of-accuracy distributions of adaptive $p^2 $ FR at different slices in the spanwise direction when simulating the transitional flow over the SD7003 wing. $ t=26 $.}	
	\label{sd_20_p2_slices}	
\end{figure}

\begin{figure}		
	\centering
	\begin{subfigure}{0.49\textwidth}
		\includegraphics[width=\textwidth]{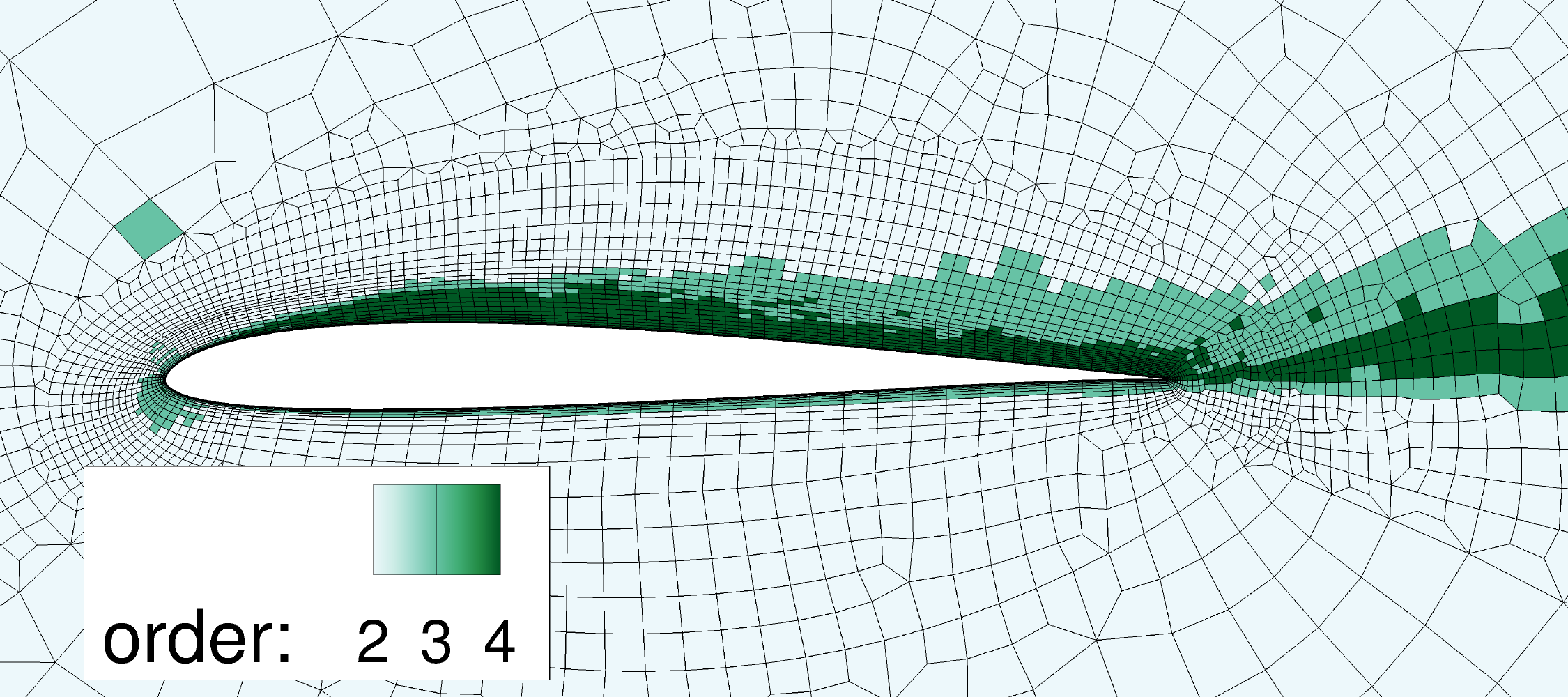}
		\subcaption{$ z=0.05 $}
		\label{sd_20_p3_slice_1}
	\end{subfigure}
	\begin{subfigure}{0.49\textwidth}
		\includegraphics[width=\textwidth]{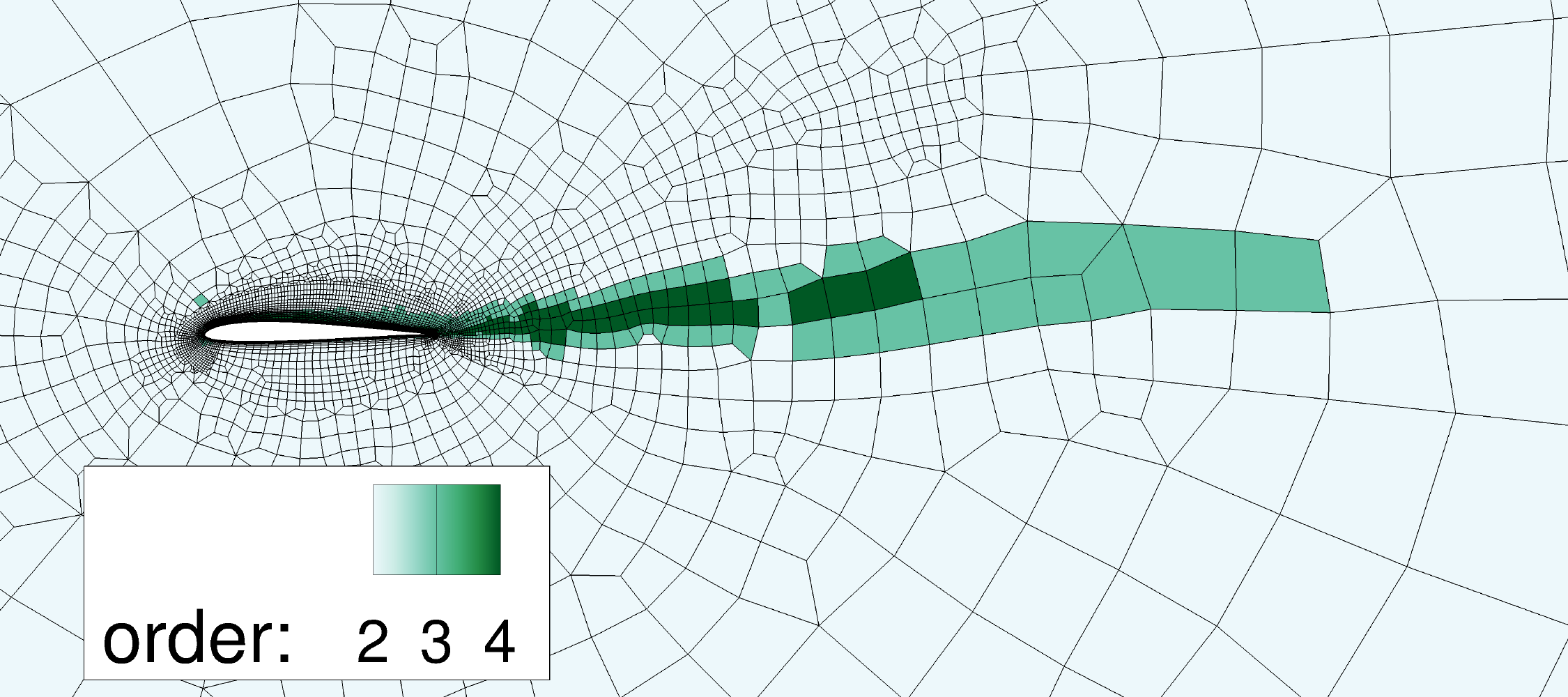} 	
		\subcaption{$ z=0.15 $}
		\label{sd_20_p3_slice_4}
	\end{subfigure}	\\
	\vspace{10pt}
	\begin{subfigure}{0.49\textwidth}
		\includegraphics[width=\textwidth]{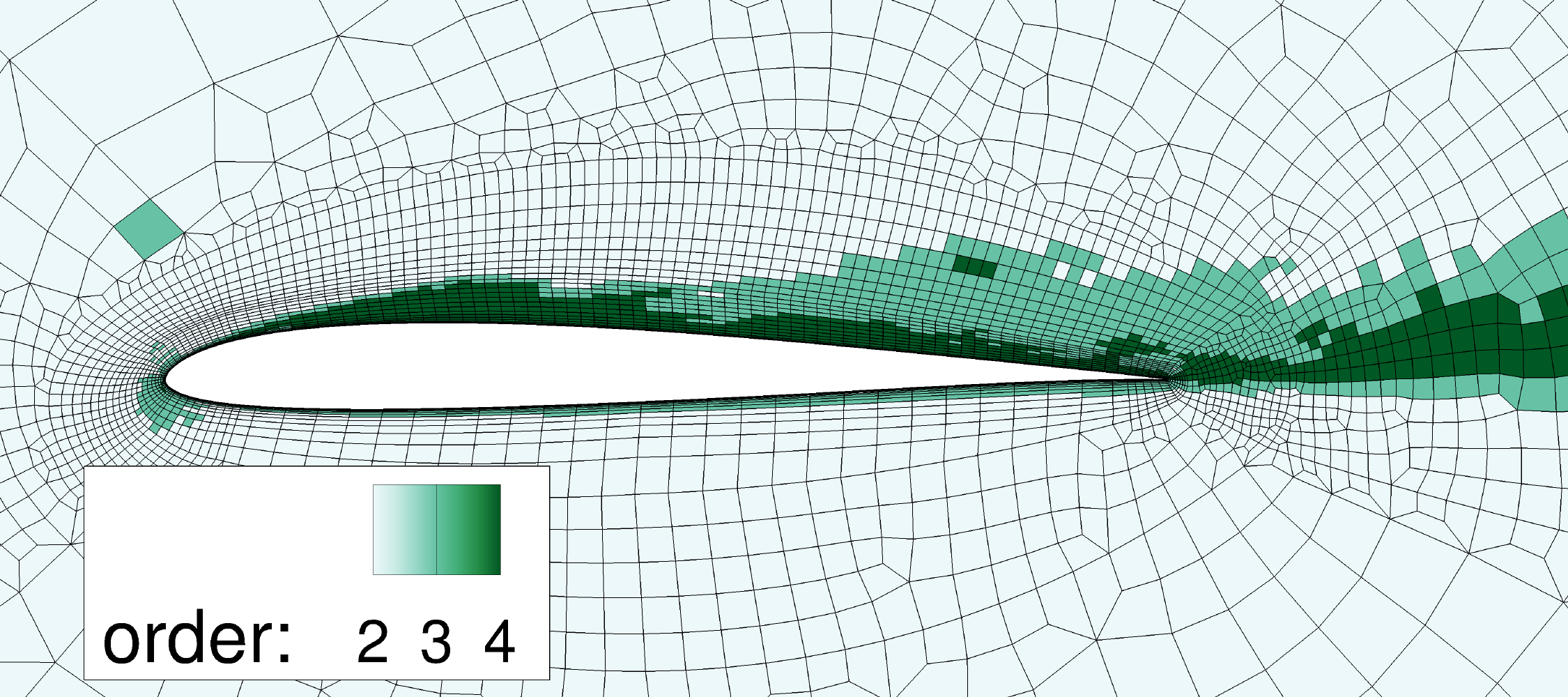} 	
		\subcaption{$ z=0.1 $}
		\label{sd_20_p3_slice_2}
	\end{subfigure}
	\begin{subfigure}{0.49\textwidth}
		\includegraphics[width=\textwidth]{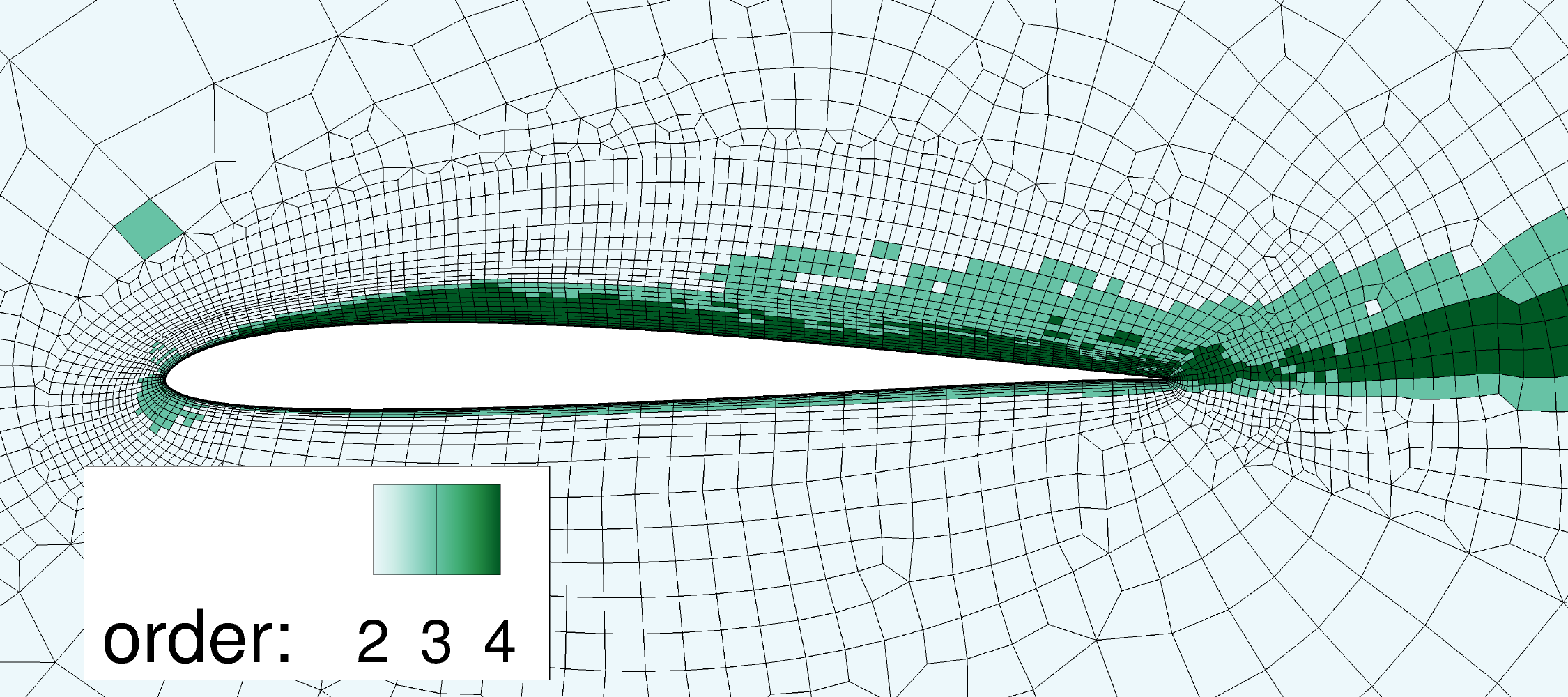} 	
		\subcaption{$ z=0.15$}
		\label{sd_20_p3_slice_3}
	\end{subfigure}
	\caption{Instantaneous order-of-accuracy distributions of adaptive $p^3 $ FR at different slices in the spanwise direction when simulating the transitional flow over the SD7003 wing. $ t=40 $.}	
	\label{sd_20_p3_slices}	
\end{figure}

\begin{figure}		
	\centering
	\begin{subfigure}{0.24\textwidth}
		\includegraphics[width=\textwidth]{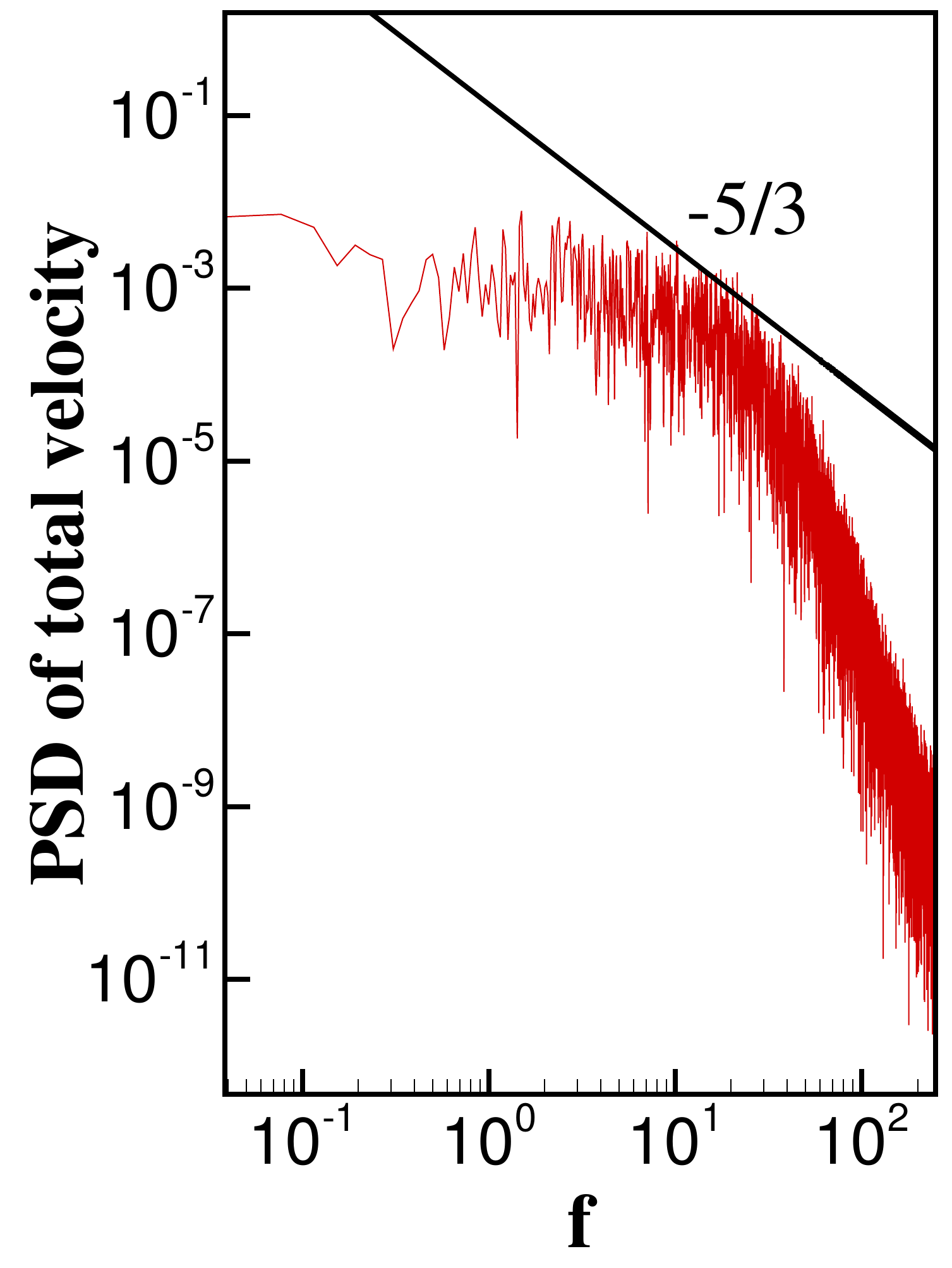}
		\subcaption{$ (0.3,0.057,0.1)^\intercal$}
		\label{sd_20_layer_pnt1_3rd}
	\end{subfigure}
	\begin{subfigure}{0.24\textwidth}
		\includegraphics[width=\textwidth]{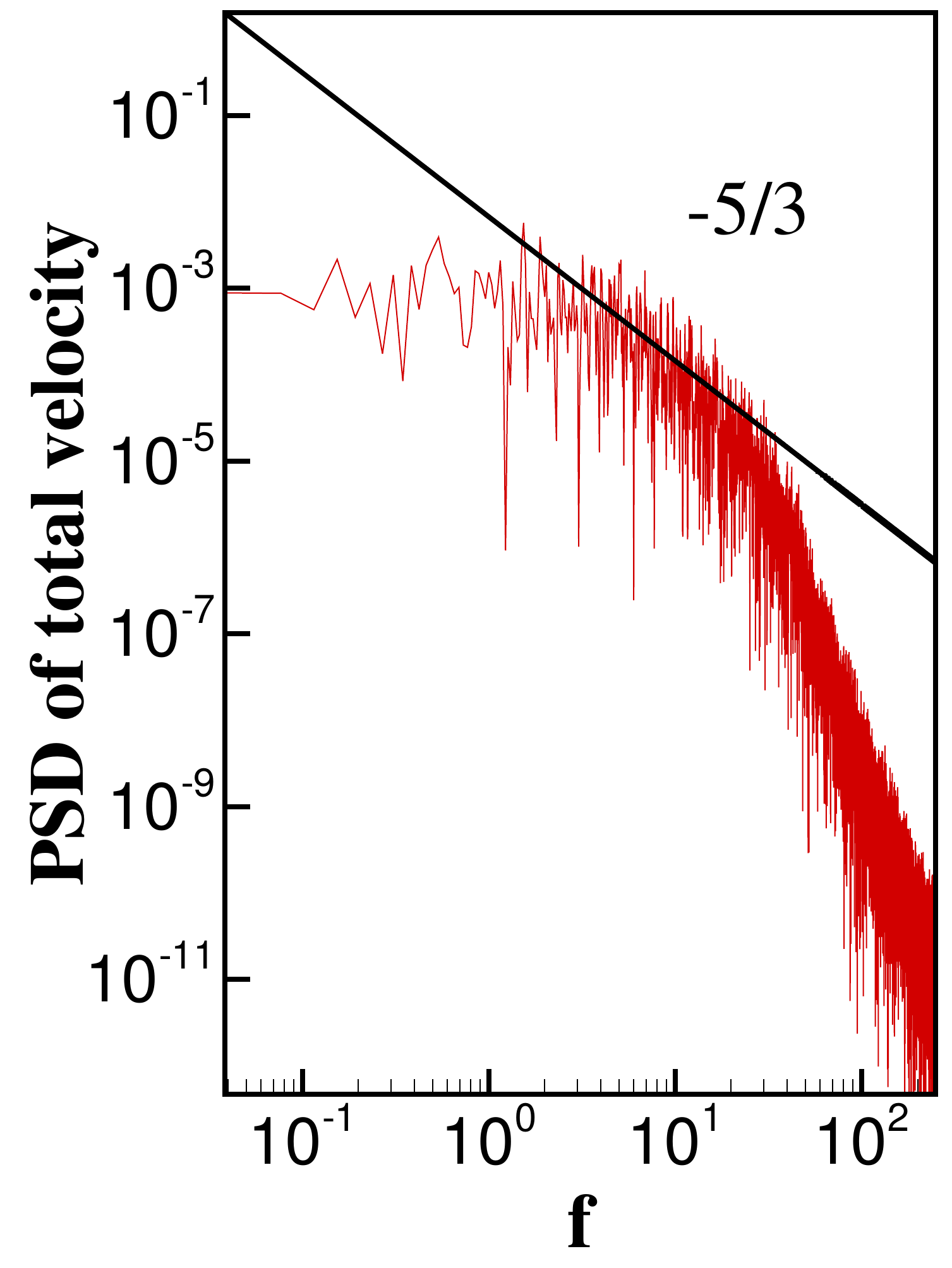} 	
		\subcaption{$ (0.5, 0.048, 0.1)^\intercal $}
		\label{sd_20_layer_pnt2_3rd}
	\end{subfigure}
	\begin{subfigure}{0.24\textwidth}
		\includegraphics[width=\textwidth]{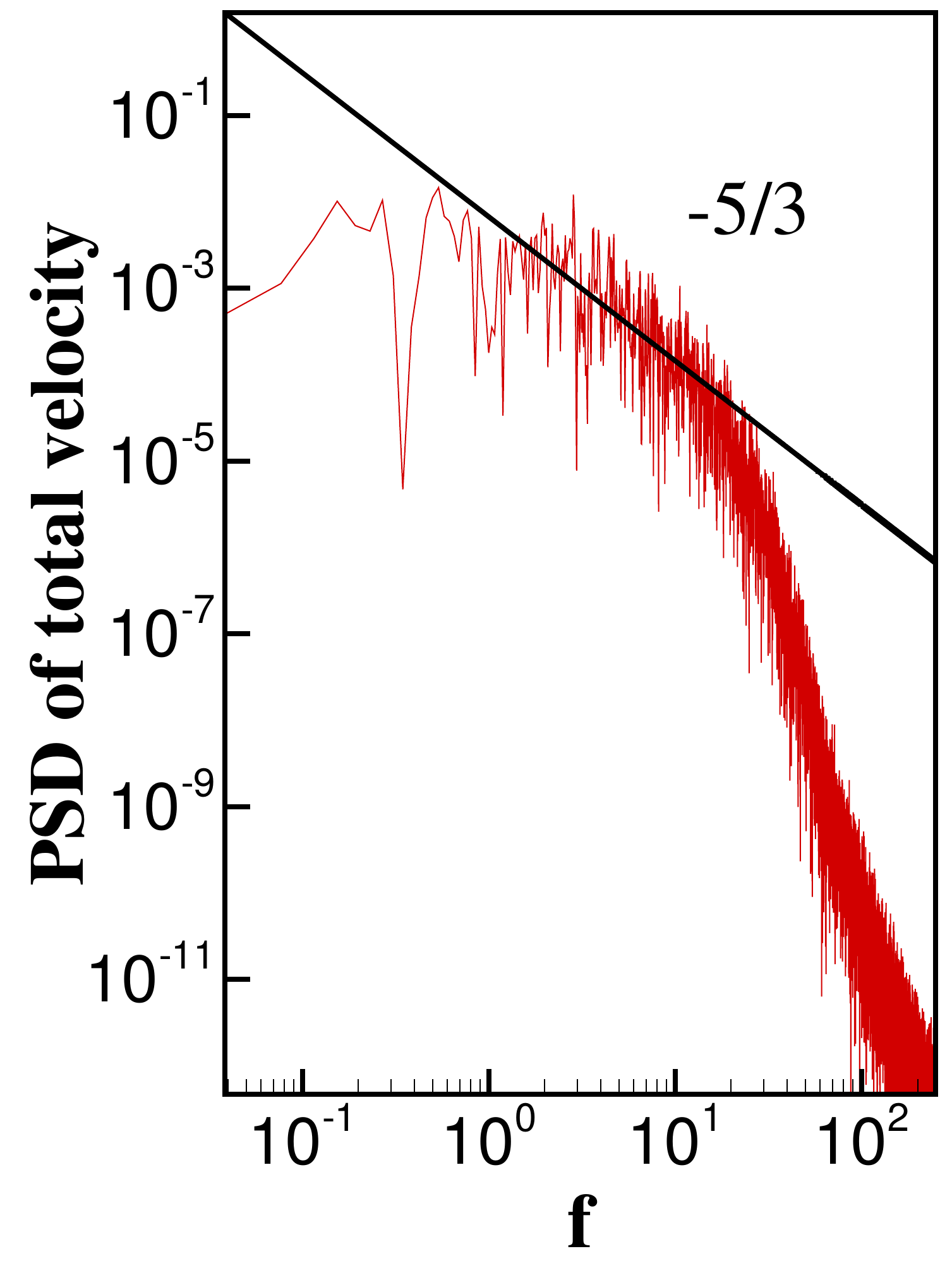} 	
		\subcaption{$ (0.7, 0.032, 0.1)^\intercal $}
		\label{sd_20_layer_pnt3_3rd}
	\end{subfigure}
	\begin{subfigure}{0.24\textwidth}
		\includegraphics[width=\textwidth]{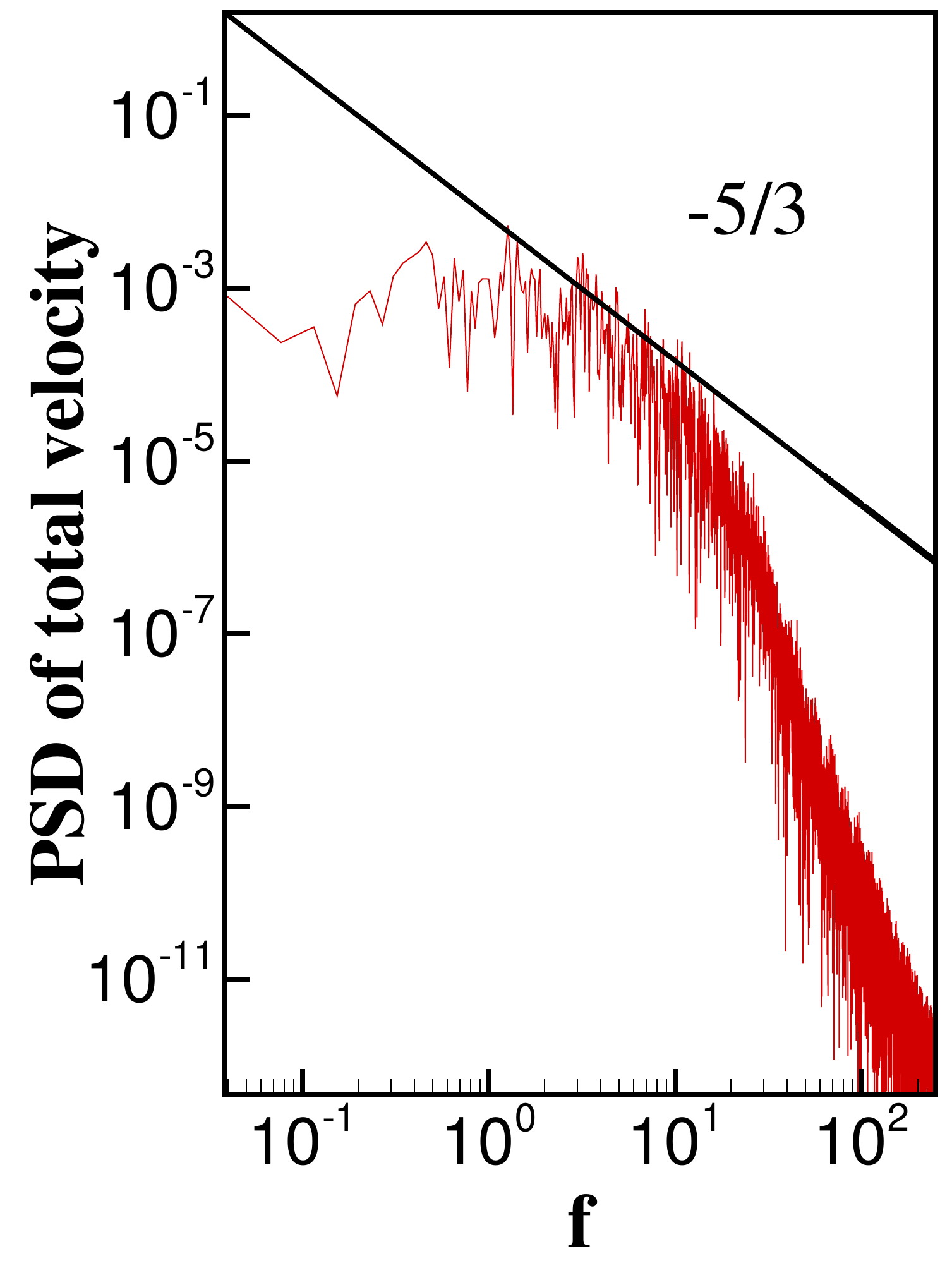} 	
		\subcaption{$ (0.9, 0.012, 0.1)^\intercal $}
		\label{sd_20_layer_pnt4_3rd}
	\end{subfigure}\\
	\vspace{10pt}
	\begin{subfigure}{0.24\textwidth}
		\includegraphics[width=\textwidth]{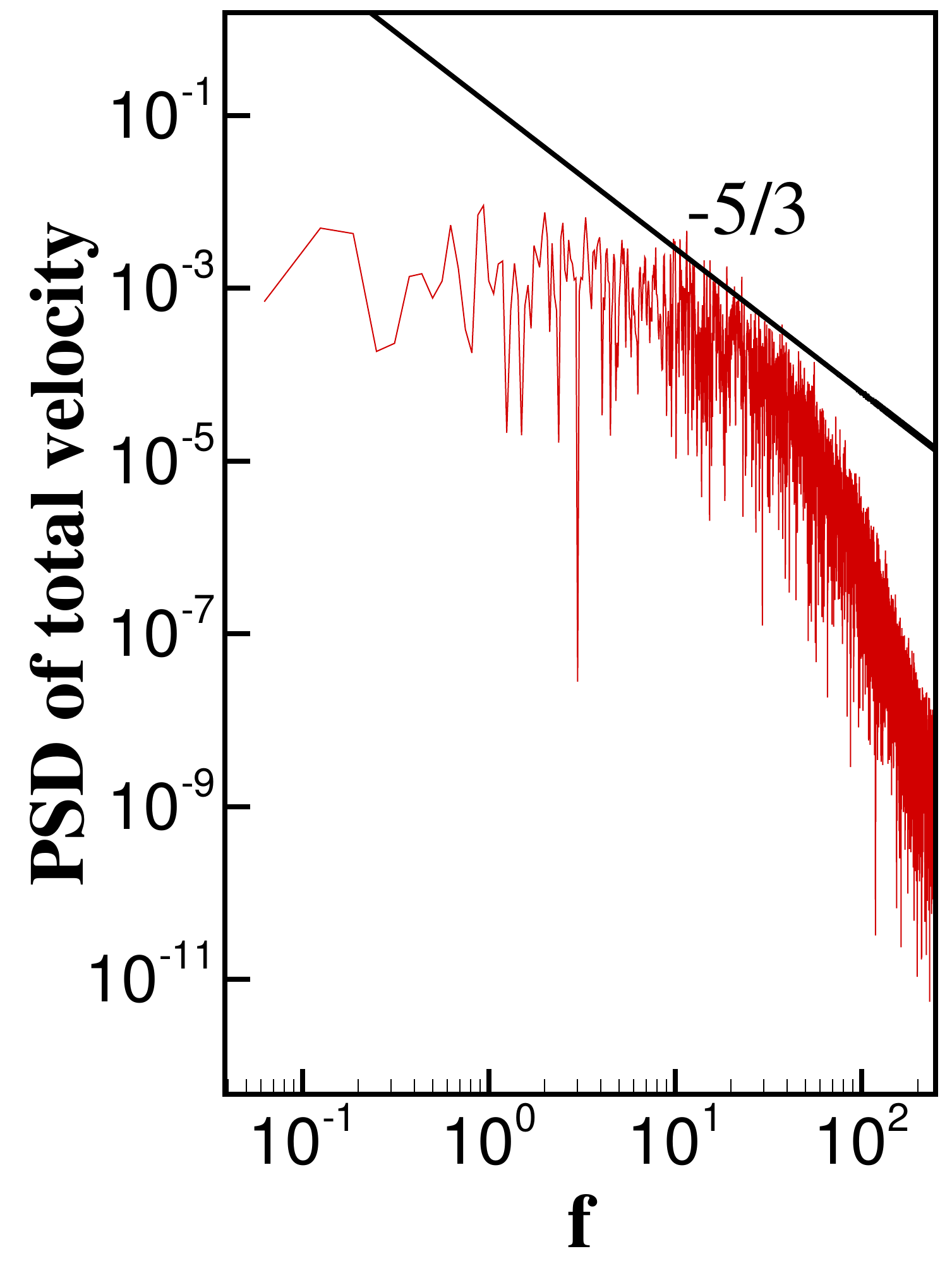}
		\subcaption{$ (0.3,0.057,0.1)^\intercal$}
		\label{sd_10_layer_pnt1_4th}
	\end{subfigure}
	\begin{subfigure}{0.24\textwidth}
		\includegraphics[width=\textwidth]{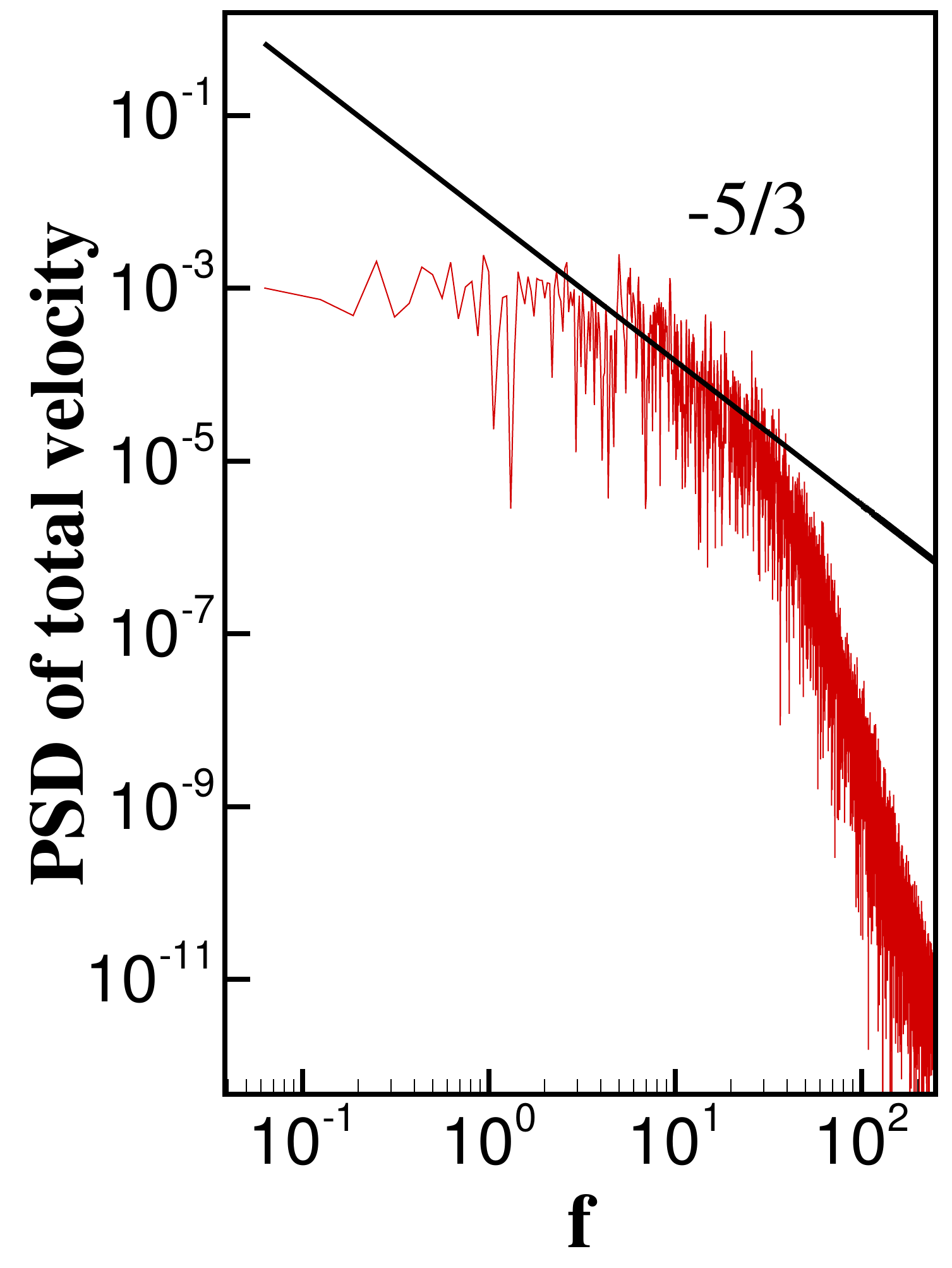} 	
		\subcaption{$ (0.5, 0.048, 0.1)^\intercal $}
		\label{sd_10_layer_pnt2_4th}
	\end{subfigure}
	\begin{subfigure}{0.24\textwidth}
		\includegraphics[width=\textwidth]{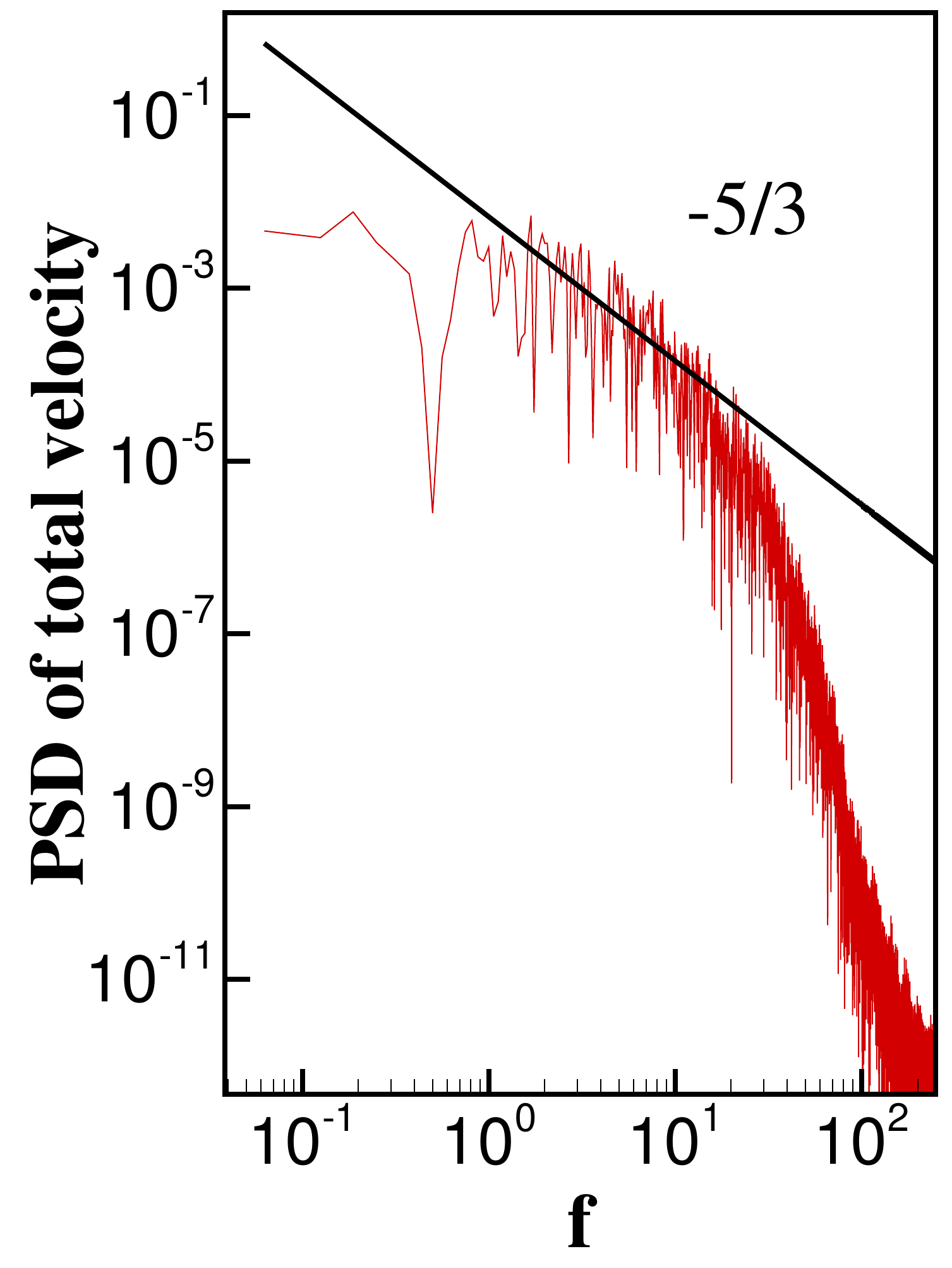} 	
		\subcaption{$ (0.7, 0.032, 0.1)^\intercal $}
		\label{sd_10_layer_pnt3_4th}
	\end{subfigure}
	\begin{subfigure}{0.24\textwidth}
		\includegraphics[width=\textwidth]{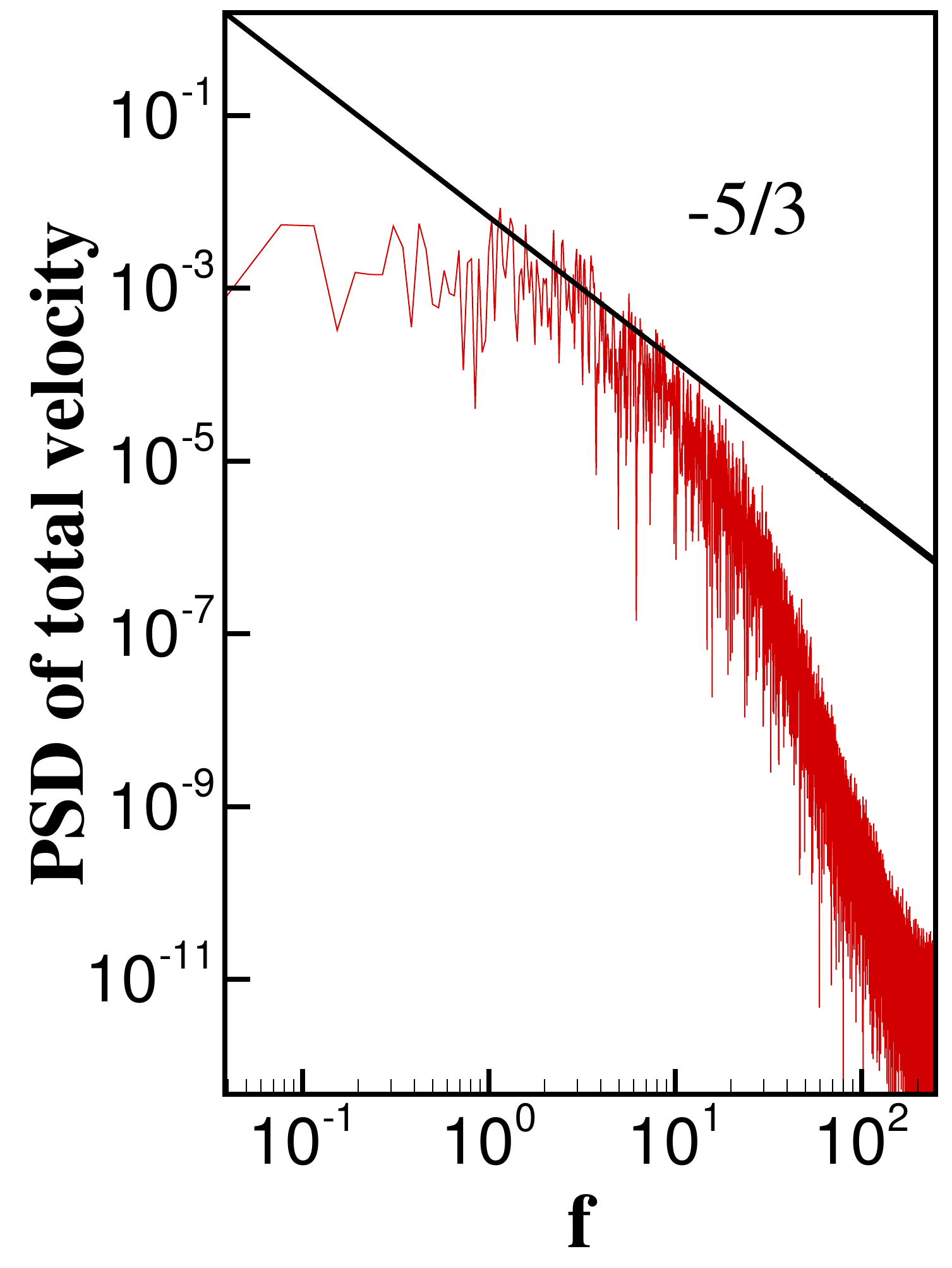} 	
		\subcaption{$ (0.9, 0.012, 0.1)^\intercal $}
		\label{sd_10_layer_pnt4_4th}
	\end{subfigure}
	\caption{Power spectral density of the total velocity at different locations in the wake region. (a)--(d) $ p^3 $ FR with no adaptation and (e)--(h) adaptive $ p^3 $ FR with $ (\nu_{max},\nu_{min})=(0.1,0.01) $. A line of slope $ -5/3 $ is  added to every graph as a reference.}	
	\label{sd_psd}	
\end{figure}

Two snapshots of the instantaneous isosurfaces of the $ Q $-criterion, where $ Q=500 $, are presented in Figure~\ref{qiso_sd_20_layer} for the simulation conducted on the 20-layer mesh. One visible difference is that more finer structures are resolved using adaptive $ p^3 $ FR. From Figure~\ref{sd_20_p2_slices} and Figure~\ref{sd_20_p3_slices}, we observe that polynomials of degree $ p>1 $ are clustered in regions near the stagnation point, turbulent boundary layers as well as the wake region. The order-of-accuracy distributions at different slices in the spanwise direction are not exactly the same since the $ p $-adaptation is conducted locally. Overall, the feature-based adaptation method can give an feature-tracking $ p $-distribution. Unlike the transitional flow over the cylinder, the choice of $ (\nu_{max},\nu_{min})=(0.1,0.001)$ actually clusters all the polynomials of degree $ p>1 $ in a small domain and only a few high-order elements can be found in the wake region far away from the wing.
The power spectral density of the total velocity at four locations close to the suction side of the wing, i.e., $ (0.3,0.057,0.1)^\intercal $  $ (0.5, 0.048, 0.1)^\intercal $, $ (0.7, 0.032, 0.1)^\intercal $, and $ (0.9, 0.012, 0.1)^\intercal $ are illustrated in Figure~\ref{sd_psd}. The first point is near the end of the separation bubble, where the transition from laminar flow to turbulent flow takes place. The slopes of PSDs are generally steeper than $ -5/3 $ at high frequencies at the first point in all simulation. At the rest three points, PSDs align with the reference lines in a certain range of high frequencies, which agrees with the results in~\cite{bassi2016development}.

\begin{figure}		
	\centering
	\begin{subfigure}{0.49\textwidth}
		\includegraphics[width=\textwidth]{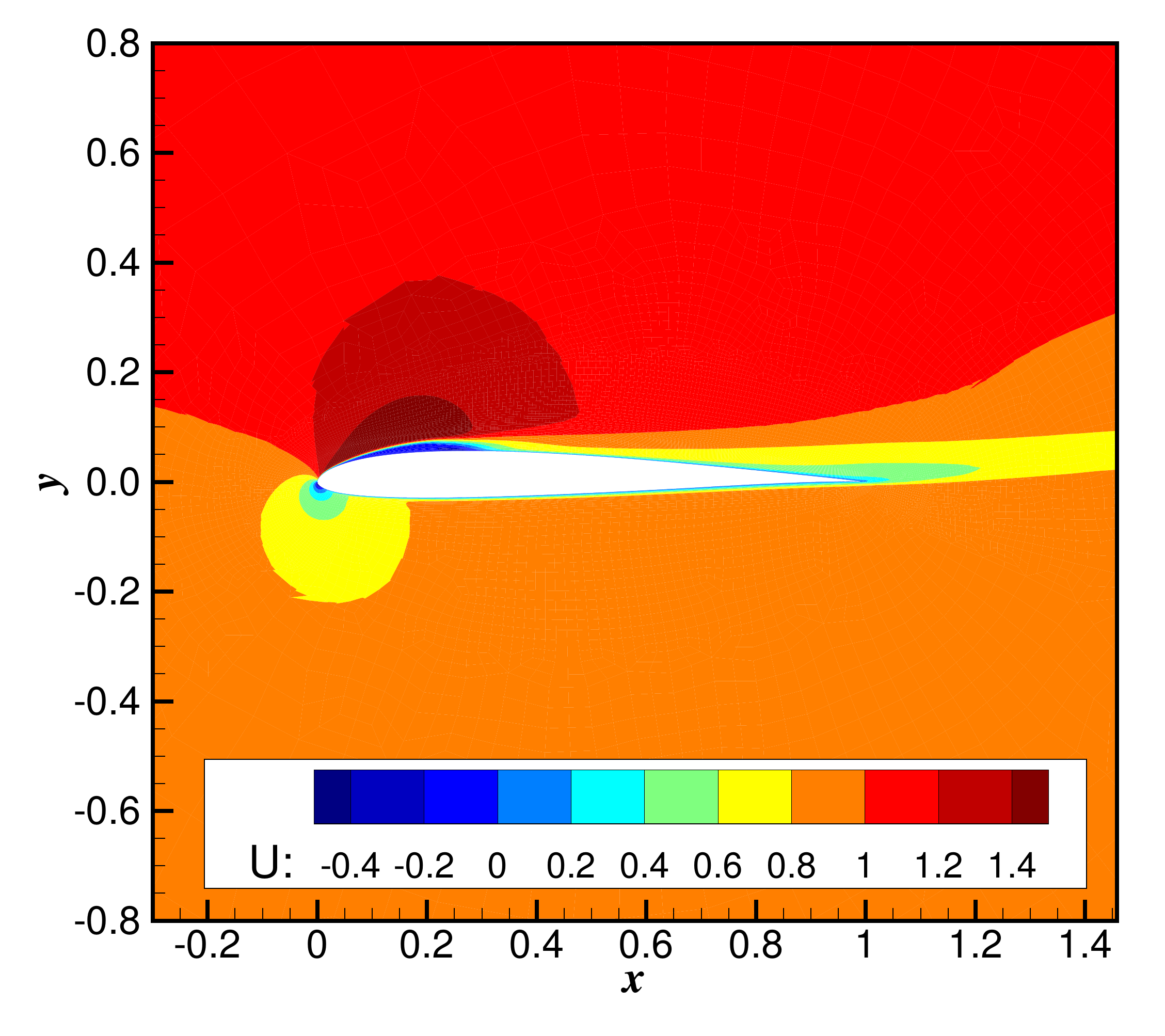}
		\subcaption{$ p^2 $ FR with adaptation}
		\label{sd_p2_20}
	\end{subfigure}
	\begin{subfigure}{0.49\textwidth}
		\includegraphics[width=\textwidth]{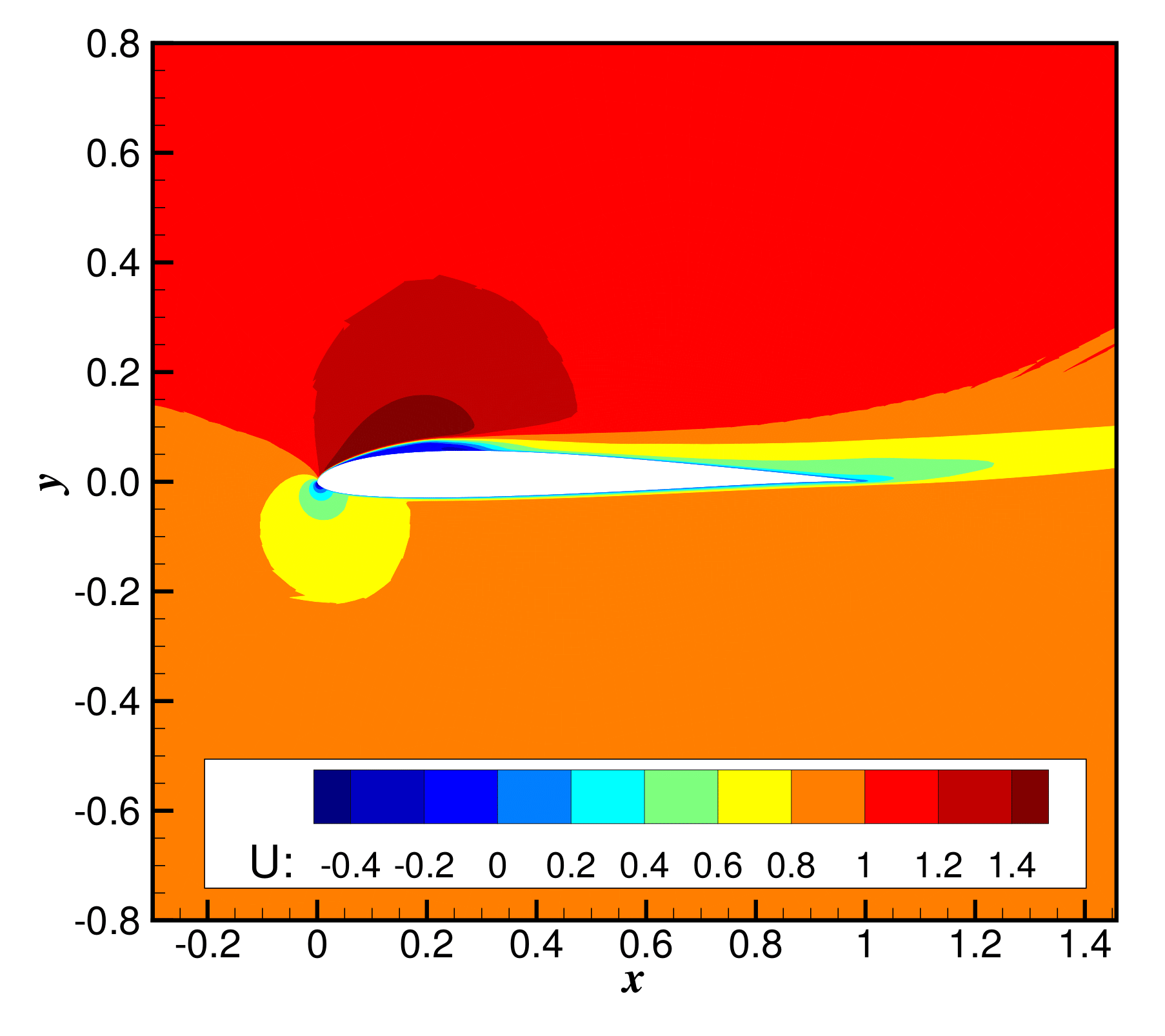} 	
		\subcaption{$ p^3 $ FR with adaptation}
		\label{sd_p3_10}
	\end{subfigure}
	\caption{Averaged velocity component in the $ x $ direction on the 20-layer mesh.}	
	\label{sd7003_ave}	
\end{figure}

\begin{figure}		
	\centering
	\begin{subfigure}{0.49\textwidth}
		\includegraphics[width=\textwidth]{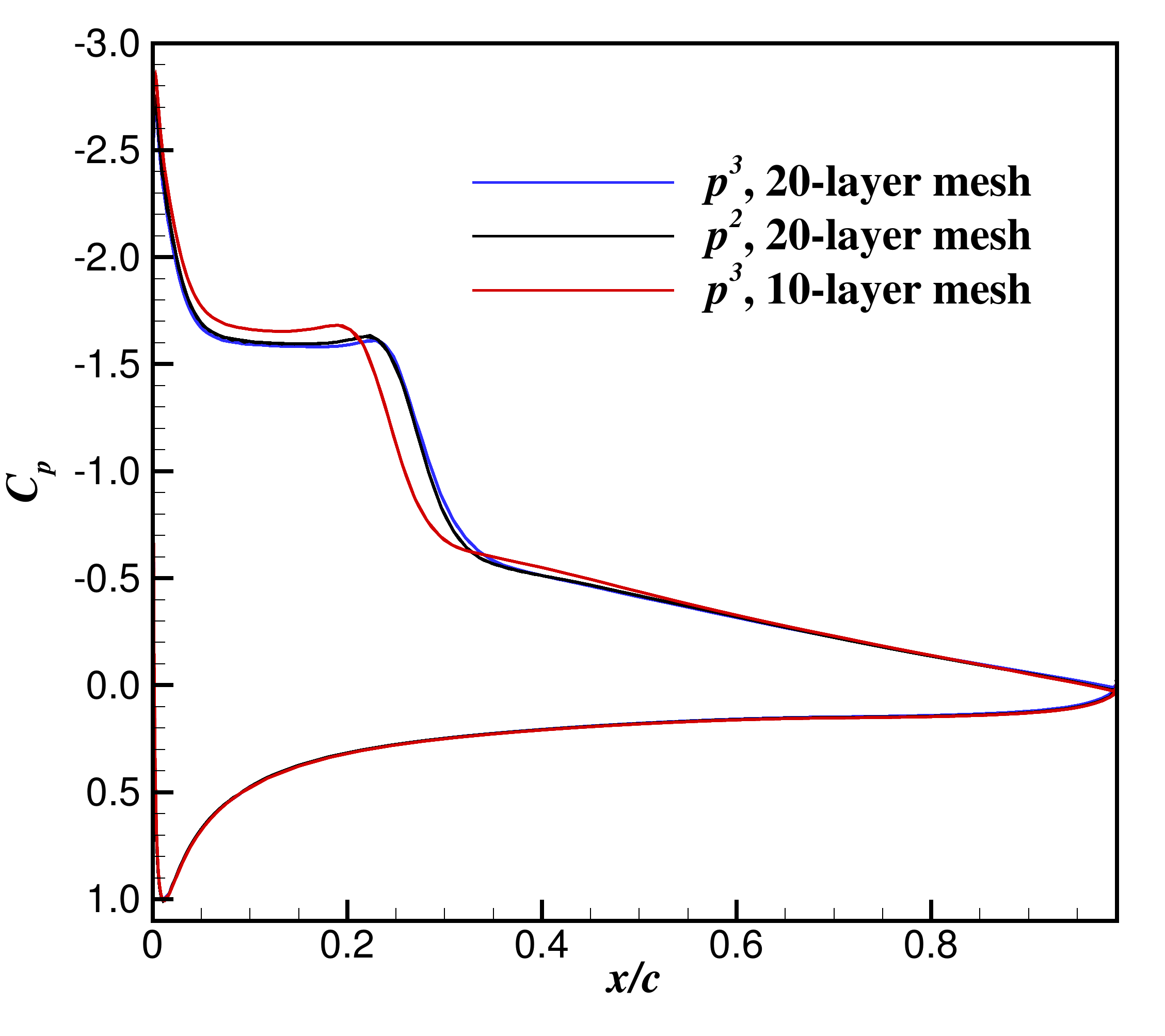}
		\subcaption{ $ C_p $}
		\label{sd_cp}
	\end{subfigure}
	\begin{subfigure}{0.49\textwidth}
		\includegraphics[width=\textwidth]{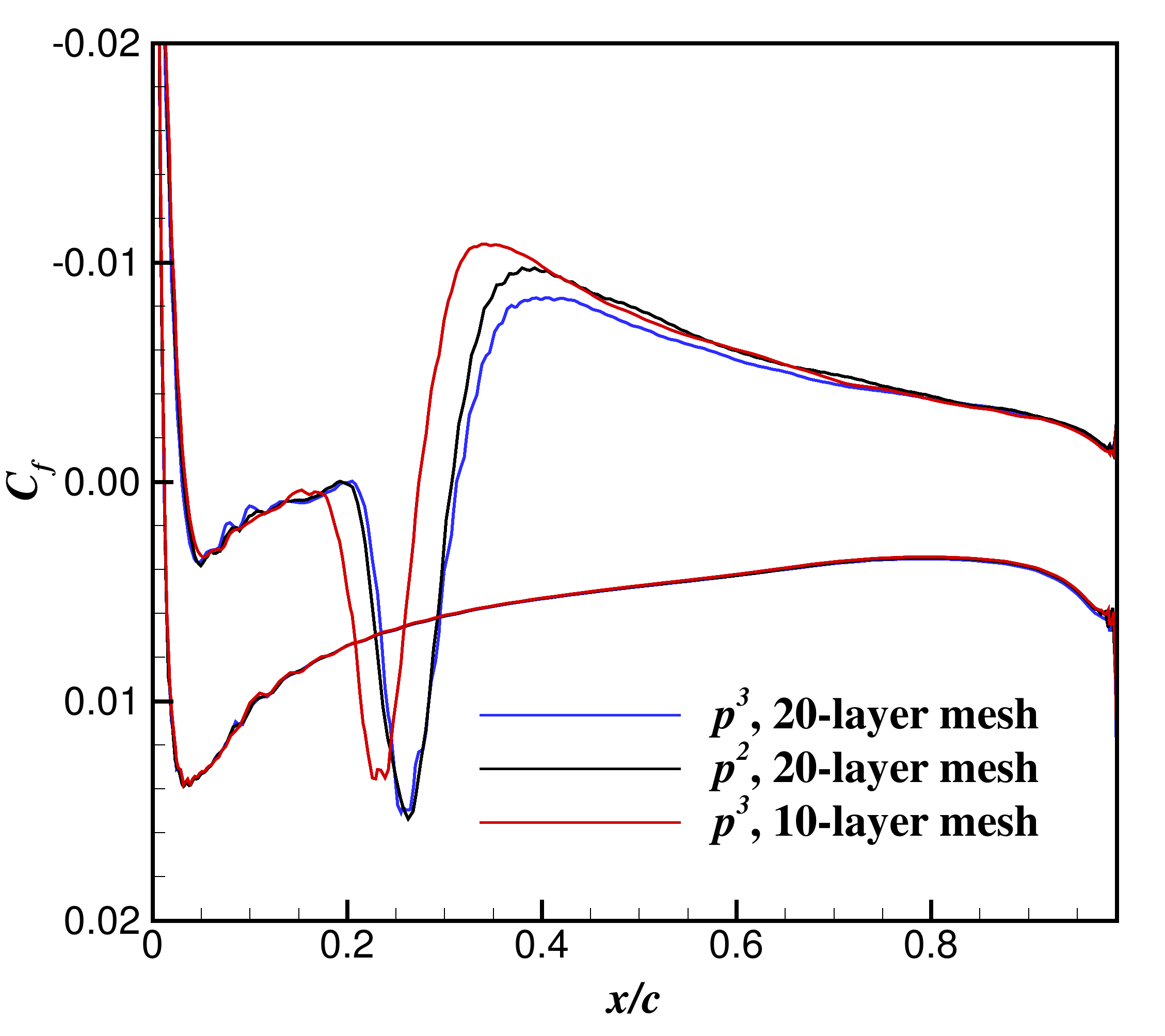} 	
		\subcaption{$ C_f $}
		\label{sd_cf}
	\end{subfigure}
	\caption{Time-averaged surface pressure coefficient $ C_p $ and surface friction coefficient $ C_f $ on the SD7003 wing.}	
	\label{sd7003_surface}	
\end{figure}

The mean field of the averaged velocity component $ u $ in the $ x $ direction are presented in Figure~\ref{sd7003_ave}. Predictions of the time-averaged flow features, namely, lift coefficient $ C_l $, drag coefficient $ C_d $, separation point $ x_s $, and reattachment point $ x_{re} $, are documented  in Table~\ref{sd_predictions}. The reduction of the total number of solution points $ n_{sp}^{tot} $ at specific time instances are also provided in Table~\ref{sd_predictions}.  The time-averaged surface pressure coefficient $ C_p $ and surface friction coefficient $ C_f $ on the SD7003 wing are illustrated in Figure~\ref{sd7003_surface}. 
All numerical experiments over-predict the drag when compared to the experimental result~\cite{selig1995summary}. Simulation of the incompressible Navier--Stokes equations done by Bassi et al.~\cite{bassi2015linearly} gave larger $ C_d $ than those of the compressible Navier--Stokes equations when $ \text{Ma}=0.1 $. The numerical results of our current work have a decent agreement with those in~\cite{beck2014high}.
For simulation conducted on the 20-layer mesh,  the adaptive $ p^2 $ FR has 1,644,414 solution points at $ t_1=26 $; the adaptive $ p^3 $ FR has 2,570,297 solution points at $ t_2=42 $. The reductions compared to $ p $-uniform $ p^2 $ and $ p^3 $ FR are 44.40\% and 63.34\%, respectively. The reduction of solution points on the 10-layer mesh using $ p $-adaptive $ p^3 $ FR is 60.57\% at $ t_1=26 $. 
From the reduction in the total number of solution points, we can speculate a similar reduction in the computational cost or run time. We note that insufficient resolution of the physical scales will lead to failure of the under-resolved turbulence simulation due to insufficient grid resolution in the 10-layer mesh. However, with a  small dissipation introduced by nodal polynomial filtering, the force prediction can be accurate to 0.01 and the length of separation bubble is only slightly shorter. It is hard to know whether one under-resolved turbulence simulation will fail due to aliasing errors. Therefore, in the practice of performing under-resolved turbulence simulation using high-order methods, we would recommend to employ proper de-aliasing techniques.

\begin{table}
	\centering
	\caption{Predictions of the transitional flow using the $ p $-adaptive solver. Rows 1--3 are from current work using $ p $-adaptive FR. Rows 4--9 are previous numerical results using $ p $-uniform high-order methods. Row 10 presents the results from experiment. The abbreviation Inc. stands for ``incompressible".} 
	\label{sd_predictions}
	\footnotesize
	\setlength\tabcolsep{1.5pt}
	\begin{tabular}{rrrrrrrr}
	\hline
	
	&Spatial discretization & Ma & $ C_l $& $ C_d $ &$ x_s $ & $ x_{re} $ & Reduction of $ n_{sp}^{tot} $\\
	\hline
	1&$ p $-adaptive, $ p^2 $ FR (20-layer)& 0.1& 0.9289& 0.0459&0.0321&0.3075&44.40\% at $ t_1=26 $\\
	2&$ p $-adaptive, $ p^3 $ FR (20-layer)& 0.1& 0.9270& 0.0470&0.0301&0.3123&63.34\% at $ t_2=42 $\\
	3&$ p $-adaptive, $ p^3 $ FR (10-layer)& 0.1& 0.9316& 0.0419&0.0336&0.2735&60.57\% at $ t_1=26 $\\	
	\hline
	4&$ p^4 $ FR (Vermeire et al.~\cite{vermeire2017utility})& 0.2&0.941&0.049& 0.045&0.315&\\
	5&$ p^3 $ DG (Beck et al.~\cite{beck2014high})&0.1 &0.923& 0.045&0.027 &0.310& \\
	6&$ p^7 $ DG (Beck et al.~\cite{beck2014high})& 0.1&0.932&0.050&0.030&0.336& \\
	7&$ O(h^6) $ FD (Galbriath \& Visbal~\cite{galbraith2008implicit})&0.1&0.91&0.043&0.04&0.28& \\
	8&$ p^3 $ DG (Bassi et al.~\cite{bassi2015linearly})&Inc.&0.962&0.042&0.027&0.268& \\
	9&$ p^4 $ DG (Bassi et al.~\cite{bassi2015linearly})&Inc.& 0.953  &0.045&0.027 &0.294& \\	
	\hline
	10&Experiment (Selig et al.~\cite{selig1995summary}) &&0.92&0.029&&&\\
	\hline		
	\end{tabular}	
\end{table}

\clearpage
\section{Conclusion}\label{CFW}
In this work, a dynamically load balanced parallel $ p $-adaptive implicit high-order flux reconstruction method is developed and applied to under-resolved turbulence simulation. 
The parallel mesh partitioning API in ParMETIS, i.e., \verb|ParMetis_V3_AdaptiveRepart()|, is utilized for efficient parallel mesh partitioning. A collect-and-distribute strategy is used to redistribute the working variables to different processes. We have discussed the impact of weight calculation for each element on the parallel efficiency in the context of matrix-free implementation of the ESDIRK method. We investigate different weights related to the cost of residual evaluation, Jacobian matrix and preconditioner evaluation, and GMRES iterations. 
For $ p\le3 $, we recommend $ \omega_e = n_{sp} $, and as $ p $ grows larger, $ \omega_{e} = n_{sp}^{k}$, where $k> 1 $, is more preferable. Overall, a significant reduction in the run time and total number of solution points can be achieved via $ p $-adaptation for turbulence simulation and favorable results can be obtained.

When the adaptive solver is applied to solving the transitional flow over an infinite cylinder, due to the presence of large flow separation, the featured-based solver can result in a large domain where the polynomial degrees are refined. One can adjust the adaptation criteria towards wall-resolving  to save computational cost; however, the accuracy would be compromised.
When the flow separation is small, e.g., transitional flow over the SD7003 wing at a small angle of attack, the feature-based $ p $-adaptation method is able to confine the $ p $-refined region close to the wing, thus significantly reducing the cost while providing good predictions. 
We also show, with the SD7003 case, that insufficient mesh resolution can lead to instabilities triggered by aliasing errors of high-order methods in under-resolved turbulence simulation. A proper de-aliasing technique can overcome this issue and provide acceptable predictions.

The framework of dynamically load balanced $ p $-adaptive implicit high-order methods developed in this study paves the way towards robust and efficient ILES of turbulent flows at higher Reynolds numbers with the high-order FR/CPR method. The dynamic load balancing technique presented here can be easily extended to other types of high-order collocation methods.

\section*{Acknowledgments}
Wang and Yu gratefully acknowledge the support of the Office of Naval Research through
the award N00014-16-1-2735, and the faculty startup support from the department of
mechanical engineering at the University of Maryland, Baltimore County (UMBC). The hardware used in the computational studies is part of the
UMBC High Performance Computing Facility (HPCF).
The facility is supported by the U.S. National Science Foundation
through the MRI program
(grant nos.~CNS-0821258, CNS-1228778, and OAC-1726023)
and the SCREMS program (grant no.~DMS-0821311),
with additional substantial support from UMBC.

\bibliographystyle{ieeetr}
\bibliography{p_adaptive}

\end{document}